%% file: paper.tex
\documentclass[iop, numberedappendix]{emulateapj}
\shorttitle{Vertical shear instability}
\shortauthors{M-K.\ Lin, A.\ N.\ Youdin}

\usepackage{amsmath}
\usepackage{bm}
\newcommand{\p}{\partial}
\newcommand{\zmax}{z_\mathrm{max}}

\newcommand{\He}{\operatorname{He}}
 
\newcommand{\ii}{\mathrm{i}}

\newcommand{\zhat}{\hat{z}}
\newcommand{\khat}{\hat{k}}

\newcommand{\dd}{\delta}
\newcommand{\real}{\operatorname{Re}}
\newcommand{\imag}{\operatorname{Im}}

\defcitealias{lin12}{L12} 
\defcitealias{nelson13}{N13}
\defcitealias{barker15}{BL15}
\defcitealias{mcnally14}{MP14}
\defcitealias{goldreich67}{GS67}

\def \OmK {\Omega_{\rm K}}

\begin{document}

\title{Cooling Requirements for the Vertical Shear Instability 
  in Protoplanetary Disks}
\author{Min-Kai Lin\altaffilmark{1} \& Andrew N Youdin}
\affil{Department of Astronomy and Steward Observatory, \\
University of
  Arizona, 933 North Cherry Avenue, Tucson, AZ 85721, USA}
\altaffiltext{1}{Steward Theory Fellow}
\email{minkailin@email.arizona.edu, youdin@email.arizona.edu}

\begin{abstract}
The vertical shear instability (VSI) offers a potential hydrodynamic mechanism for angular momentum transport in protoplanetary disks (PPDs). The VSI is driven by a weak vertical gradient in the disk's orbital motion, but must overcome vertical buoyancy, a strongly stabilizing influence in cold disks, where heating is dominated by external irradiation.  Rapid radiative cooling reduces the effective buoyancy and allows the VSI to operate.  We quantify the cooling timescale $t_c$ needed for efficient VSI growth, through a linear analysis of the VSI with cooling in vertically global, radially local disk models. We find the VSI is most vigorous for rapid cooling with $t_c<\OmK^{-1}h|q|/(\gamma -1)$ in terms of the Keplerian orbital frequency, $\OmK$; the disk's aspect-ratio, $h\ll1$; the radial power-law temperature gradient, $q$; and the adiabatic index, $\gamma$.  For longer $t_c$, the VSI  is much less effective because growth slows and shifts to smaller length scales, which are more prone to viscous or turbulent decay.  We apply our results to PPD models where $t_c$ is determined by the opacity of dust grains.  We find that the VSI is most effective at intermediate radii, from $\sim5$AU to $\sim50$AU with a characteristic growth time of $\sim30$ local orbital periods. Growth is suppressed by long cooling times both in the opaque inner disk and the optically thin outer disk.  Reducing the dust opacity by a factor of 10 increases cooling times enough to quench the VSI at all disk radii.  Thus the formation of solid protoplanets, a sink for dust grains, can impede the VSI.

\end{abstract}

\section{Introduction}\label{intro}
Understanding how disks transport mass and angular momentum underlies 
many problems in astrophysics, including star and planet formation  
\citep[][]{armitage10}.  The turbulence associated with many transport mechanisms
is particularly important for dust evolution and planetesimal formation \citep{yl07, chiang10}. 

Magneto-hydrodynamic (MHD) turbulence driven by
the magneto-rotational instability \citep[MRI,][]{balbus91} has long been the most
promising transport mechanism in low mass disks with weak self-gravity. 
However, many parts of protoplanetary  
disks (PPDs) are cold, have low levels of ionization, and do not support the MRI 
\citep{blaes94,salmeron03}. Recent simulations
suggest that significant portions of PPDs fail to develop MHD
turbulence \citep[e.g.][]{simon13, lesur14,bai15,gressel15}. 

A purely hydrodynamic mechanism could circumvent difficulties with non-ideal MHD, but must overcome
the strong centrifugal stability imposed by the positive radial specific angular
momentum gradient in nearly Keplerian disks \citep{balbus96}.
One possible route to hydrodynamic turbulence is the vertical shear
instability (VSI, \citealp{urpin98, urpin03, nelson13}, hereafter \citetalias{nelson13}).  The
basic mechanism of the VSI  in disks was first 
identified in the context of differentially rotating stars (\citealp{goldreich67}, hereafter \citetalias{goldreich67}, \citealp{fricke68}).   
The VSI arises when vertical shear, i.e.\ a variation in orbital motion along the rotation axis, 
destabilizes inertial-gravity waves, which are oscillations 
with rotation and buoyancy as restoring forces. 
Vertical shear
occurs wherever the disk is baroclinic, i.e.\ when constant 
density and constant pressure surfaces are misaligned.  Baroclinicity, and thus vertical shear, is 
practically unavoidable in astrophysical disks, except at special locations like the
midplane. 

To overcome centrifugal stabilization, the VSI triggers motions which are vertically elongated 
and radially narrow.  
Vertical elongation taps the free energy of the vertical shear \citep{umurhan13}, but is also subject to 
the stabilizing effects of vertical buoyancy if the disk is stably
stratified. To overcome vertical buoyancy, the VSI requires a short cooling time
(\citetalias{goldreich67}; \citetalias{nelson13}). Rapid radiative 
cooling, i.e.\ a short thermal relaxation timescale, brings a
moving fluid element into thermal equilibrium with 
its surroundings, thereby diminishing buoyancy. 
An isothermal equation of state implies instantaneous thermal relaxation, and is the ideal context for studying the VSI
(\citealp{urpin03}; \citetalias{nelson13}; \citealp{mcnally14}, hereafter \citetalias{mcnally14}; \citealp{barker15},
hereafter \citetalias{barker15}).    

Alternatively, vertical buoyancy can be eliminated by strong internal
heating, i.e.\ by the onset of convection, so that the disk becomes 
neutrally stratified in the vertical direction
\citepalias{nelson13,barker15}. However, realistic PPDs should be vertically
stably stratified in the outer regions, beyond $\sim 1$--5 AU, where
heating is  irradiation dominated \citep{bitsch15}.  Even in the inner
disk, strong vertical buoyancy is possible if accretion heating is
weak or is concentrated in surface layers \citep{gammie96,lesniak11}.  

Understanding the VSI in real disks therefore requires
considering finite, non-zero cooling times, $t_{\rm c}$. Non-linear
hydrodynamical simulations with a prescribed  $t_{\rm c}$ find that VSI
turbulence in stably stratified disks requires rapid cooling  
with $t_{\rm c}$ shorter than orbital timescales
 \citepalias{nelson13}. When the cooling time is short enough, 
 the VSI can drive moderately strong transport, with Reynolds
stresses of $\alpha \sim 10^{-3}$ times the mean pressure \citepalias{nelson13}. Simulations
with realistic radiative transfer, in lieu of a fixed $t_{\rm
  c}$, find that the VSI in irradiated disks drive transport with
$\alpha \sim 10^{-4}$  in a  $\sim 2$ -- 10AU PPD model \citep{stoll14}.  

Studying the linear growth of the VSI is necessary for 
understanding how it can ultimately drive turbulent transport.
The pioneering linear analyses of the VSI considered vertically 
local disturbances \citep{urpin98,urpin03}.  A vertical global analysis 
(e.g.\ \citetalias{nelson13,mcnally14,barker15}) 
is essential for understanding how vertically elongated disturbances 
interact with the disk's vertical structure.  Moreover a vertically global analysis
allows more direct comparisons with modern numerical simulations.  This work 
generalizes previous vertically global analyses by including a finite cooling time.


This paper is organized as follows. In \S\ref{vsi_require} we explain, without derivation,  
our main result for the critical cooling timescale, beyond which VSI growth is 
suppressed. 
 We develop our disk model in \S\ref{setup} and derive linearized
 perturbation equations in  \S\ref{linear}.
 Section \ref{analytical} contains our main analytic results, leading to 
 the derivation of the critical cooling time in \S\ref{iso_vsi_beta_crit}.
 In \S\ref{numerical} we analyze linear VSI growth, and numerically confirm the critical cooling time.
We apply our results to PPDs in \S\ref{application}, with cooling times derived from dust opacities.
We discuss caveats and extensions in \S\ref{caveats}.
 We conclude in \S\ref{summary} with a summary.  Some technical and background 
 developments are explained in the appendices.

\section{Why must cooling be so fast?}\label{vsi_require}     
In a stably stratified thin disk, the VSI  
requires a thermal timescale $t_c$ significantly shorter than the
disk dynamical timescale, 
\begin{align}
  t _c \ll \OmK^{-1},
\end{align}
where $\OmK$ is the Keplerian frequency (N13).    
This requirement --- that the cooling time be much shorter than an orbital period, 
which in turn is much smaller than the relevant oscillation period of vertically elongated gravity waves  --- is quite stringent.
It highlights the fact that vertical buoyancy is strongly 
stabilizing. Rapid radiative damping is thus required to weaken
buoyancy and allow the weak vertical shear to drive instability.   

To roughly quantify the required smallness of $t_c$, we consider a
vertically isothermal disk with aspect-ratio $ h$, 
 radial temperature profile $T \propto r^q$ and adiabatic index
$\gamma>1$ so the disk is stably stratified.  For a PPD, $ h \sim 0.05$,
$q\sim -0.5$ and $\gamma\simeq 1.4$.  

\begin{figure*}
  \includegraphics[width=\linewidth]{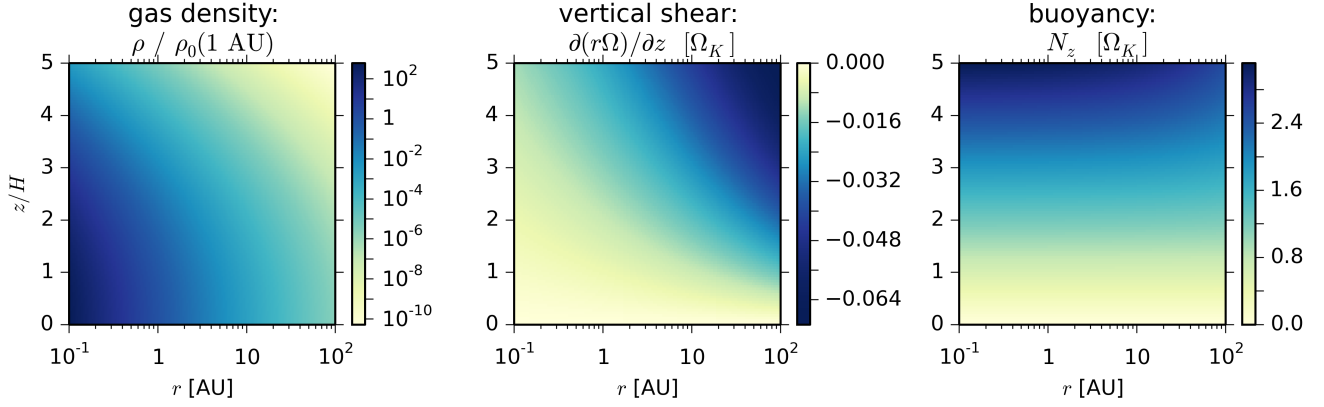}
  \caption{The radial and vertical structure of a `minimum mass' PPD model showing gas density (left, relative to the midplane density at 1 AU), vertical shear rate (middle) and
    buoyancy frequency (right).
    \label{eqm_structure} 
  }
\end{figure*}

For a thin disk ($h\ll 1$), the vertical shear rate varies with height, $z$, from the midplane as 
\begin{align}\label{vshear_thin}
  r \frac{\p \Omega }{\p z} \simeq  h q \OmK \left(\frac{z}{2 H}\right),
\end{align}
where $H=hr$ is the characteristic disk scale height. 

This destabilizing shear competes with the stabilizing vertical
buoyancy frequency, $N_z$.  In a thin disk,     
\begin{align}\label{nz_thin}
  N_z^2 \simeq \left(\frac{\gamma-1}{\gamma}\right) \left(\frac{z}{H}\right)^2
  \OmK^2.  
\end{align}
Fig. \ref{eqm_structure} maps the vertical shear rate and buoyancy
frequency (as well as the gas density) 
in a fiducial PPD model, see \S\ref{toy_relax} for details. 

Vertical shear is 
generally weak compared to buoyancy, suggesting stability. In fact, 
without any cooling, the Solberg-Hoiland criteria 
confirms that vertical buoyancy is strong
enough to ensure (axisymmetric) stability, see \S\ref{solberg}.  With radiative cooling, 
thermal fluctuations decay which, combined with pressure equilibration, reduces the effective buoyancy.

How short must $t_c$ be for vertical shear to prevail?  We start with 
 the Richardson number $\mathrm{Ri} = N_z^2/(r\p_z\Omega)^2$, a 
ratio of buoyant to shear energies.  Though not precisely applicable to our problem,
non-rotating shear flows are stable if $\mathrm{Ri} > 1/4$ (\citealp{chandrasekhar61}, 
also see \citealp{ys02, lee10} for applications to thin dust layers in PPDs).
Following \cite{urpin03} and \cite{townsend58},  we reduce the buoyant energy by the 
ratio of cooling to forcing timescales, $t_c |r \p_z\Omega| \leq 1$ (with no reduction in buoyant energy 
expected when this inequality fails).  With this reduction, the Richardson-like criterion for 
shear instability becomes  \citep{urpin03}
\begin{align}\label{bcrit_est}
  t_c \lesssim \frac{\left|r\p\Omega/\p z\right|} {N_z^2}.
\end{align}
If we interpret Eq.\ \ref{bcrit_est} as a local criterion, then we see that 
the maximum cooling time which permits instability decreases with height as $1/|z|$;  
the stabilizing effect of vertical buoyancy increases more rapidly away from
the midplane than the destabilizing effect of vertical shear.  
This finding is relevant to the radiative damping of so called `surface 
modes', which we consider in \S\ref{numerical}.

We are more interested, however, in the global instability criterion for disturbances at all heights.
Our key result 
is the cooling requirement 
\begin{align}\label{prelim_bcrit}
  t_c <  &\frac{h |q| }{\gamma - 1}\OmK^{-1}
\end{align}   
for VSI growth that we derive in \S\ref{iso_vsi_beta_crit}.  We can simply 
(but non-rigorously) obtain Eq. \ref{prelim_bcrit}
by evaluating  Eq. \ref{bcrit_est} at $|z|=\gamma H/2$. 
 
\begin{figure}
  \includegraphics[width=\linewidth,clip=true,trim=0cm 1.7cm 0cm
  0.73cm]{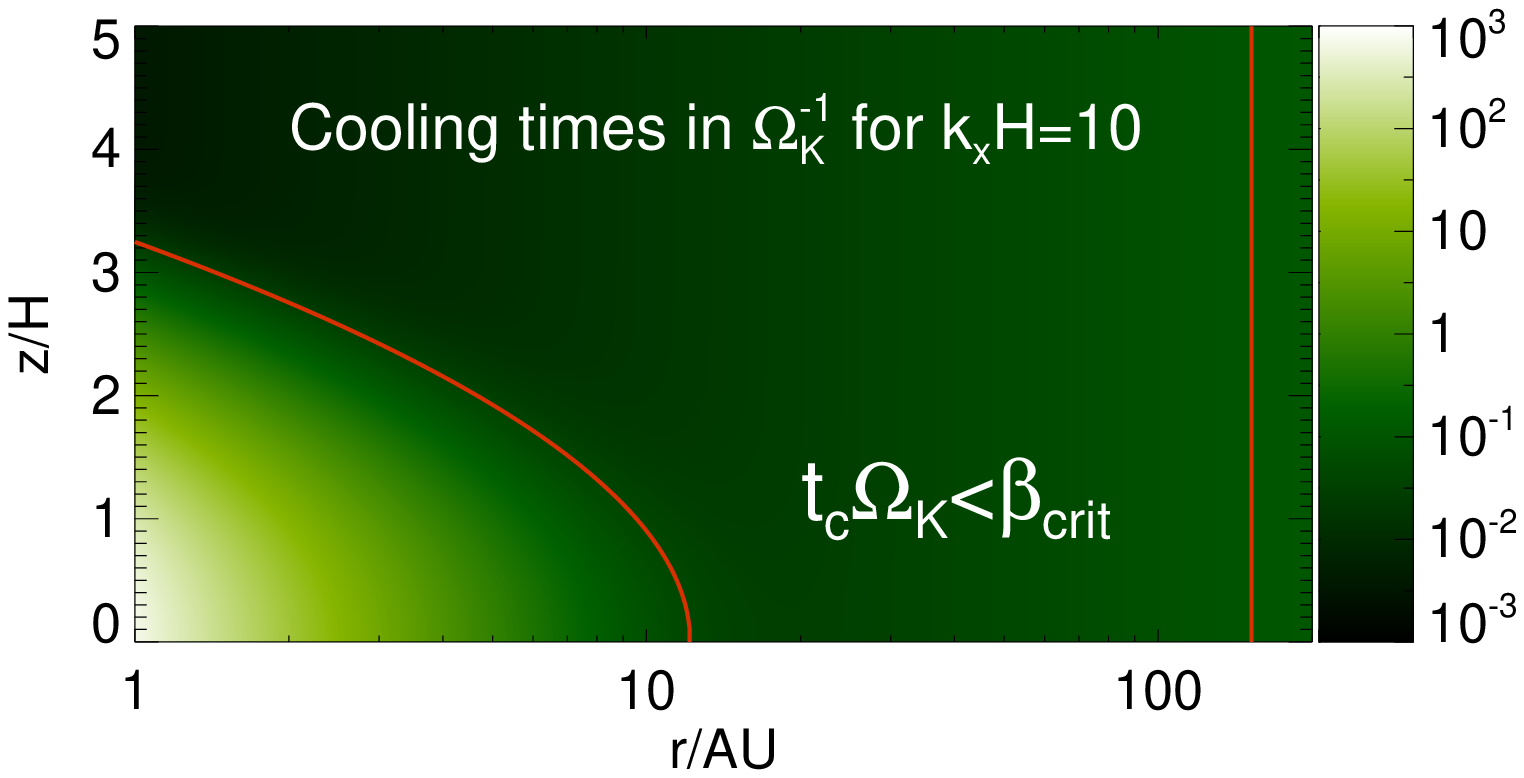}
  \includegraphics[width=\linewidth,clip=true,trim=0cm 0.46cm 0cm
  0.73cm]{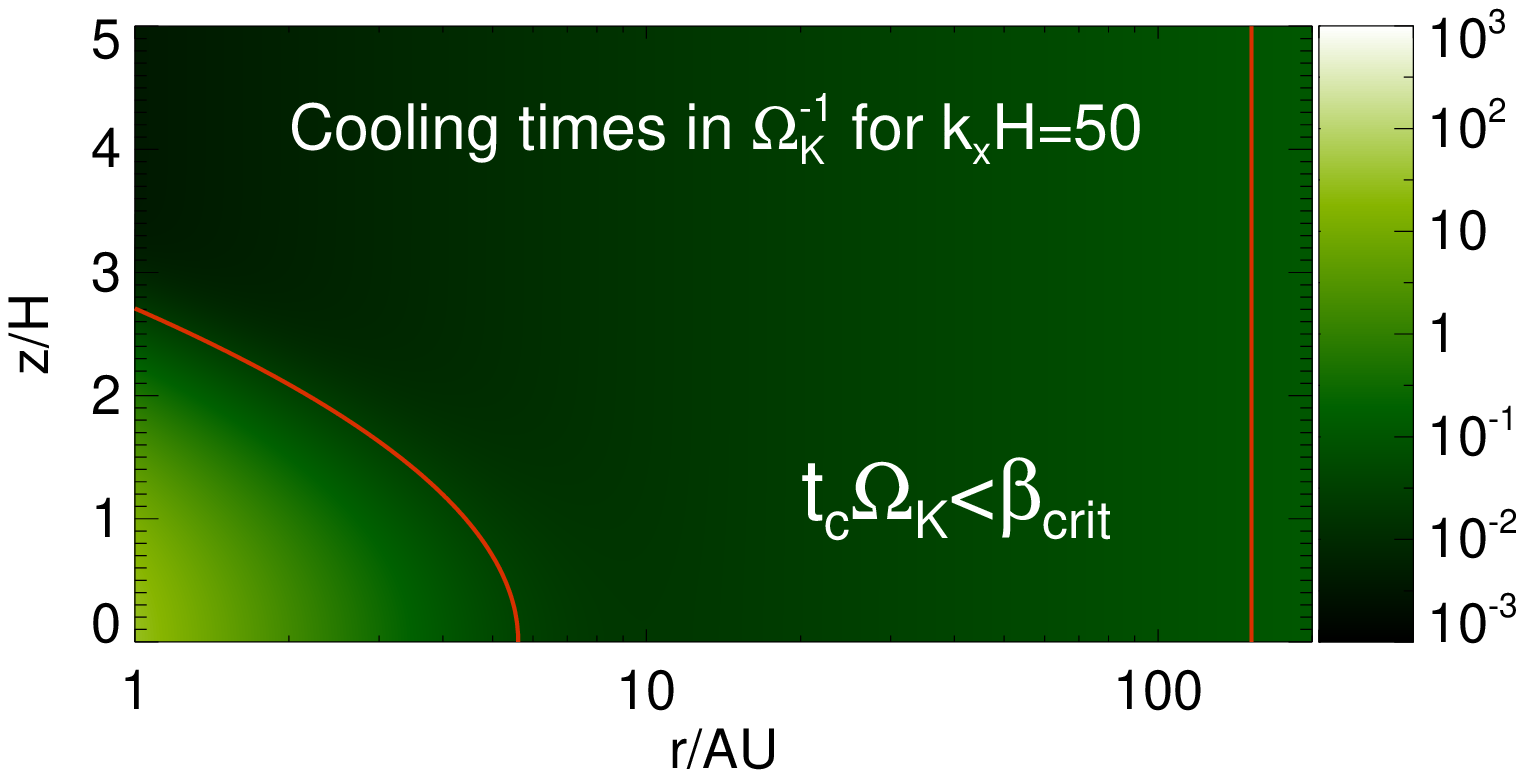}
  \caption{Estimate of the thermal timescale $t_c$  in a fiducial 
    PPD model, the Minimum Mass Solar Nebula. Here, $\beta_\mathrm{crit}=h|q|/(\gamma-1)$  
    is a maximum dimensionless cooling timescale below which the VSI
    is expected to operate effectively. This corresponds to   
    the region bounded by the red lines. 
    \label{bcrit_mmsn2d} 
  }
\end{figure}

We now consider whether, and where, thermal timescales in realistic PPDs
are short enough to support VSI growth.  Fig. \ref{bcrit_mmsn2d} plots the
 thermal timescales in our fiducial PPD model, see \S\ref{toy_relax} for details.
 The regions between the red lines satisfy the cooling requirement of Eq.\ \ref{prelim_bcrit}.
 Even without a detailed analysis we can expect VSI growth between $\sim 10$ -- 150AU.
 The inner disk is complicated by the fact that the optically thick
 midplane has longer cooling times.  
 Thus the cooling criterion fails in the midplane but is satisfied
 above.  This case requires the  
 more detailed analysis of \S\ref{application}.
 
  Fig. \ref{bcrit_mmsn2d} also compares two different radial wavenumbers, 
  $k_x = 10/H$ and $50/H$.  In the inner disk, the higher wavenumber
  mode cools faster, due to radiative diffusion. 
  Thus in the inner disk, shorter wavelength VSI modes are more likely
  to grow. In the outer disk, the cooling time is the  
  same for different $k_x$ and $z$, because optically thin cooling is
  independent of both lengthscale and density. 
  Thus the outer barrier to growth, beyond $\sim 150$ AU in this
  model, is independent of  wavelength.  
This example highlights the fact that optically thin cooling sets a
lower limit to the cooling time; a fact that is sometimes obscured
when  
the radiative diffusion approximation is made at the outset.

\input{setup}

\input{linear_iso}

\input{linear_adia}

\input{numerical}

\input{application} 
\input{summary}

\appendix
\input{appendix}

\bibliographystyle{apj}
\bibliography{ref}

\end{document}

%% file: setup.tex
\section{Governing equations and disk models}\label{setup}
An inviscid, non-self-gravitating disk orbiting a central star of mass $M_*$
obeys the three-dimensional fluid equations:
\begin{align}
  &\frac{\p\rho}{\p t} + \nabla\cdot{\left(\rho\bm{v}\right)} = 0,\label{full_mass}\\
  &\frac{\p\bm{v}}{\p t} + \bm{v}\cdot\nabla\bm{v} =
  -\frac{1}{\rho}\nabla P - \nabla\Phi_*,\\
  &\frac{\p P}{\p t} + \bm{v}\cdot\nabla P  = - \gamma P
  \nabla\cdot\bm{v} - \Lambda 
  , \label{full_energy}
\end{align}
where $\rho$ is the mass density, $\bm{v}$ is the velocity field (the
rotation frequency being $\Omega=v_\phi/r$), $P$
is the pressure, $\gamma$ is the constant adiabatic index, and $\Phi_*
= -GM_*(r^2 + z^2)^{-1/2}$ is the gravitational potential of the
central star with $G$ as the gravitational constant. The cylindrical
co-ordinates $(r,\phi, z)$ are centered on the star. 
In the energy equation (Eq.\ \ref{full_energy}) the sink term
$\Lambda$ includes non-adiabatic cooling and (negative) heating. 
The gas temperature $T$ obeys the ideal gas 
equation of state,  
\begin{align}
P = \frac{\mathcal{R}}{\mu}\rho T,
\end{align}
where $\mathcal{R}$ is the gas constant and $\mu$ is the mean molecular
weight.    

\subsection{Thermal relaxation}
We model radiative effects as thermal
relaxation with a timescale $t_c$, i.e.,
\begin{align}\label{thermal_relax}
  \Lambda  \equiv \frac{\rho\mathcal{R}}{\mu}\frac{T - T_\mathrm{eq}}{t_c},
\end{align}
where $T_\mathrm{eq}=T_\mathrm{eq}(r,z)$ is the equilibrium temperature.  We 
define the dimensionless cooling time 
\begin{align}\label{beta_def}
  \beta \equiv \OmK t_c
\end{align}
relative to the Keplerian frequency, $\OmK=\sqrt{GM_*/r^3}$.  We 
use the terms `thermal relaxation' and `cooling' interchangeably
throughout this paper.   

For most of this paper we will take $\beta$ as a constant input
parameter, so that we can vary the thermodynamic response of the
disk from isothermal ($\beta\ll 1$) to adiabatic ($\beta \gg 1$) in a
controlled manner. However, in \S\ref{application} 
we will consider a more realistic model for $\beta$ in
PPDs with a dust opacity (as in Fig.\  \ref{bcrit_mmsn2d}).  
 
 


\subsection{Baroclinic disk equilibria}\label{eqm}
The equilibrium disk is steady and axisymmetric with density,
pressure and rotation profiles $\rho(r,z)$, $P(r,z)$ and
$\Omega(r,z)$, respectively. These functions satisfy  
\begin{subequations}\begin{align}
  0 &= \frac{1}{\rho}\frac{\p P}{\p z} + \frac{\p \Phi_*}{\p z},\label{vert_equilibrium}\\
  r\Omega^2&= \frac{1}{\rho}\frac{\p P}{\p r} + \frac{\p \Phi_*}{\p
    r}.\label{hori_equilibrium} 
\end{align} \end{subequations}
A unique solution requires additional assumptions about disk structure.  We choose
\begin{subequations}\begin{align}
 \rho(r,0) &\equiv \rho_0(r) = \rho_{00}\left(\frac{r}{r_0}\right)^p, \\
 T(r,0) &\equiv T_0(r) = T_{00}\left(\frac{r}{r_0}\right)^q, \\
 P(r,z) & = K(r)\rho^{\Gamma}(r,z). \label{eqm_press}
\end{align} \end{subequations}
where $r_0$ is a fiducial radius and $p$ and $q$ are the standard power-law indices
for midplane density and temperature, respectively. Without loss of
generality we take $q<0$, as is typical in PPDs.   
The polytropic index $\Gamma$ 
parametrizes the vertical stratification with $\Gamma = 1$ ($\Gamma = \gamma$)
describing vertically isothermal (adiabatic) disks, respectively.  The ideal gas
law requires $K = \rho_0^{1-\Gamma}\mathcal{R} T_0/\mu$.



We further define a modified sound speed
\begin{align}
	c_s \equiv \sqrt{\Gamma P /\rho } , 
\end{align}
which in general differs from the  isothermal, $c_\mathrm{iso} =
c_s/\sqrt{\Gamma}$, and adiabatic $c_\mathrm{ad} =
c_s\sqrt{\gamma/\Gamma}$, sound speeds. 
We also introduce the characteristic scale-height, $H = c_s(r,0)/\OmK$
and disk aspect-ratio
\begin{align}
   h(r) \equiv 
  \frac{H}{r} \propto r^{(1+q)/2}
\end{align}
We are interested in thin disks with $ h \ll 1$.

\subsubsection{Density structure}
The equilibrium density field follows from vertical hydrostatic equilibrium (Eq.\ \ref{vert_equilibrium}).
The solution for $\Gamma\neq1$ is
\begin{align}\label{eqm_dens}
  &\left(\frac{\rho}{\rho_0}\right)^{\Gamma-1} = 1 +
  \frac{\left(\Gamma-1\right)}{ h^2}\left(\frac{1}{\sqrt{1+z^2/r^2}}-1\right).
\end{align}
Note that for $\Gamma > 1 +  h^2$ there is a disk surface $H_s$
where $\rho(r,H_s)=0$; and $H_s \simeq \sqrt{2/(\Gamma-1)}H$ for $h\ll 1$.

The vertically isothermal case, $\Gamma = 1$, can be calculated either 
as a special case or by taking the limit of Eq. \ref{eqm_dens} as $\Gamma\to 1$,    
\begin{subequations}\begin{align}
 \frac{\rho(r,z)}{\rho_0(r)} &=
  \exp{\left[\frac{1}{ h^2}\left(\frac{1}{\sqrt{1+z^2/r^2}}-1\right)\right]},   \\
  & \simeq \exp {\left( - {z^2 \over 2 H^2}\right)} \quad\text{for } |z|\ll r \label{rhoisothin}. 
  \end{align}\end{subequations}
Eq.\ \ref{rhoisothin} is the approximate form of the density field in the thin-disk limit.  
We will primarily focus on 
vertically isothermal disks, the relevant case for passively irradiated
PPDs \citep{chiang97}.

 \subsubsection{Rotation and vertical shear profiles}\label{vshear_def}
The equilibrium rotation field, $\Omega(r,z)$ follows from the density field and centrifugal balance (Eq.\ \ref{hori_equilibrium}),
giving 
\begin{align}\label{eqm_rot}
  \frac{\Omega^2(r,z)}{\OmK^2(r)}=1 +
  \frac{p+q}{\Gamma} h^2(r) 
  +\frac{s}{\Gamma} \left(1-\frac{1}{\sqrt{1+z^2/r^2}}\right), 
\end{align}
for all $\Gamma$, where $s\equiv q+p(1-\Gamma)$.  We also refer to the
epicyclic frequency $\kappa$ defined via
\begin{align}\label{kap2_def}
\kappa^2 \equiv \frac{1}{r^3}\frac{\p}{\p r}\left(r^4\Omega^2\right) = \frac{1}{r^3}\frac{\p j^2}{\p r}, 
\end{align}
where $j$ is the specific angular momentum. 

The vertical shear rate follows from Eq. \ref{eqm_rot}, 
\begin{align}
  r\frac{\p\Omega^2}{\p z} = \frac{sz}{\Gamma r}  \frac{\OmK^2}{
    \left(1+z^2/r^2\right)^{3/2}} . \label{vertical_shear_ex} 
\end{align}
For $\Gamma = 1$, notice that $s/\Gamma = q$ is the only disk parameter that affects vertical shear.
Vertical shear increases linearly with height near the midplane, and $|\p_z \Omega^2|$ is maximized at
$\mathrm{min}(r/\sqrt{2},H_s)$, which is expected to limit the maximum VSI growth rate. 

A more general expression for vertical shear holds for vertical polytropes 
(which satisfy Eq.\ \ref{eqm_press}), with no assumed radial structure:
\begin{align}
  r\frac{\p\Omega^2}{\p z} = - \rho^{\Gamma-1}\frac{\p\ln\rho}{\p
    z}\frac{dK}{dr} = - \frac{g_z}{\Gamma}\frac{d\ln K}{dr}, \label{vertical_shear}
\end{align}
where $g_z = \p_z P/\rho$.



\subsection{Entropy gradients and vertical buoyancy}
The gradients of specific entropy, $S\equiv
C_P\ln{P^{1/\gamma}/\rho}$, in our disk models are
\begin{subequations}
\begin{align}
  &\frac{\p S}{\p r} = C_P\left[\frac{s}{\gamma r} +
    \left(\frac{\Gamma}{\gamma}-1\right)\frac{\p \ln{\rho}}{\p
       r}\right], \label{dSdr}\\
  &\frac{\p S}{\p z} = C_P
  \left(\frac{\Gamma}{\gamma}-1\right)\frac{\p \ln{\rho}}{\p z} = {C_P g_z \over c_s^2} {\Gamma - \gamma \over \gamma},\label{dSdz}
\end{align} 
\end{subequations}
 where $C_P$ is the heat capacity at
constant pressure. 

The vertical buoyancy frequency $N_z$ is 
\begin{align}
  N_z^2 \equiv - C_P^{-1}g_z \frac{\p S}{\p z} = {g_z^2 \over c_s^2}{\gamma - \Gamma \over \gamma}. 
    \label{nzsq_def}
\end{align}
We only consider convectively stable disks with with $\Gamma <
\gamma$ and  $N_z^2\geq0$. 

\subsection{Stability without cooling}\label{solberg}
In the absence of cooling, axisymmetric stability is ensured
if both Solberg-Hoiland criteria  are satisfied:
\begin{subequations}\begin{align}
  \kappa^2 - \frac{1}{\rho C_P}\nabla P \cdot \nabla S &> 0,\\
  -\frac{\p P}{\p z}\left(\frac{\p j^2}{\p r}\frac{\p S}{ \p z} -
    \frac{\p j^2}{\p z}\frac{\p S}{\p r} \right)&>0 \label{second_SH} 
\end{align}\end{subequations}
\citep{tassoul78}. 
For rotationally-supported, thin disks 
the first criterion is easily satisfied. 
Thus, we consider the second
criterion. 
Without loss of generality (since Eq. \ref{second_SH} is even in $z$), we consider $z>0$
so that $\p_zP<0$. With $\p_r j^2\simeq r^3\OmK^2$ and
using Eq. \ref{vertical_shear} for $\p_zj^2$  we find that
\begin{align}\label{solberg2}
  \gamma - \Gamma > \frac{ h^2}{\Gamma}
  \left(\frac{\rho}{\rho_0}\right)^{\Gamma-1} \left[s^2
    -s\left(\gamma-\Gamma\right)\frac{\p\ln{\rho}}{\p\ln{r}} \right]
\end{align} 
implies stability. For typical model parameters, the right hand
side is $O(h^2) \lesssim 10^{-2}$. The left hand side is positive and order
unity 
  in our disk models (with $\gamma-\Gamma\geq0.1$)
  implying strong stability to convection, see Eq.\ \ref{nzsq_def}. 
  This stable stratification is
expected for irradiated disks  \citep{chiang97}. 
Thus, adiabatic disturbances in a standard, irradiated disk
 are stable to a disk's vertical shear,
explaining why the VSI requires cooling. 

\section{Linear problem}\label{linear} 
This section presents the two sets of equations we use to study the
linear development of the VSI.  Both sets are radially local and vertically global 
with a finite cooling time.
The first set, presented in \S\ref{sec:radlocal}, is more general and used 
for numerical calculations.  The second set, in \S\ref{sec:simplified}, makes
additional approximations  about disk structure and wave frequency.  This simplified
set is used to obtain analytic results.

\subsection{Radially local approximation}\label{sec:radlocal}
We consider axisymmetric perturbations to the above disk equilibria.    
The growth of the VSI is strongest for short radial wavelengths, as compared
 to the disk radius \citepalias{nelson13,barker15}.  We thus perform a
 standard two step process to obtain linearized equations in the radially local 
 approximation (e.g., \citealp{goldreich67}, who also consider vertically localized perturbations).  
 
 We first expand all fluid variables 
 into the equilibrium value plus an Eulerian perturbation denoted by $\delta$, e.g.\ 
 $\delta \rho$ for the perturbed density, and drop all non-linear perturbations.  Second, we perform
 a Taylor expansion about a fiducial radius $r_0$ with the local radial coordinate
 $x = r - r_0$, keeping only the leading order terms in $x/r_0$.  We also relabel 
 (trivially for axisymmetric perturbations) the azimuthal direction as $y$.  
 
 Perturbations take the form of a radial plane wave with arbitrary vertical dependence, 
 e.g.\
 \begin{align}
  \delta\rho \rightarrow \delta\rho(z)\exp{\left(\ii k_x x - \ii\upsilon
      t\right)},    
\end{align}
where $\upsilon = \omega + \ii \sigma$ is a (generally) complex frequency, 
with $\omega$ and $\sigma$ being
the real frequency and growth rate, respectively, and $k_x$ is a real 
radial wavenumber.  We take $k_x>0$ without loss of generality, 
assume $k_xr_0 \gg 1$ and neglect curvature terms. 
Henceforth all unperturbed fluid variables, including 
gradients such as $\p_z \Omega$, refer to  equilibrium values at $(r_0, z)$.

We further define the perturbation variables  $W \equiv \delta P /\rho$ 
and $Q \equiv c_s^2\delta\rho/\rho$.  With this procedure the linearized
system of equations are
\begin{subequations}\label{lin_all}\begin{align}
  \ii\upsilon \frac{Q}{c_s^2}  &=  \ii k_x \delta v_x + \frac{d\dd
    v_z}{dz} + \frac{\p\ln{\rho}}{\p z}\delta v_z + \zeta
  \frac{\p\ln{\rho}}{\p r} \delta v_x,\label{lin_mass}\\
  \ii\upsilon \delta v_x  &= \ii k_x W - 2\Omega\delta v_y -
  \zeta\frac{1}{\rho}\frac{\p P}{\p r}\frac{Q}{c_s^2}\label{lin_vx}\\
   \ii \upsilon\delta v_y &= r_0\frac{\p \Omega}{\p z}\delta v_z +
  \frac{\kappa^2}{2\Omega}\delta v_x, \label{lin_vy}\\
   \ii\upsilon\delta v_z &= \frac{dW}{dz} +
  \frac{\p\ln{\rho}}{\p z}\left(W-Q\right), \label{lin_vz}\\
  \ii \upsilon W &= c_s^2\frac{\p\ln{\rho}}{\p z}\delta v_z +
  c_s^2\frac{\gamma}{\Gamma} \left(\ii k_x\delta v_x + \frac{d\delta
      v_z}{dz}\right) \notag\\
  &+ \frac{1}{t_c}\left(W - \frac{Q}{\Gamma}\right) +
  \zeta\frac{1}{\rho}\frac{\p P}{\p r}\delta v_x.\label{lin_energy}
\end{align}\end{subequations}
This is a set of ordinary differential equations (ODEs) in
$z$.  
Solutions to these 
are presented in \S\ref{numerical} and \S\ref{application}.   The coefficient $ \zeta = 1$ is 
introduced simply to label terms with explicit radial gradients of the equilibrium state, 
which again are evaluated at $(r_0, z)$. For clarity, hereafter we drop the subscript $0$.    
We will consider the effects of ignoring these radial gradients below.

%% file: linear_iso.tex
\subsection{Reduced equation for vertically isothermal disks}\label{sec:simplified}

Our simplified model starts with  Eqs.\ \ref{lin_all} 
and makes the following additional simplifications:

\begin{enumerate}
 
 \item We set $\Gamma = 1$, focusing on vertically isothermal disks.
  
\item  We set $\zeta = 0$,  neglecting terms with an explicit dependence on the 
radial structure of the equilibrium disk.  Vertical shear, which implicitly depends on the radial
temperature gradient, is retained.  This fully-radially-local approximation is also made in the
 `vertically global shearing box' of \citetalias{mcnally14}. 

\item We make the low frequency approximation, assuming $|\upsilon^2|\ll \kappa^2,\,\Omega^2$. 
Similar to the incompressible \citepalias{goldreich67} or anelastic \citepalias{nelson13,barker15} approximations,
the low frequency approximation filters acoustic waves in favor of
inertial-gravity waves \citep{lubow93}.  

\item We make the Keplerian approximation, setting $\Omega \rightarrow \OmK$ 
and $\kappa \rightarrow \OmK$, but retaining the vertical dependence 
in the crucial vertical shear term,
  $\p_z\Omega$. 
  
\item We consider thin disks with $h\ll 1$ 
  and use the Gaussian approximation, Eq.\ \ref{rhoisothin}, for the equilibrium density field.  

\end{enumerate}

 In terms of the dimensionless variables
\begin{align}\label{nondim}
  \hat{z} = z/H,\quad \hat{k}=k_xH, \quad \hat{\upsilon} =\hat{\omega} +
  \ii\hat{\sigma}= \upsilon/\OmK,
\end{align}
where $\hat{\omega}$ and $\hat{\sigma}$ are real, the above
approximations lead to a single second order ODE, 
\begin{align}
  \delta v_z^{\prime\prime} - z A\delta v_z^\prime +
  (B - C\zhat^2)\delta v_z = 0,\label{nearly_iso_explicit}
\end{align}
where $^\prime$ denotes $d/d\zhat$ and 
\begin{subequations}\begin{align}
  &A \equiv 1 + \ii h q \hat{k},\\
  &B \equiv \hat{\upsilon}^2\left(\chi + \hat{k}^2\right) -
  \left(\chi + \ii h q \hat{k}\right),\\
  &C \equiv \left(1-\chi\right)\left(\hat{k}^2 - \ii
    h q\hat{k}\right), 
\end{align}\end{subequations}
where
\begin{align}\label{chi}
\chi = \frac{1-\ii\hat{\upsilon}\beta}{1-\ii\hat{\upsilon}\beta\gamma}.
\end{align}
The derivation of Eq. \ref{nearly_iso_explicit}  is detailed in
Appendix \ref{adia_improve}. 
 

 In Appendix \ref{global_corr}  we discuss the limits of these approximations.
In particular we point out that the fully-radially-local approximation is only valid for
short cooling times, $\beta \lesssim O(1)$, which is the 
  regime in which the VSI operates, as demonstrated below. 
  On the other hand, for 
  $\beta\gtrsim O(1)$, this approximation can introduce artificial
  instability due to the non-self-consistent neglect of global radial gradients (see \S\ref{analytic_adia}).

%% file: linear_adia.tex
\section{Analytic results with finite cooling times}\label{analytical}  

Previous analytic studies of the VSI have largely focused on isothermal
perturbations, with infinitely rapid cooling, as discussed in the introduction.  We further explore
this $\beta = 0$ limit in Appendix \ref{iso_discuss} both to further develop intuition for this 
idealized case and to establish a connection with previous 
works.  However our main interest is the effect of finite cooling times, which
we explore analytically in this section.  

In \S\ref{relax_pert} we show that even an infinitesimal increase in the cooling time,
starting from $\beta = 0$, slows the growth rate of the VSI.  In \S\ref{disp_relax} we 
derive exact solutions to the simplified VSI model developed above (Eq.\ \ref{nearly_iso_explicit}).
In \S\ref{iso_vsi_beta_crit} we reach our main result, the maximum cooling time above which 
VSI growth is suppressed.

%

\subsection{Introducing a small but finite
  cooling time}\label{relax_pert}
We are interested in finite thermal relaxation
timescales $\beta > 0$, but it is instructive to
first ask  the more analytically tractable question: how 
do the eigenfrequencies and eigenfunctions change when we change
$\beta$ from zero to a small but finite value? For 
sufficiently small $\beta$ we expect the solution to only differ
slightly from a case with $\beta\equiv 0$. We thus perturb a solution
for $\beta=0$ to see the effect of finite cooling.  

For definiteness, let us consider the simple solution 
\begin{align}
  \beta\equiv 0, \quad \delta v_z = 1,\quad \hat{\upsilon}^2 = \frac{1 +
  \ii h q \hat{k}}{1+\hat{k}^2}, \label{pert_basic} 
\end{align}
which solves Eq. \ref{nearly_iso_explicit} since $\delta
v_z=$constant and $\chi=1$ so that $B=C=0$. 
This is the lowest order VSI mode, the `fundamental corrugation mode',
where the vertical velocity is constant throughout the disk
column. Hereafter we shall simply call it the fundamental
mode. 

The fundamental mode has been observed to dominate numerical simulations
\citep[\citetalias{nelson13}]{stoll14}, and are the ones we find to typically 
dominate in numerical calculations with increasing $\beta$
(\S\ref{therm_relax_eff}) for moderate    
values of $\khat$, as well as in PPDs 
with a realistic estimate for thermal timescales
(\S\ref{application}). We will see later that consideration of low
order modes in fact provides a useful way to characterize the effect
of increasing the thermal timescale on the VSI. 

We linearize Eq. \ref{nearly_iso_explicit} about the above
solution for $\beta\equiv0$ and write 
\begin{align}\label{nearly_iso_pert}
  \beta \to 0 + \delta\beta,\, \delta v_z\to \delta v_z+\delta
  v_{z1},\,\hat{\upsilon} \to \hat{\upsilon} + \delta\hat{\upsilon}, 
\end{align}
which implies 
\begin{align}
  \chi \to 1 + \delta\chi = 1 + \ii \hat{\upsilon}\left(\gamma-1\right)\delta\beta,
\end{align}
with $\delta v_z$ and $\hat{\upsilon}$ given by Eq. \ref{pert_basic}. We
may then seek  
\begin{align}
  \delta v_{z1} = a \zhat^2 + b,
\end{align}
where $a$, $b$ are constants. 

We insert Eq. \ref{nearly_iso_pert} into
Eq. \ref{nearly_iso_explicit}, keeping only first order terms, and
solve for $\delta\hat{\upsilon}$ using the above expressions for $\delta\chi$
and $\hat{\upsilon}$. We find imaginary part of $\delta\hat{\upsilon}$
is 
\begin{align}
  \delta\hat{\sigma} =
  -\frac{\left(\gamma-1\right)\hat{k}^2 \left(\hat{k}^2 -
      2h^2q^2\hat{k}^2 - h^2q^2\right)}{2\left(1+h^2 q^2
      \hat{k}^2\right)\left(1+\hat{k}^2\right)^2}\delta\beta.  
\end{align}
Since $h \ll 1$, introducing finite cooling $\delta\beta>0$
implies $\delta\hat{\sigma} < 0$, i.e. stabilization, since $\gamma>1$. 

A finite cooling time allow buoyancy
forces to stabilize vertical motions in sub-adiabatically stratified
disks. The $\zhat^2$ dependence in $\delta v_{z1}$ makes  
sense because it becomes significant at large $|\zhat|$, i.e. away from
the midplane where the effect of buoyancy first appears as $\beta$ is
increased from zero (since $N_z^2\propto z^2$ for a thin disk, see
Eq. \ref{nzsq_def}).   

\subsection{Explicit solutions and dispersion relation}\label{disp_relax}
We now solve Eq. \ref{nearly_iso_explicit} explicitly. We first write  
\begin{align}
  \delta v_z(\zhat) =
  g(\zhat)\exp\left(\frac{\lambda\zhat^2}{2}\right), \label{adia_ansatz}
\end{align}
where $\lambda$ is a constant to be chosen for convenience. Inserting
Eq. \ref{adia_ansatz} into Eq. \ref{nearly_iso_explicit} gives
\begin{align}
  0 = g^{\prime\prime} - \hat{z}\left(A - 2\lambda\right)g^\prime + \left(B +
    \lambda\right)g
  +\left(\lambda^2 - \lambda A - C\right)\zhat^2 g.
\end{align}
We choose $\lambda$ to make the coefficient of $\zhat^2g$
vanish, and impose the vertical kinetic energy density
$\rho|\delta v_z|^2\propto |g|^2 \exp{\left(\real\lambda -
    1/2\right)\zhat^2}$ to remain finite as $|\zhat|\to\infty$. 
Then assuming $g(\zhat)$ is a polynomial, we require  
\begin{align}
  \real\lambda < \frac{1}{2}, 
\end{align}
which amounts to choosing 
\begin{align}
  \lambda = \frac{1}{2}\left(A - \sqrt{A^2 + 4C}\right).  
\end{align}

Eq. \ref{nearly_iso_explicit} becomes 
\begin{align}
  0 = g^{\prime\prime} - \hat{z}\left(A - 2\lambda\right)g^\prime +
  \left(B + \lambda\right)g.
\end{align}
We seek polynomial solutions 
\begin{align}
  g(\zhat) = \sum_{m=0}^M b_m \zhat^m,
\end{align}
which requires
\begin{align}
  B(\hat{\upsilon}) + \lambda(\hat{\upsilon}) =
  M\left[A-2\lambda(\hat{\upsilon})\right].\label{adia_disp0} 
\end{align}
For ease of analysis, we re-arrange Eq. \ref{adia_disp0} and square it
to obtain a polynomial in $\hat{\upsilon}$,  
\begin{align}
  0 = \sum_{l=0}^{6}c_l\hat{\upsilon}^l,\label{relax_disp}
\end{align}
where the coefficients $c_l$ are given in Appendix \ref{relax_coeff}.
The dispersion relation $\hat{\upsilon}=\hat{\upsilon}(\khat)$ is
complicated and generally requires a numerical solution. However,
simple results may be obtained in limiting cases which we consider
below.  

\subsection{The critical cooling time derived}\label{iso_vsi_beta_crit}
Here we estimate the maximum thermal relaxation timescale 
$\beta_c$ that allows growth of the VSI. In Appendix \ref{disp_neut_limit}
we derive the following relation between $\beta_c$ and the wave
frequency $\hat{\omega}_c$, \emph{assuming} marginal stability to the
VSI:  
\begin{subequations}\begin{align}
    \left(\hat{\omega}_c\khat\right)^4  =& \left(\hat{\omega}_c\khat\right)^2 
  + M(M+1)\left(1-h^2q^2\khat^2\right) ,\label{relax_cond_simp1}  \\
  \left(\hat{\omega}_c\khat\right)^2 = & \frac{\beta_c}{h q} (\gamma-1)(1+2M)^2
  \left(\hat{\omega}_c\khat\right) \label{relax_cond_simp2}\\
  &- 2M(M+1) . \notag
\end{align}\end{subequations}
These equations consider $\khat^2\gg 1$ and 
assumes $|\hat{\omega}\khat| $ is $O(1)$ and $\beta_c\ll 1$.   Note that 
Eq. \ref{relax_cond_simp1} reduces to the  
dispersion relation for low-frequency inertial waves in the absence of
vertical shear (see Appendix \ref{stable_novshear} for details). 
 
We are most interested in the longest cooling time which allows growth
for any $M$. In Appendix \ref{max_cool}, we find that if
$|hq\khat|\leq 1$, so that marginal stability exists, then $\beta_c$
decreases with increasing $M$.    
The VSI criterion is then set by $\beta_c$ for the
fundamental mode ($M = 0$), for which the exact solution to 
Eq. \ref{relax_cond_simp1} --- \ref{relax_cond_simp2} is
$\hat{\omega}_c\khat = \beta_c(\gamma-1)/hq = -1$. We thus find the
cooling criterion for VSI growth is    
 \begin{align}\label{iso_vsi_cond}
   \beta <   \beta_\mathrm{crit}  \equiv
   \frac{h|q|}{\gamma-1} . 
 \end{align}
 The thin disk approximation, $h \ll 1$, indicates that $\beta_\mathrm{crit} \ll
 1$, as assumed to obtain Eq. \ref{relax_cond_simp1}---\ref{relax_cond_simp2}. 
 We heuristically explain $\beta_\mathrm{crit}$ in \S\ref{vsi_require}, and numerically
 test its validity in \S\ref{bcrit_num_test}.

%% file: numerical.tex
\section{Numerical results for fixed cooling times}\label{numerical}
This section investigates the linear growth of the VSI with fixed
cooling times, $\beta\OmK^{-1}$, that are independent of height.  
We solve the radially local model of Eqs.\ \ref{lin_all} with
$\zeta=1$, i.e.\ retaining  radial gradients of the background disk
structure. The equilibrium disk structure given is in \S\ref{eqm}. This
numerical study does not make the simplifying approximations  
used in \S\ref{analytical}. 

We solve the linear problem by expanding the 
perturbations in Chebyshev polynomials $T_l$ up to $l=512$
and discretizing the equations on a grid with
$z\in[-\zmax,\zmax]$. Our standard boundary conditions at the vertical
boundaries is a free surface,  
\begin{align}\label{bc}
  \Delta P \equiv \delta P + \bm{\xi}\cdot\nabla P= 0 \quad \text{at } z=\pm\zmax,
\end{align}
where $\bm{\xi}$ is the Lagrangian displacement with meridional 
components $\xi_{x,z} = \ii\delta v_{x,z}/\upsilon$. In some cases we
impose a rigid boundary with $\delta v_z(\pm\zmax)=0$. 

The above discretization procedure
converts the linear system of differential equations to a set of 
algebraic equations,  which we solve using LAPACK matrix
routines\footnote{Available at \url{http://www.netlib.org/lapack/}.}.  

\begin{figure}
  \includegraphics[width=\linewidth,clip=true,trim=0cm 0cm 0cm
  0cm]{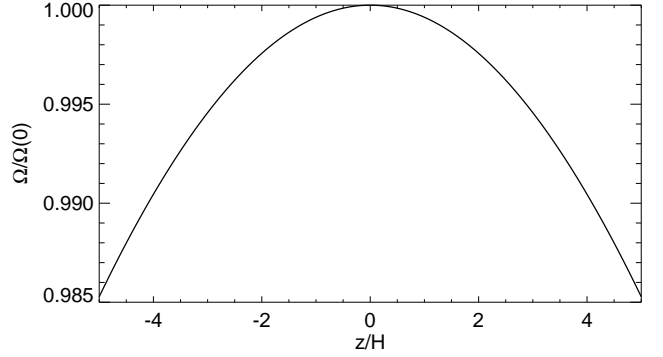} 
  \caption{Equilibrium rotation profile $\Omega(z)$,
    normalized by its mid-plane value, for  the fiducial disk model with $\Gamma=1.011$
    and $(p,q, h)=(-1.5,-1,0.05)$. 
    \label{omega_z} 
  }
\end{figure}

Unless otherwise stated, we adopt fiducial disk parameters 
$(\gamma, \Gamma) = (1.4, 1.011)$, $(p,q, h)=(-1.5,-1,0.05)$, and 
vertical domain size $\zmax=5H$. This setup is effectively vertically
isothermal since $c_s^2(\zmax) \simeq 0.9c_s^2(0)$.\footnote{While this model
has a zero density surface at $z = \pm H_s \simeq \pm 16.4 H$ (see Eq.\ \ref{eqm_rot}), the surface is 
outside the numerical domain.} In Fig. \ref{omega_z}
we plot the corresponding equilibrium rotation profile, which shows
that $\Omega$ decreases slightly away from the midplane.   From Eq.\ \ref{iso_vsi_cond},
we expect that $\beta < \beta_\mathrm{crit} = 0.125$ is needed for VSI growth in our fiducial model. 

\begin{figure}
  \includegraphics[width=\linewidth]{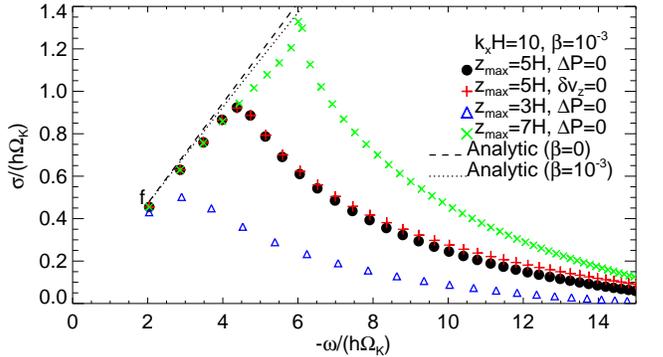}
  \caption{Growth rate, $\sigma$, vs.\ oscillation frequency, $\omega$, for
  VSI modes with wavenumber $k_x = \khat/H=10/H$ in the fiducial disk model
    with rapid cooling, $\beta=10^{-3}$. 
We investigate the effect of vertical box size, from $\zmax=3H$
(\emph{blue triangles}) to $7H$ (\emph{green} x's)
    with a free surface boundary ($\Delta P=0$).  For $\zmax=5H$  (black dots) we also compare free surface (\emph{black dots}) and rigid (\emph{red plusses}), finding little difference.
      The fundamental mode is marked by `f'. Lines show the analytic predictions for
      isothermal perturbations ($\beta = 0$, \emph{dashed}) and for $\beta = 10^{-3}$ (\emph{dotted}).
      See text for discussion.
    \label{compare_modes_iso_kx10} 
  }
\end{figure}

\subsection{Rapid thermal relaxation}\label{vertiso_pertiso} 
We first calculate the VSI in a disk with rapid thermal relaxation by
setting $\beta=10^{-3}$.  Since $\beta \ll \beta_\mathrm{crit}$, this case
gives similar results to the well studied case of isothermal perturbations
 \citepalias[e.g.][]{nelson13,mcnally14,barker15}.\footnote{We confirmed that smaller
 $\beta$ values gave similar results.} This case allows us to explore and test our
finite cooling time model in a familiar context.

\begin{figure}
  \includegraphics[width=\linewidth,clip=true,trim=0cm 1.75cm 0cm
  0cm]{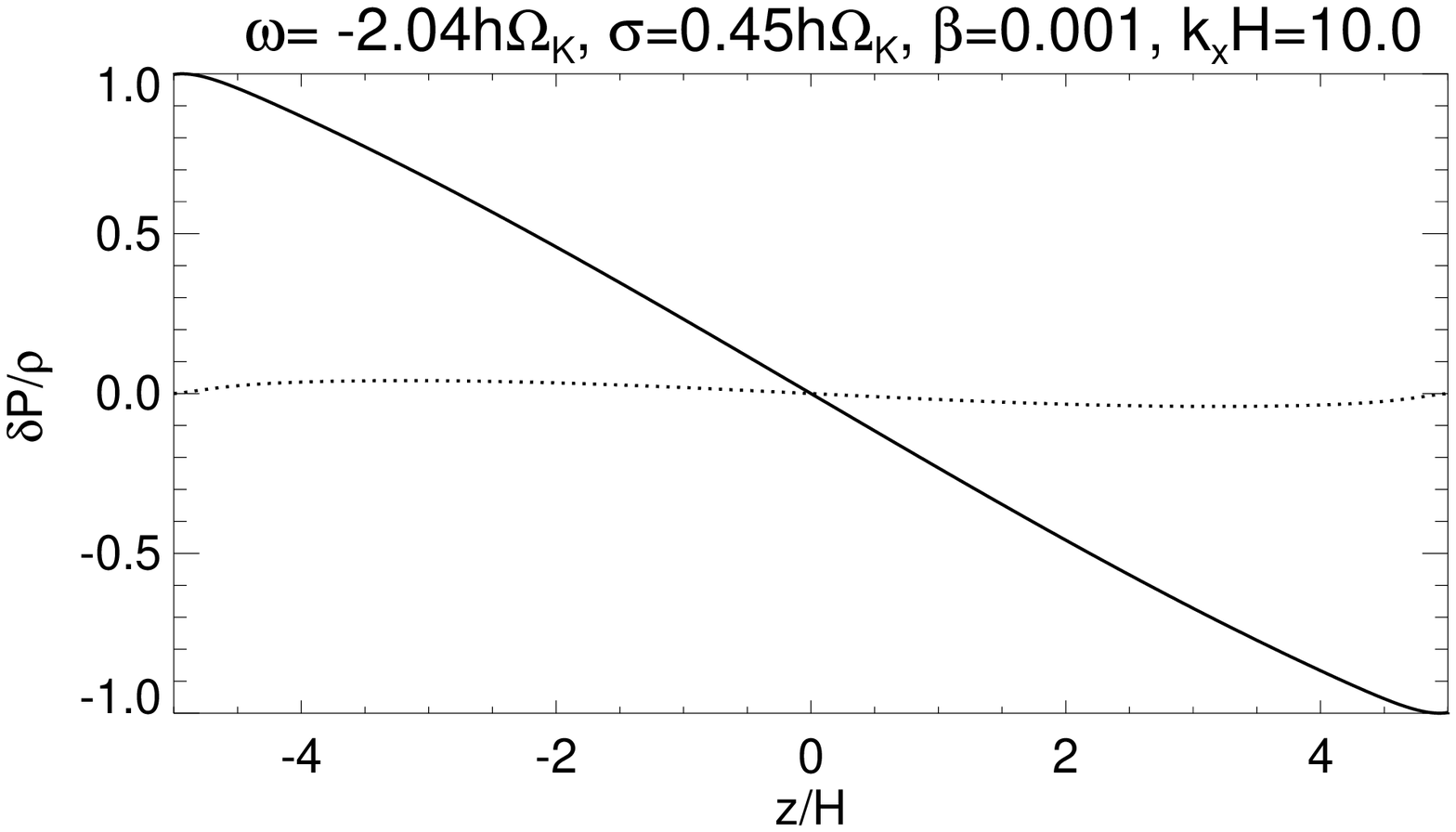} 
  \includegraphics[width=\linewidth,clip=true,trim=0cm 0cm 0cm
  1cm]{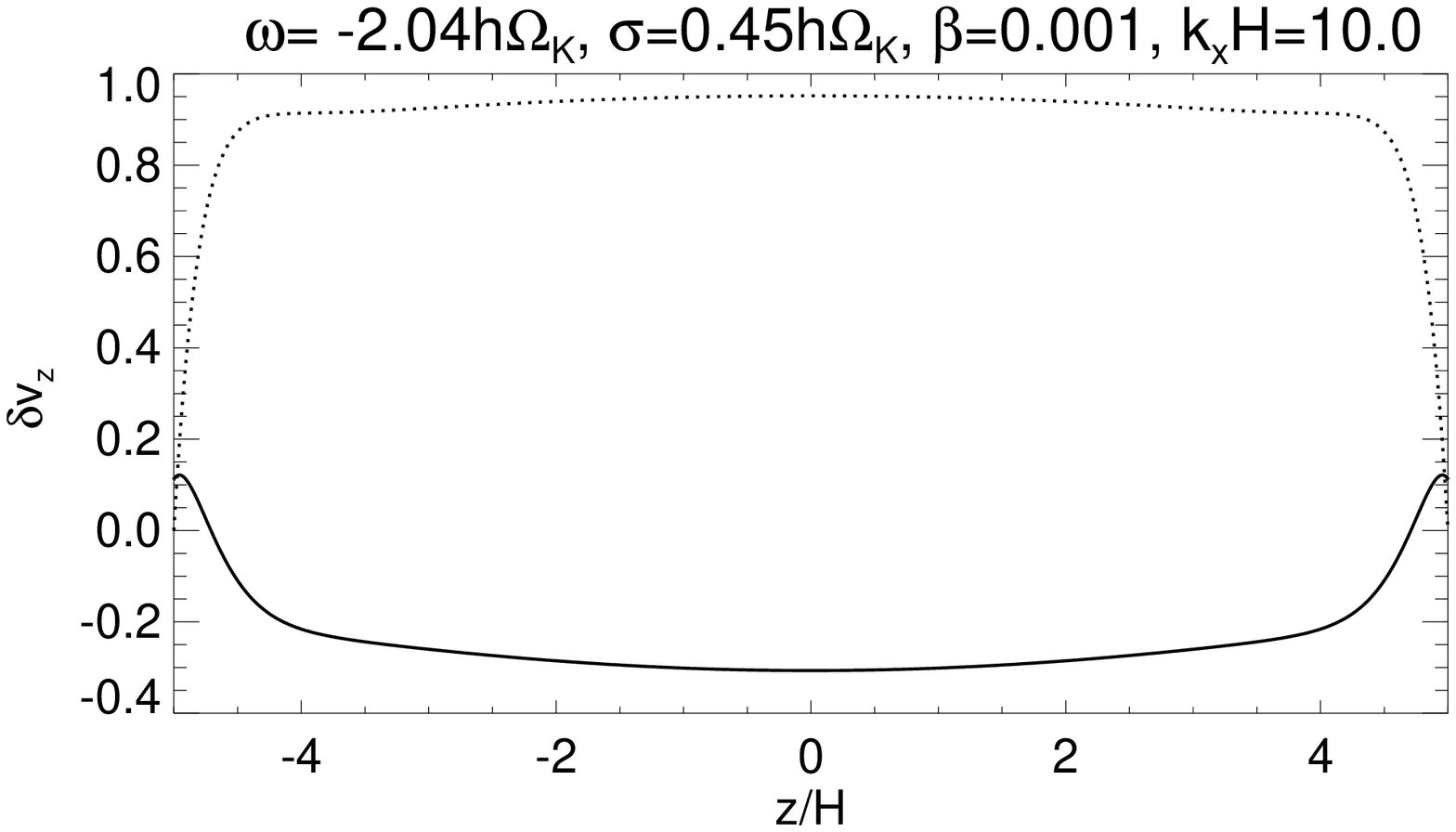}
  \caption{Pseudo-enthalpy perturbation $W$ (top) and vertical velocity
    perturbation $\delta v_z$ of the fundamental VSI with 
    $\khat=10$ in the fiducial disk with rapid thermal relaxation $\beta=10^{-3}$. The real  
    (imaginary) parts of $W$ and $\delta v_z$ are plotted as solid 
    (dotted) lines.
    \label{lowfreq_eigenfunc}
  }
\end{figure}

\begin{figure}
  \includegraphics[width=\linewidth]{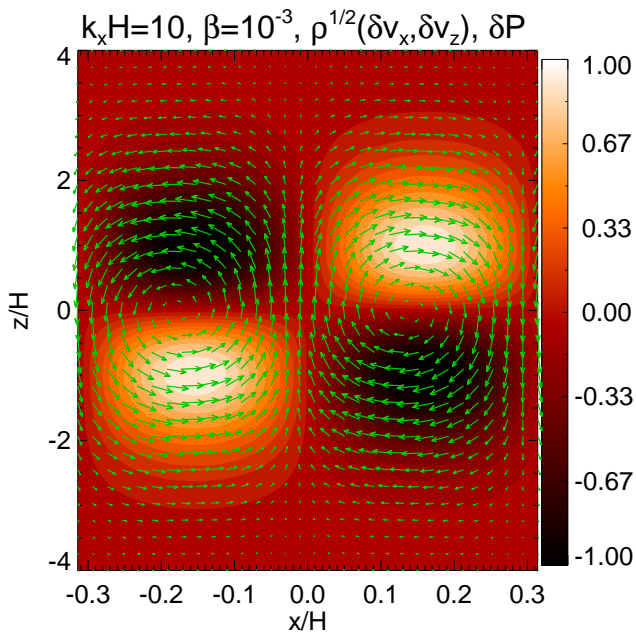}
  \caption{Visualization of the fundamental VSI mode in
    Fig. \ref{lowfreq_eigenfunc}. For this plot the mode has been
    transformed back to real space. Thus, the color scale corresponds to the
    \emph{real} 
    pressure perturbation scaled by its maximum value.
    Arrows show the \emph{real} meridional flow  multiplied by
    $\sqrt{\rho}$. 
  The horizontal axis has been stretched 
    for clarity.  
    \label{lowfreq_eigenfunc_2d}
  }
\end{figure}

\subsubsection{Case study: $\khat=10$}\label{k10}
Fig. \ref{compare_modes_iso_kx10} maps the growth rates $\sigma$ and
wave frequencies $\omega$ of unstable 
modes for $\khat=10$.  Each symbol 
corresponds to a different order mode, i.e.\ a different number of vertical node crossings.  Lower order modes
have smaller $|\omega|$ for fixed radial wavenumber, as expected for
inertial-gravity waves (see \S\ref{stable_novshear}).  

We compare our numerical results to the analytic dispersion relations for 
$\beta = 0$ and $10^{-3}$, from Eqs.\ \ref{simple_growth} and
\ref{relax_disp}, respectively.   The analytic models also have
discrete modes, which lie on the continuous curves that are plotted
for clarity in Fig. \ref{compare_modes_iso_kx10}.   While the analytic
models have many simplifications, they crucially lack an artificial
vertical surface (as they assume an infinite vertical domain).   

For our fiducial case (black dots), the lower order and lower
frequency modes closely follow the analytic prediction with growth
rates increasing with $|\omega|$. 
However, for $|\omega| \gtrsim 4 h \OmK$ the growth rate declines with
$|\omega|$.  This break from the analytic prediction is not due the
choice of boundary condition, as we demonstrate by considering rigid
boundaries (red plusses). 

Rather, the decline in growth rate for high order models is due to the
vertical box size, as seen by comparing the $\zmax/H = 3, 5$ and 7
cases in Fig. \ref{compare_modes_iso_kx10}. Increasing $\zmax$ give
better agreement with the analytic theory, which have $\zmax \to  
\infty$. Larger boxes include regions of larger vertical shear (see
Eqs. \ref{vshear_thin} and \ref{vertical_shear_ex}), which higher
order modes can tap to give larger growth rates. For the $\zmax = 7
H$ case, we begin to see another branch of modes with the highest
growth rates at $\omega \sim -6 h\OmK$. This branch contains the
`surface modes' described below.   

The need for vertically extended domains to capture the largest VSI
growth rates is problematic, especially for hydrodynamic simulations.
This  complication is mitigated by at least two factors.  First, we
show below that with slower cooling, the growth of higher order modes
is preferentially damped, consistent with the analysis in
\S\ref{iso_vsi_beta_crit}.  Second, the fastest growing modes may not
dominate transport when they operate in very low density  surface
layers, i.e. at many $H$. Quantifying which modes will contribute
most to non-linear transport is an important issue, but beyond the
scope of this work.



Fig. \ref{lowfreq_eigenfunc} shows 
eigenfunctions $W = \delta P/\rho$ and $\delta   v_z$ for the lowest
order fundamental mode.   (In this and similar plots below, we
  normalize eigenfunctions such that its maximum amplitude is unity
  with vanishing imaginary part at the lower boundary.) 
Fig. \ref{lowfreq_eigenfunc_2d} maps the pressure perturbation and meridional flow,
scaled by $\sqrt{\rho}$ to reflect the contribution to kinetic
energy. Notice the stretched $x$ axis. Radial velocities are in fact typically much 
smaller than vertical velocities, as expected for a vertically
elongated, anelastic mode \citepalias{nelson13}. Most of the kinetic
energy is contained within $\sim 2H$ of the midplane due to the
density stratification.  

\begin{figure}
  \includegraphics[width=\linewidth]{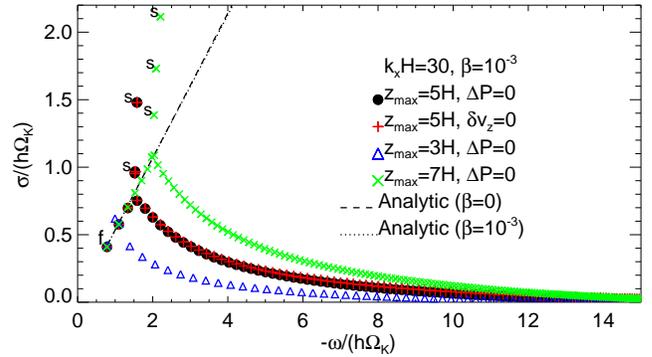}
  \caption{Same as Fig. \ref{compare_modes_iso_kx10} but for
    $\khat=30$. Examples of surface modes are 
    marked by `s'. \label{compare_modes_iso_kx30}
  }
\end{figure}

\begin{figure}
  \includegraphics[width=\linewidth,clip=true,trim=0cm 0cm 0cm
  0cm]{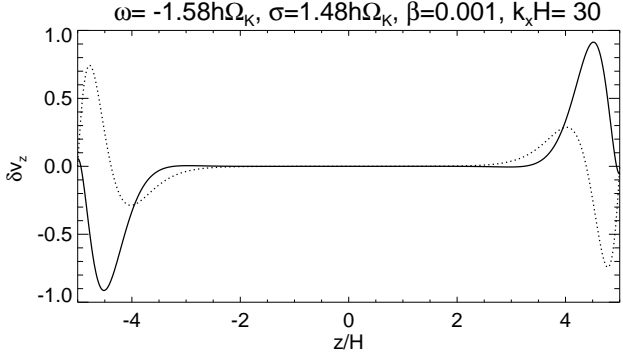}
  \caption{Example of a `surface mode' in the fiducial disk model.
    \label{lowfreq_eigenfunc_surf}
  }
\end{figure}

\subsubsection{Surface modes of the VSI}\label{surf_comment} 
The surface modes mentioned above are more prominent for larger
wavenumbers. Fig. \ref{compare_modes_iso_kx30} maps the $\khat = 30$
eigenvalues, with the surface modes labeled.  Surface modes
strongly depend on the location of the imposed vertical boundary, and
disagree with the analytic models, which lack an imposed surface. We
thus discount the physical significance of surface modes for our
model, as we explain further below.  

Surface modes are a well known feature of the VSI in finite vertical
domains \citepalias{nelson13, mcnally14}. \citetalias{barker15} show
that surface modes arise whether the surface is imposed (as in
  our models) or natural (for disk models with a zero-density surface).
 BL15 mention the interface between a disk's interior and corona
  as a physical boundary that can trigger  surface modes. 
  Since vertically isothermal disk models lack a surface, modes which
  depend on the adopted value of $\zmax$ are not physically meaningful.


The possibility of artificial surface modes seeding growth in
numerical simulations of the VSI  merits further study. Indeed
\citetalias{nelson13} find that the initial growth of perturbations
primarily occurs near the vertical boundaries. 
The motion in surface modes is indeed concentrated near the surface,
as shown in Fig. \ref{lowfreq_eigenfunc_surf},  where the density
  is low.  
Thus, their contribution to
transport might (by themselves) be weak,  following the arguments in 
\S\ref{k10}.  Moreover surface modes, like all modes with large
wavenumbers, are more prone to viscous damping, as discussed in
\S\ref{caveats_visc}. 

While the relatively large growth rates of surface modes is
tantalizing, we dismiss them in disk models that lack a physical
surface.  

%
%


\begin{figure}
  \includegraphics[width=\linewidth,clip=true,trim=0cm 1.75cm 0cm
  0cm]{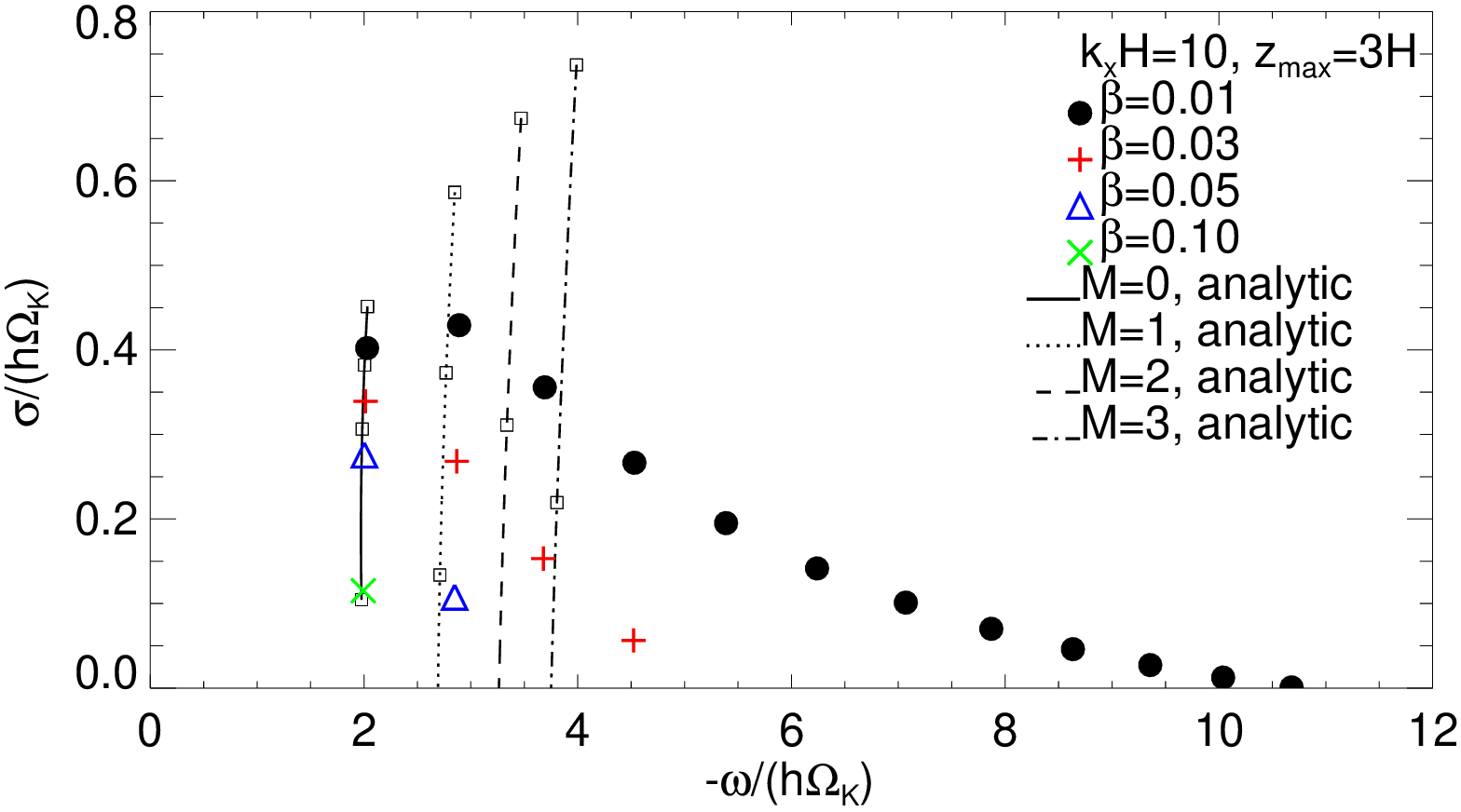} 
  \includegraphics[width=\linewidth,clip=true,trim=0cm 1.75cm 0cm
  0cm]{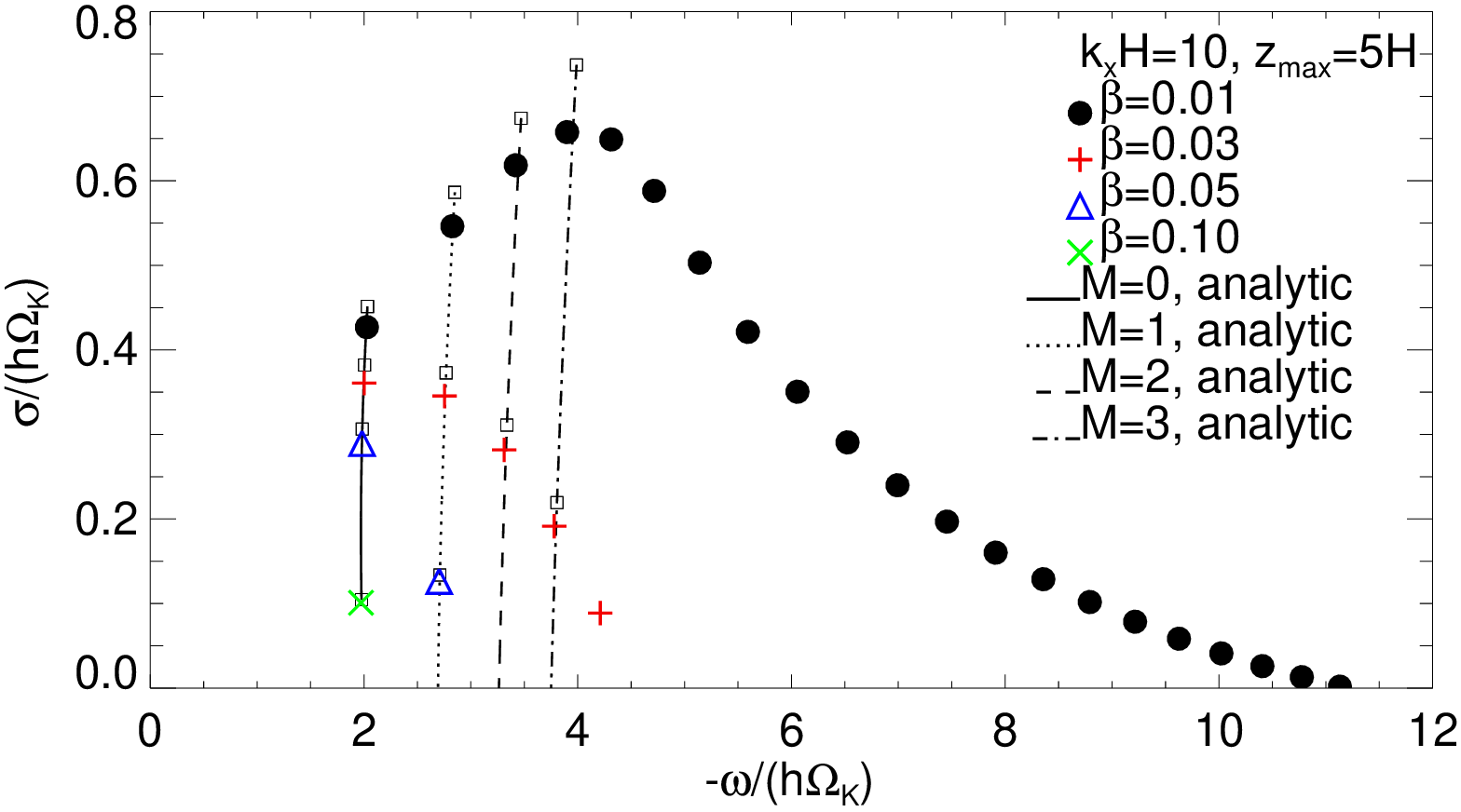}
  \includegraphics[width=\linewidth]{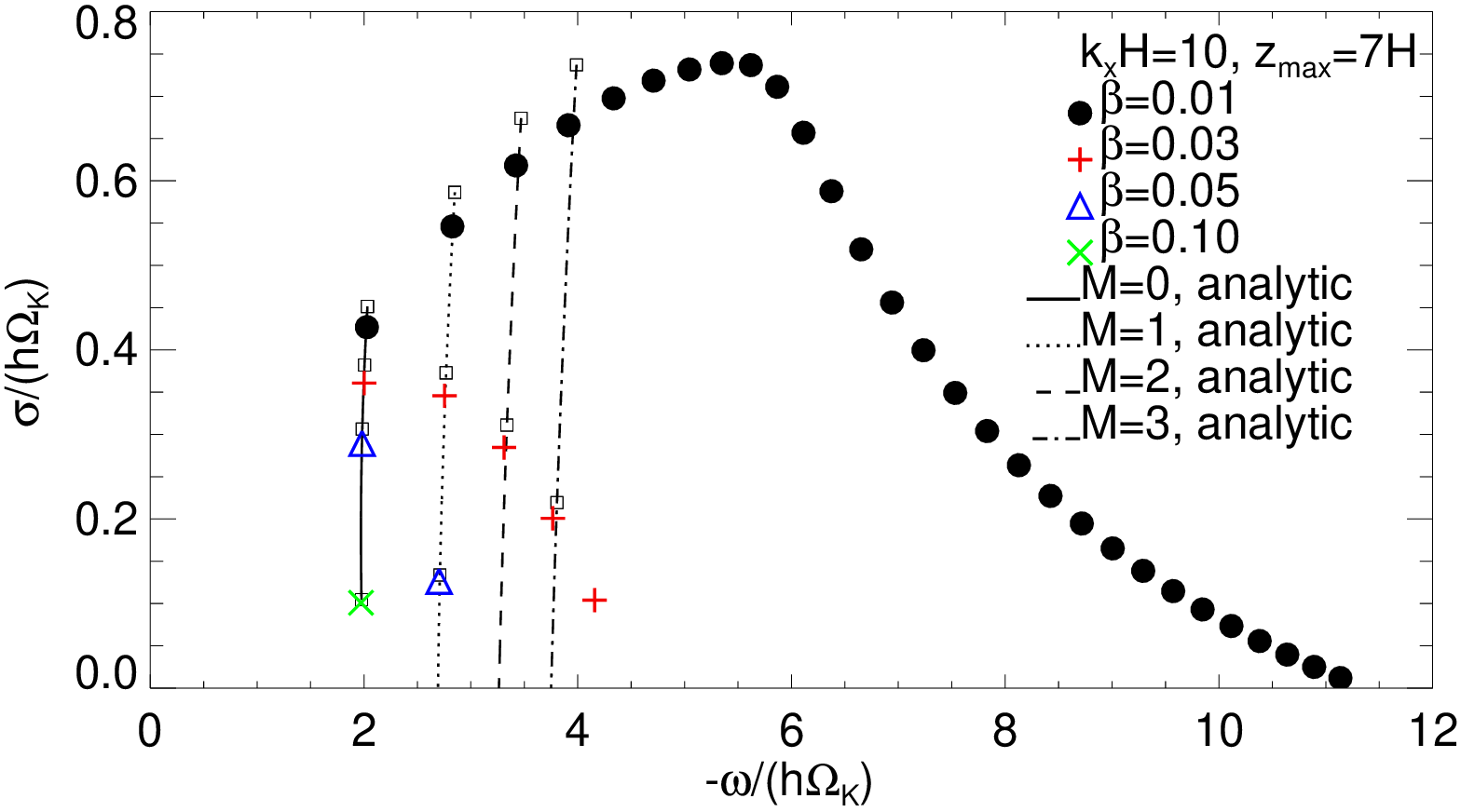}
  \caption{Unstable modes with $\khat=10$ and thermal
    relaxation timescales $\beta=0.01$ (black dots), $\beta=0.03$ (red
    crosses), $\beta=0.05$ (blue triangles) and $\beta=0.1$ (green 
    crosses); for different vertical domain sizes $\zmax=3H$ (top),
    $\zmax=5H$ (middle, the fiducial setup) and $\zmax=7H$
    (bottom). Lines are computed from
    Eq. \ref{relax_disp} for  $M=0$ (solid, fundamental mode) and
    $M=1,\,2,\,3$ (dotted, dashed, dash-dot, respectively). Along each
    line, $\beta$ increases continuously from $0.01$ to $0.1$ from top
    to bottom, and squares mark corresponding $\beta$ values with
    numerical results. 
    \label{compare_modes_cool_kx10} 
  }
\end{figure}

\subsection{Slower thermal relaxation}\label{therm_relax_eff}
We now consider the effect of longer cooling timescales by gradually
increasing  
$\beta$. We expect VSI growth for  $\beta < \beta_\mathrm{crit} =
0.125$.  We also expect that higher order modes will damp at even
lower $\beta$ values, as argued in \S\ref{iso_vsi_beta_crit}. 

Fig. \ref{compare_modes_cool_kx10} largely confirms these expectations
by plotting eigenvalues for $\khat=10$  
and $\beta$ from $0.01$ to  $0.1$.  
The analytic results from Eq. \ref{relax_disp} are now plotted as
different curves for each mode order $M$, with $\beta$ varying along
each curve. We see the standard increase in frequency, $|\omega|$,
with mode order. 

As expected, the higher order modes are preferentially damped.  For
$\beta \geq 0.03$ the fundamental ($M = 0$) mode is the fastest
growing. For $\beta = 0.1$, only the $M = 0$ mode grows. This growth
is slow since $\beta$ is near $\beta_\mathrm{crit}$. 

Fig. \ref{compare_modes_cool_kx10} also shows how cooling times affect
the dependence on $\zmax$, the size of the vertical domain.   For all
cooling times, the fundamental mode depends only weakly on $\zmax$,
and there is good agreement with analytic values.  This convergence is
reassuring given the importance of the fundamental mode at longer
cooling times.  For higher order modes and short cooling times,
however, the eigenvalues vary strongly with $\zmax$.  The disagreement
with analytic theory is strongest for the smallest domain, with $\zmax
= 3 H$.  For $\beta \geq 0.03$, the $\zmax/H = 5, 7$ and the analytic
results are well converged (aided by the fact that only $M < 5$ modes
grow).

%

\begin{figure}
  \includegraphics[width=\linewidth,clip=true,trim=0cm 1.75cm 0cm
  0cm]{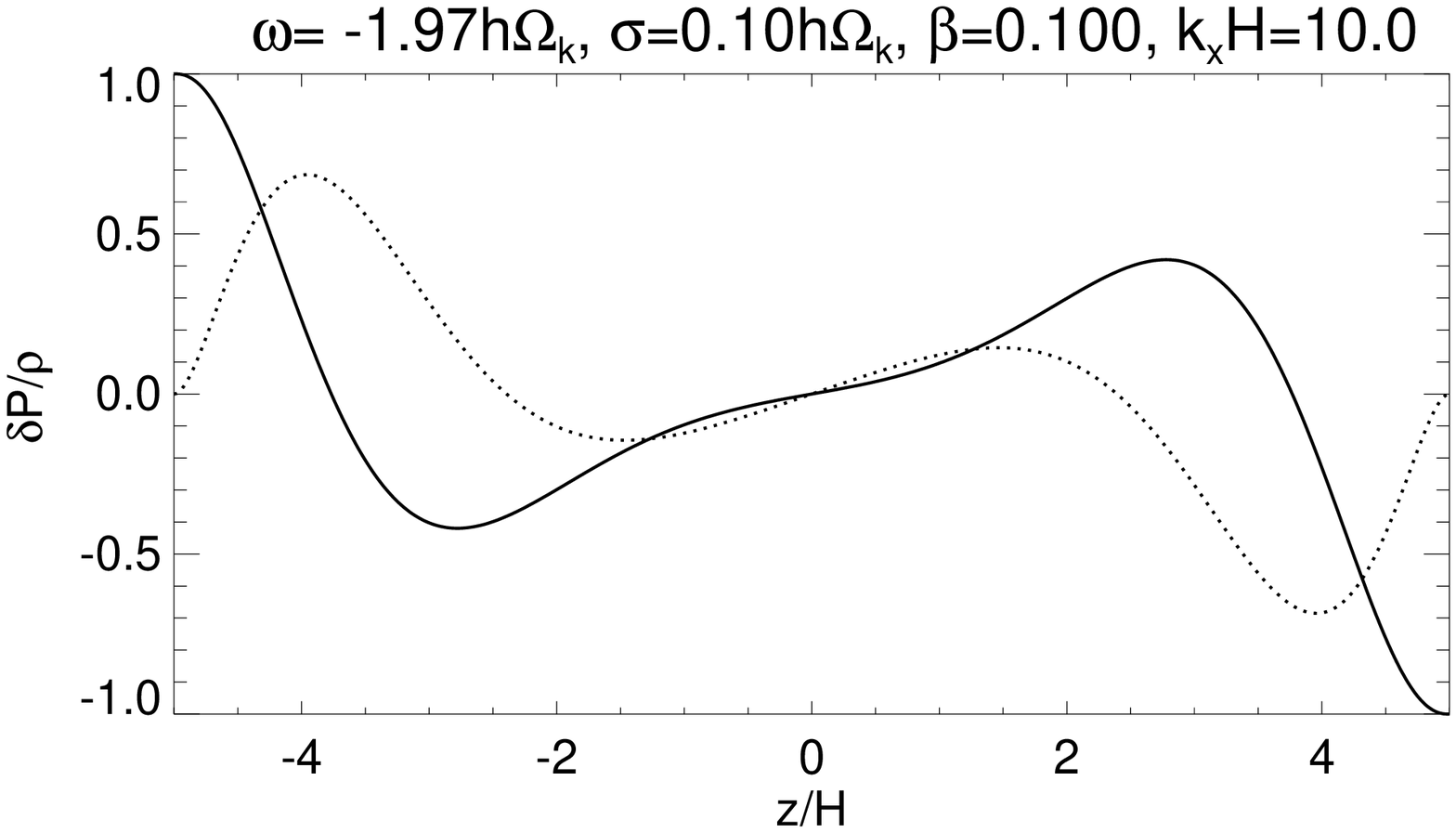} 
  \includegraphics[width=\linewidth,clip=true,trim=0cm 0cm 0cm
  1cm]{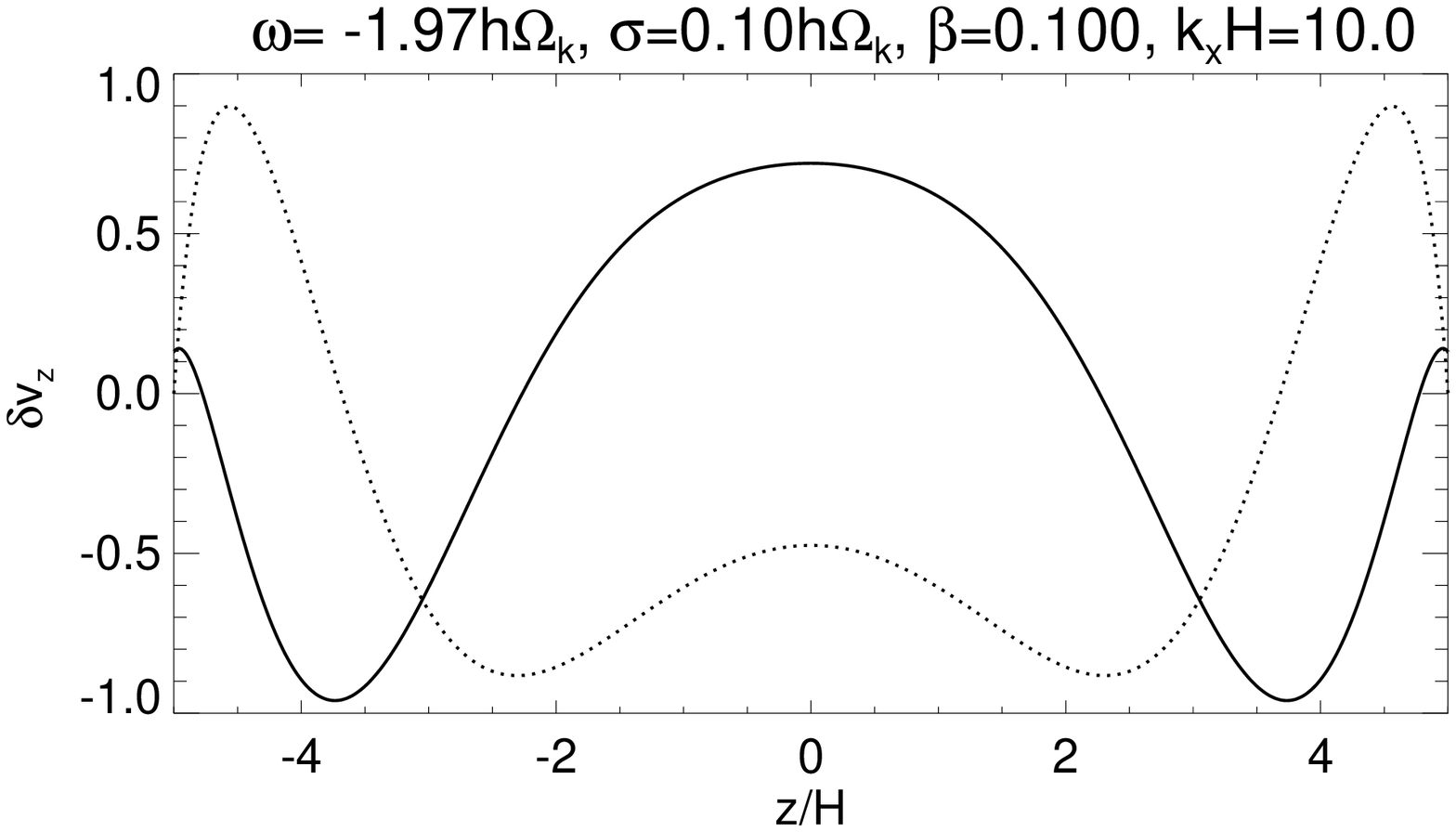}
  \caption{Same as Fig. \ref{lowfreq_eigenfunc} but with a
    thermal relaxation timescale $\beta=0.1$. 
    \label{lowfreq_eigenfunc_cool}
  }
\end{figure}

\begin{figure}
  \includegraphics[width=\linewidth]{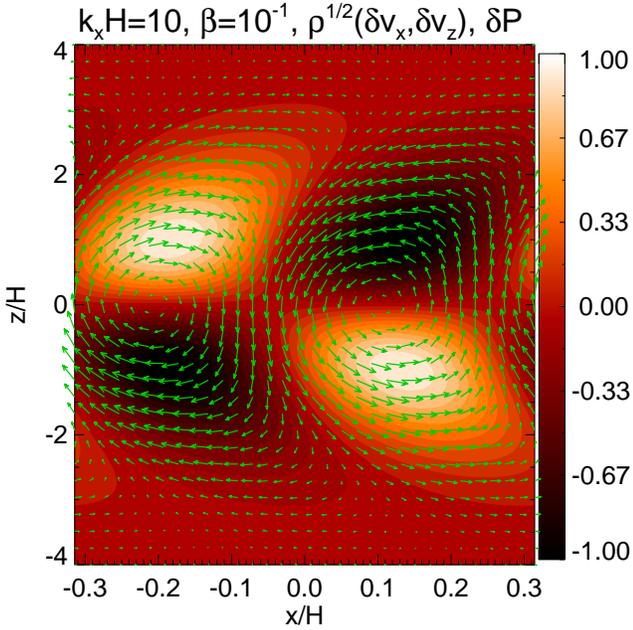}
  \caption{Same as  Fig. \ref{lowfreq_eigenfunc_2d} but with a thermal
    relaxation timescale $\beta=0.1$. 
    \label{lowfreq_eigenfunc_2d_cool}
  }
\end{figure}

Figures \ref{lowfreq_eigenfunc_cool} and \ref{lowfreq_eigenfunc_2d_cool}
illustrate the eigenfunction of the fundamental mode with $\khat=10$ and
$\beta=0.1$. Compared with the small $\beta$ case in 
Figs. \ref{lowfreq_eigenfunc} and \ref{lowfreq_eigenfunc_2d}, the eigenfuctions 
show a more complex variation of phase with height.  This dependence yields 
the `tilted' appearance of pressure field in  Fig. \ref{lowfreq_eigenfunc_2d_cool}.
Physically, the increased role of buoyancy, which increases with height, explains 
why larger $\beta$ values produce more complex vertical structure. 

\begin{figure}
  \includegraphics[width=\linewidth,clip=true,trim=0cm 1.75cm 0cm
  0cm]{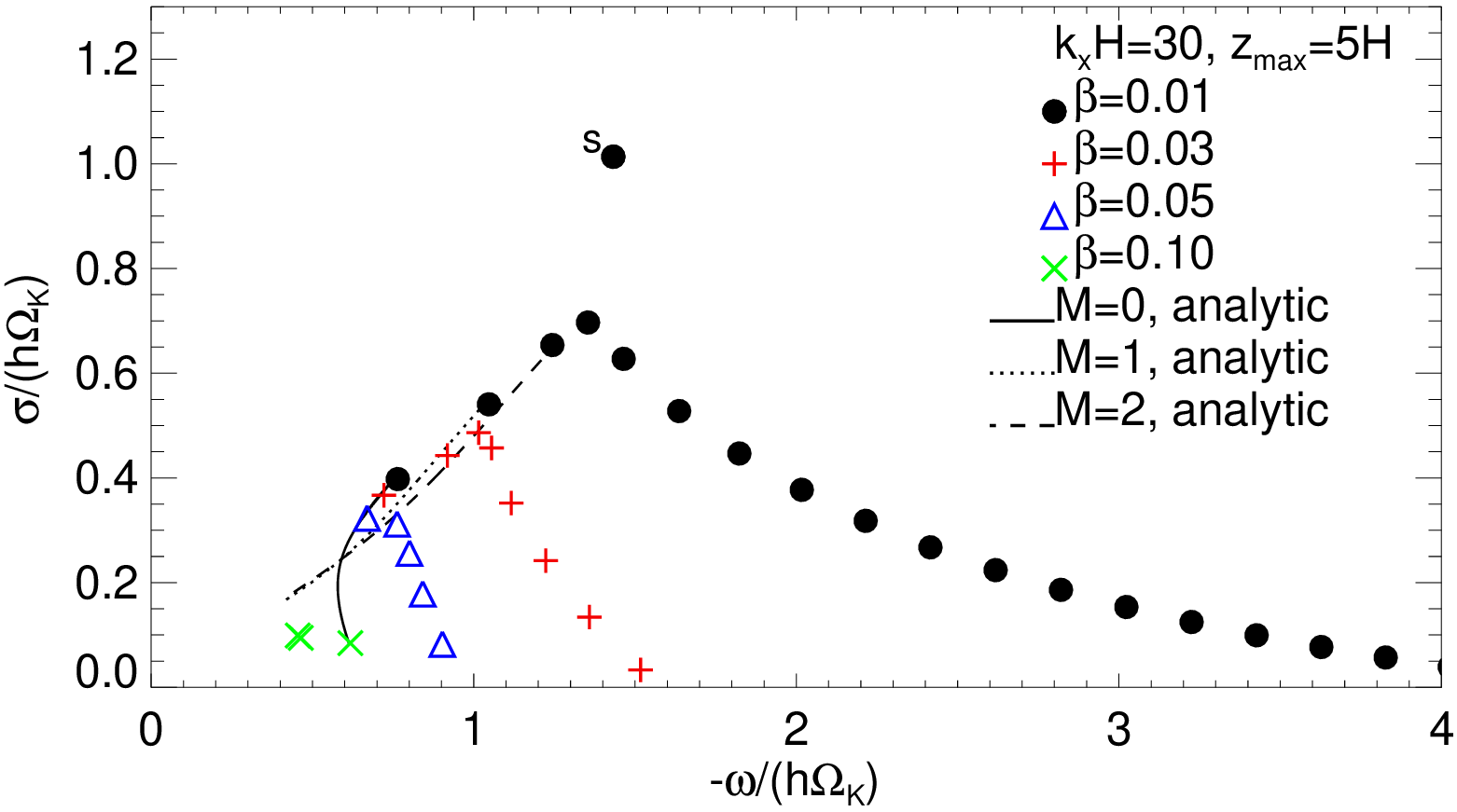}
  \includegraphics[width=\linewidth]{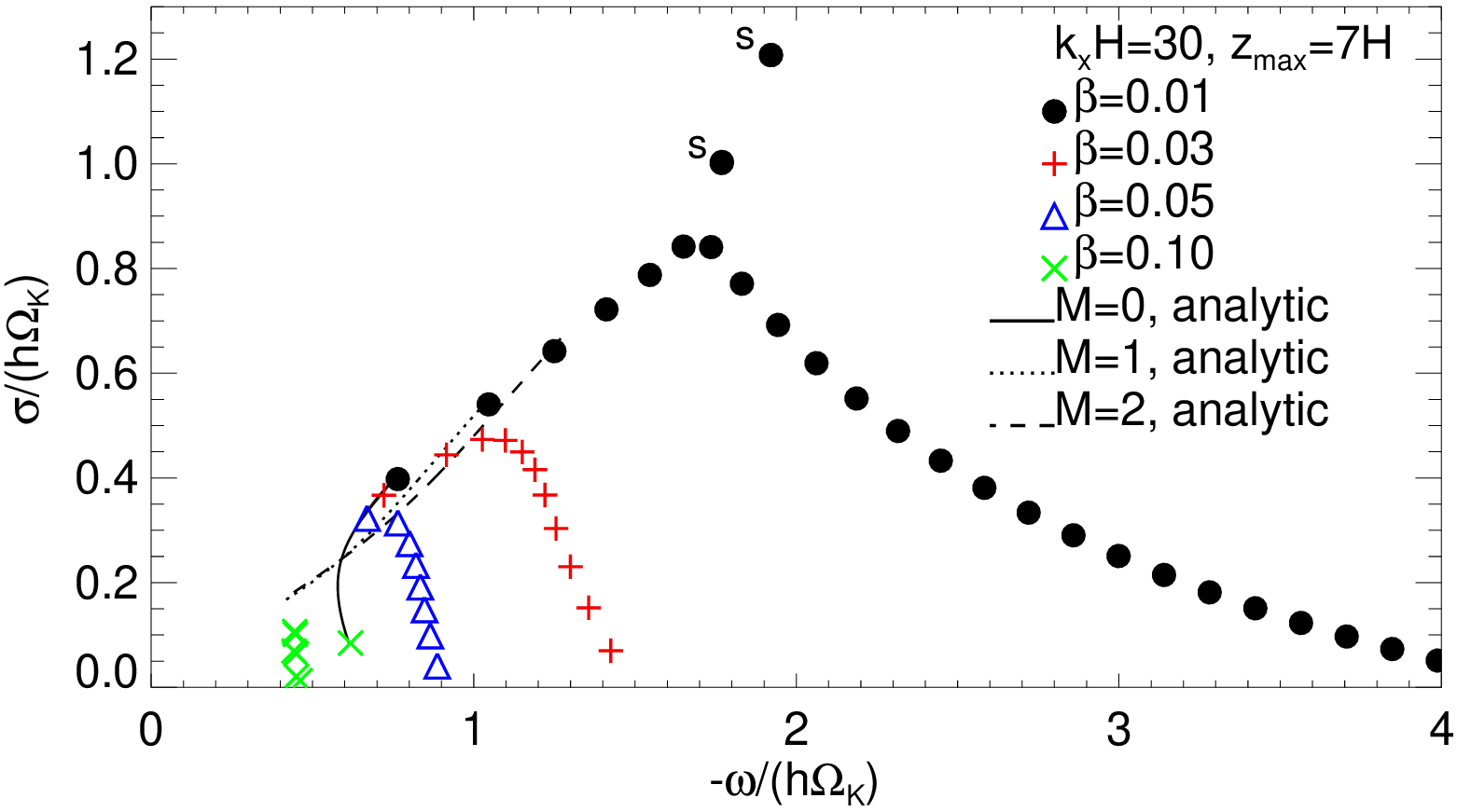}
  \caption{Same as Fig. \ref{compare_modes_cool_kx10} but for
    $\khat=30$ and domain sizes $\zmax = 5H$ (\emph{top}) and $7H$ (\emph{bottom}). Examples of surface modes are marked with `s'. 
    \label{compare_modes_cool_kx30} 
  }
\end{figure}

Fig. \ref{compare_modes_cool_kx30} shows how cooling affects higher wavenumber
modes, specifically $\khat=30$.   We only show the larger domains with $\zmax = 5H, 7H$. 
These cases again show good convergence for $\beta \geq 0.03$, which 
the smaller domain with  $\zmax = 3H$ lacks.

With slower cooling, the wave frequency shows a more complex 
dependence on mode order, both analytically and numerically. For $\beta = 0.1$,
 the fundamental mode (which lies on the $M = 0$ curve) no longer has the smallest $|\omega|$ value.  

Moreover, the fundamental mode is no longer the fastest growing mode for $\beta \geq 0.03$, or even 
for $\beta = 0.1$.  This complication is not actually surprising since $|hq\khat| = 1.5$ is no longer less than unity,
as required in the analytic derivation of \S\ref{iso_vsi_beta_crit}.  Though  inconvenient,  given that the 
fundamental mode is easy to identify and the most numerically converged, this complication is not ultimately significant
for the operability of the VSI.  

Fig. \ref{compare_modes_cool_kx30} also demonstrates that artificial surface modes are damped for $\beta \geq 0.03$.
We are thus confident that artificial surface modes should not affect our determination of the critical cooling time.

%

\begin{figure}
  \includegraphics[scale=0.4415,clip=true,trim=0cm 1.75cm 0cm
  0.9cm]{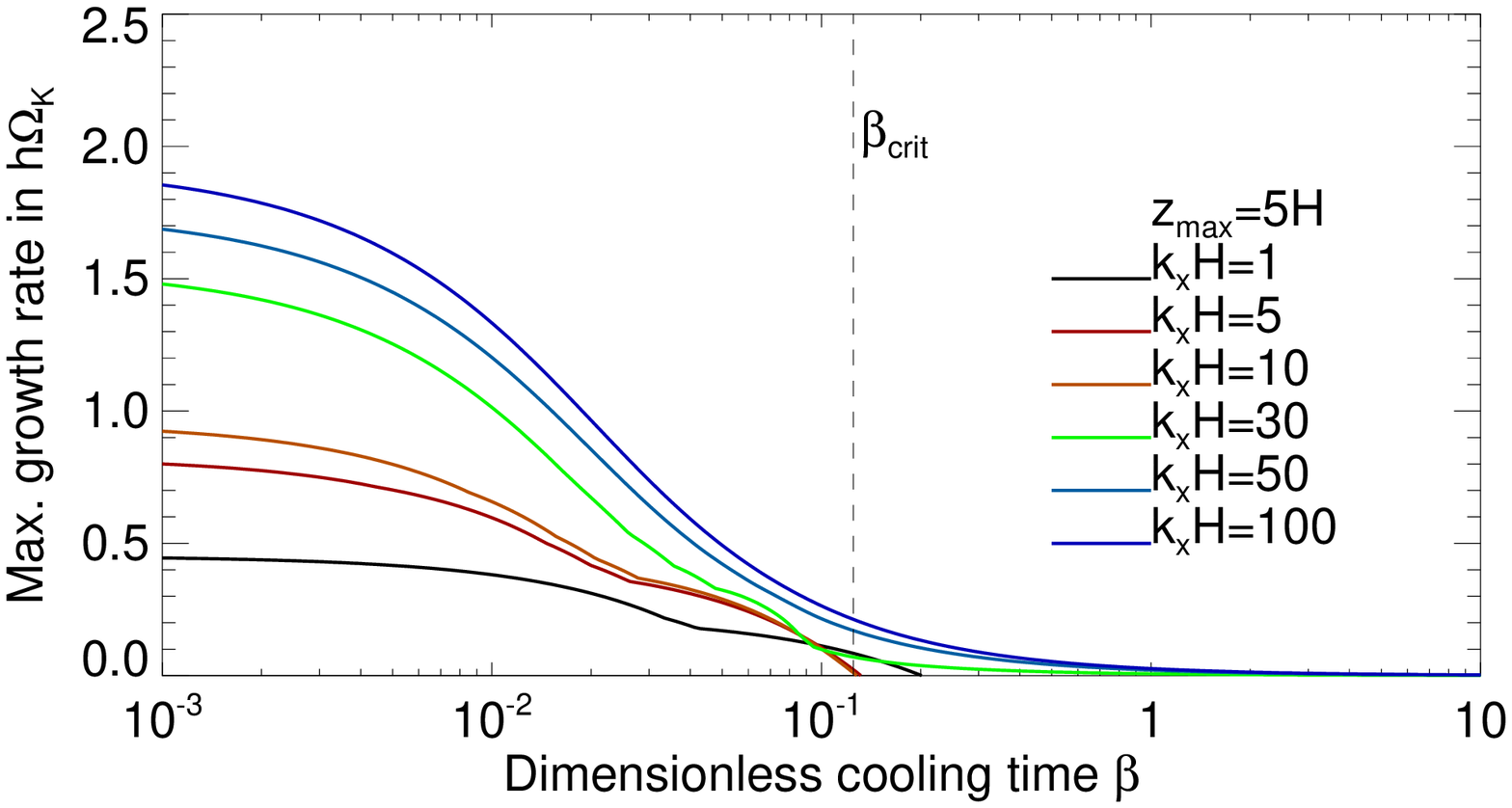} 
  \includegraphics[scale=0.4415,clip=true,trim=0cm 0.0cm 0cm
  0.9cm]{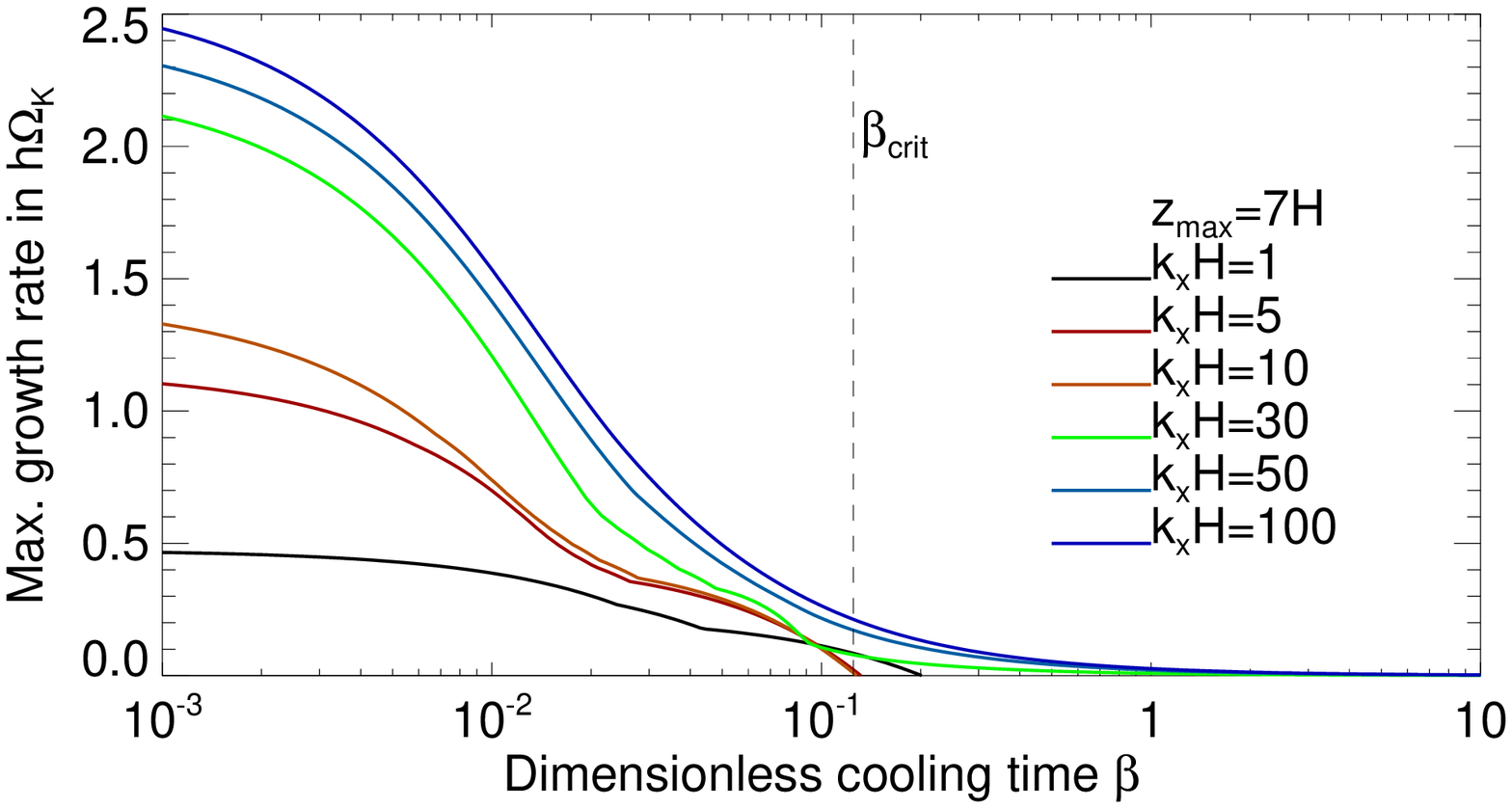}  
  \caption{Maximum VSI growth rates in the fiducial disk 
     model as a function of the thermal relaxation timescale
     $\beta$, for 
     $\zmax=5H$ (top) and $\zmax=7H$ (bottom). The vertical line is the
     critical thermal timescale $\beta_\mathrm{crit}$ obtained  
     from Eq. \ref{iso_vsi_cond}. 
     \label{bcrit_compare1}}   
 \end{figure}

\subsection{Critical thermal relaxation timescale}\label{bcrit_num_test}
Having explored the behavior of VSI modes with cooling, we turn to the numerical validity of the 
analytic cooling criterion for vertically isothermal disks, $\beta < \beta_\mathrm{crit} = h|q|/(\gamma -1)$.

Fig. \ref{bcrit_compare1} shows how VSI growth rates vary with cooling time 
in our fiducial model with $\beta_\mathrm{crit} = 0.125$.  
Curves are for a fixed horizontal wavenumber, and show the maximum growth rate for all vertical mode orders.  The 
discontinuity in some curves occurs when the fastest growth switches to a different mode order.

For $\khat = 5, 10$, the growth rate drops to zero at the expected $\beta = \beta_\mathrm{crit}$.
For longer wavelength modes, with $\khat = 1$, growth persists to slightly longer cooling times.  
This difference is not surprising since our analytic derivation assumed $\khat^2 \gg1$, but  
the change is quantitatively minor.

For shorter wavelengths, with $\khat \geq 30$, growth persists for $\beta$ significantly larger
than $\beta_\mathrm{crit}$.  This tail of growth is partly explained by the breaking of the $\khat < 1/|qh| = 20$ 
approximation, used in  the analytic derivation.  Despite the lack of a clear stability boundary 
at high $\khat$, the $\beta_\mathrm{crit}$ threshold remains useful, since growth rates drop to 
$\lesssim 10\%$ of  their maximum value at  $\beta_\mathrm{crit}$ and continue to fall for larger $\beta$. 
 Moreover, we expect that longer wavelength modes with $\khat \lesssim 20$
are more significant for disk transport, see \S\ref{caveats_visc}.

Comparing the $\zmax/H = 5, 7$ cases in Fig. \ref{bcrit_compare1}, we see that the location of the vertical
boundary has little effect on the critical cooling time.  This agreement occurs despite the fact that peak growth rates
differ as $\beta \to 0$.  We are thus confident that boundary effects, including surface modes, do not
affect our analysis of the critical cooling time.

Our results agree with the vertically isothermal simulations of \citetalias{nelson13} 
which used the same disk parameters as our fiducial model.  The expected $\beta_\mathrm{crit} = 0.125$ 
is consistent with the simulations shown in their Fig.\ 12.  \citetalias{nelson13}  
found nonlinear VSI growth for $\beta = 0.06< \beta_\mathrm{crit}$ but no growth for 
  $\beta = 0.6 > \beta_\mathrm{crit}$.\footnote{Since the dimensionless cooling time $T_{\rm relax}$ in \citetalias{nelson13}
  is normalized to the orbital period, we convert $\beta = \OmK P_{\rm orb} T_{\rm relax} = 2 \pi T_{\rm relax}$.}

\begin{figure}
  \includegraphics[width=\linewidth,clip=true,trim=0cm 0.cm 0cm
  0cm]{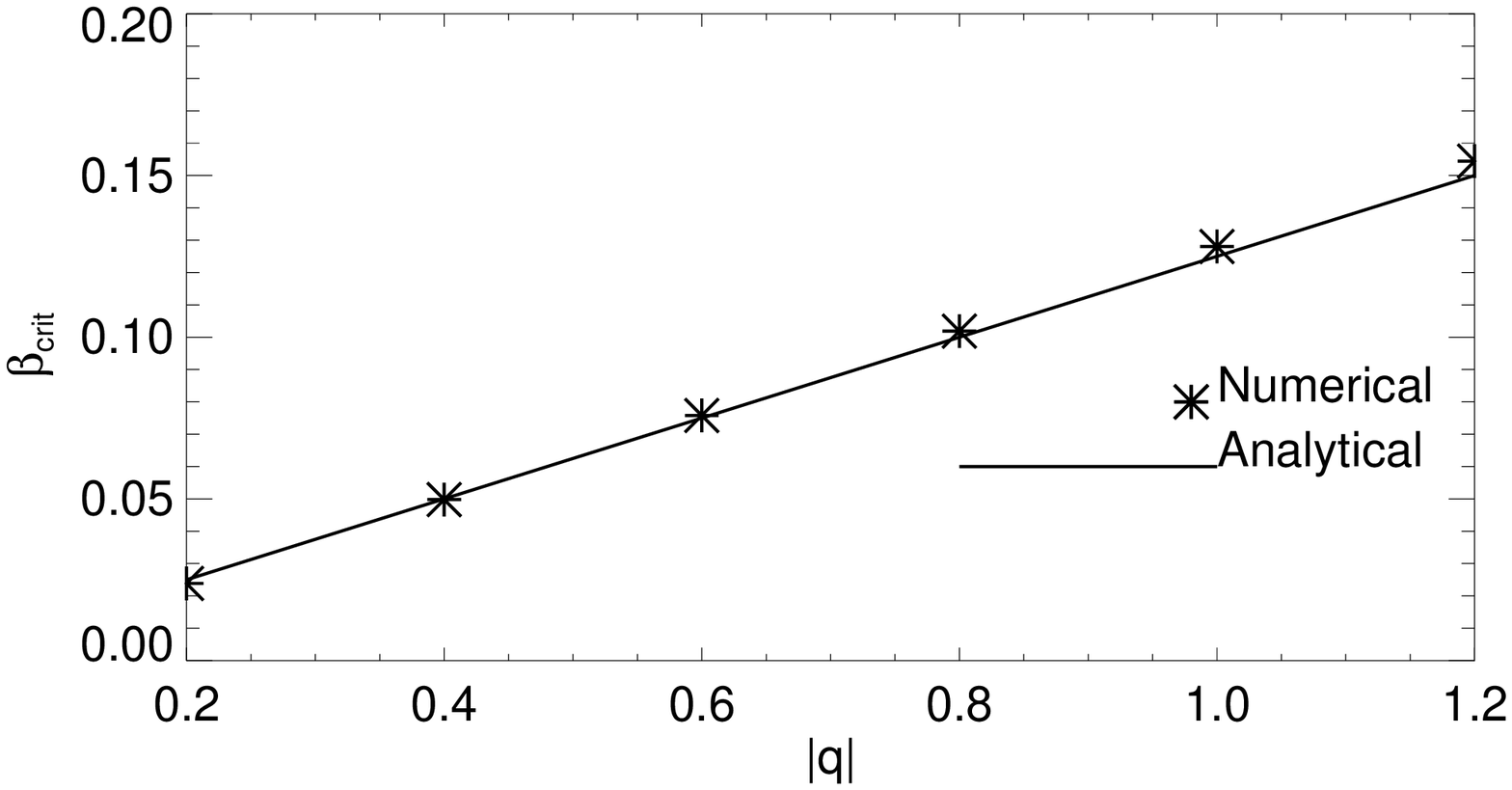} 
  \includegraphics[width=\linewidth,clip=true,trim=0cm 0.0cm 0cm
  0.8cm]{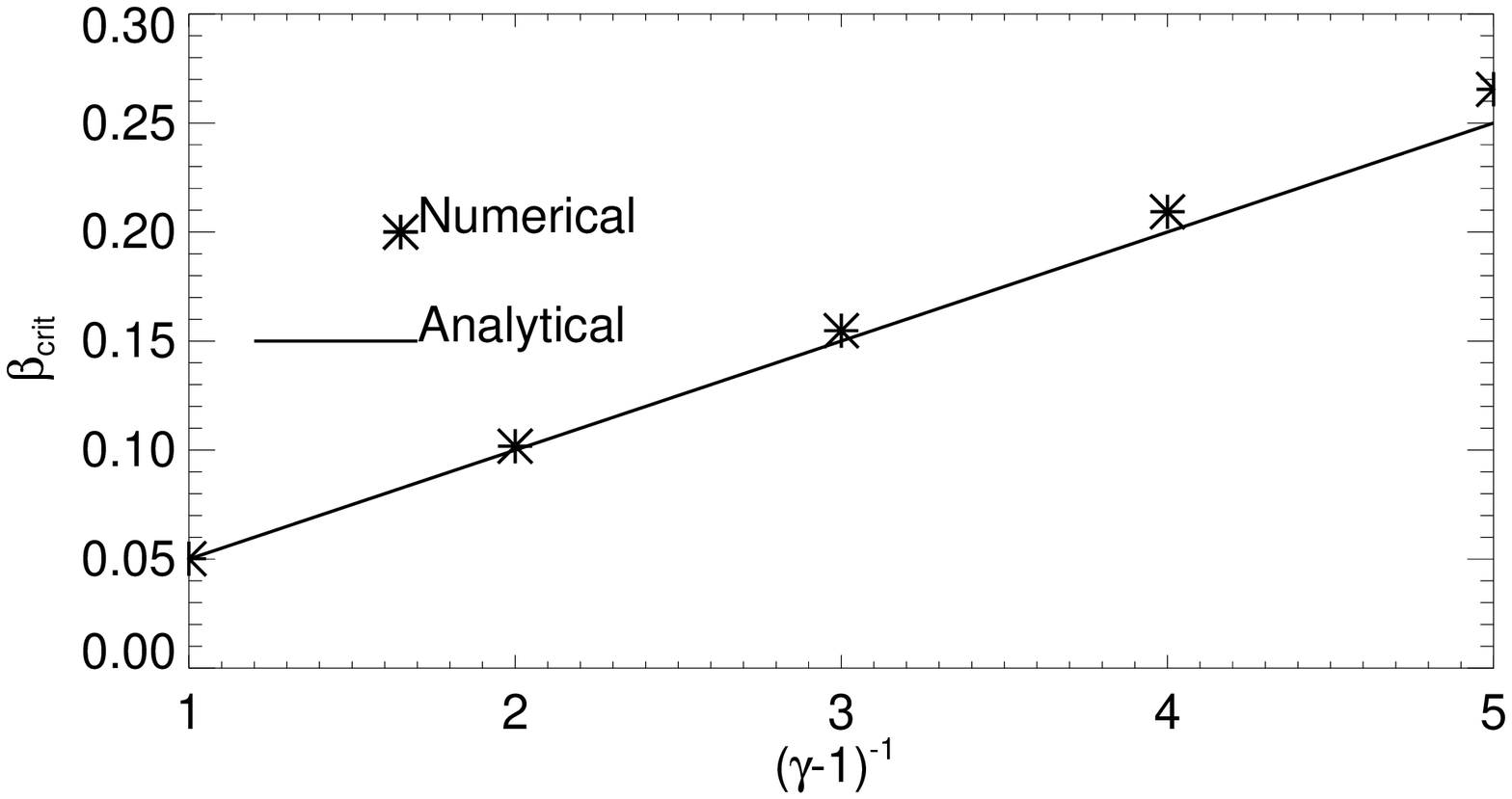}
  \includegraphics[width=\linewidth,clip=true,trim=0cm 0.0cm 0cm
  0.8cm]{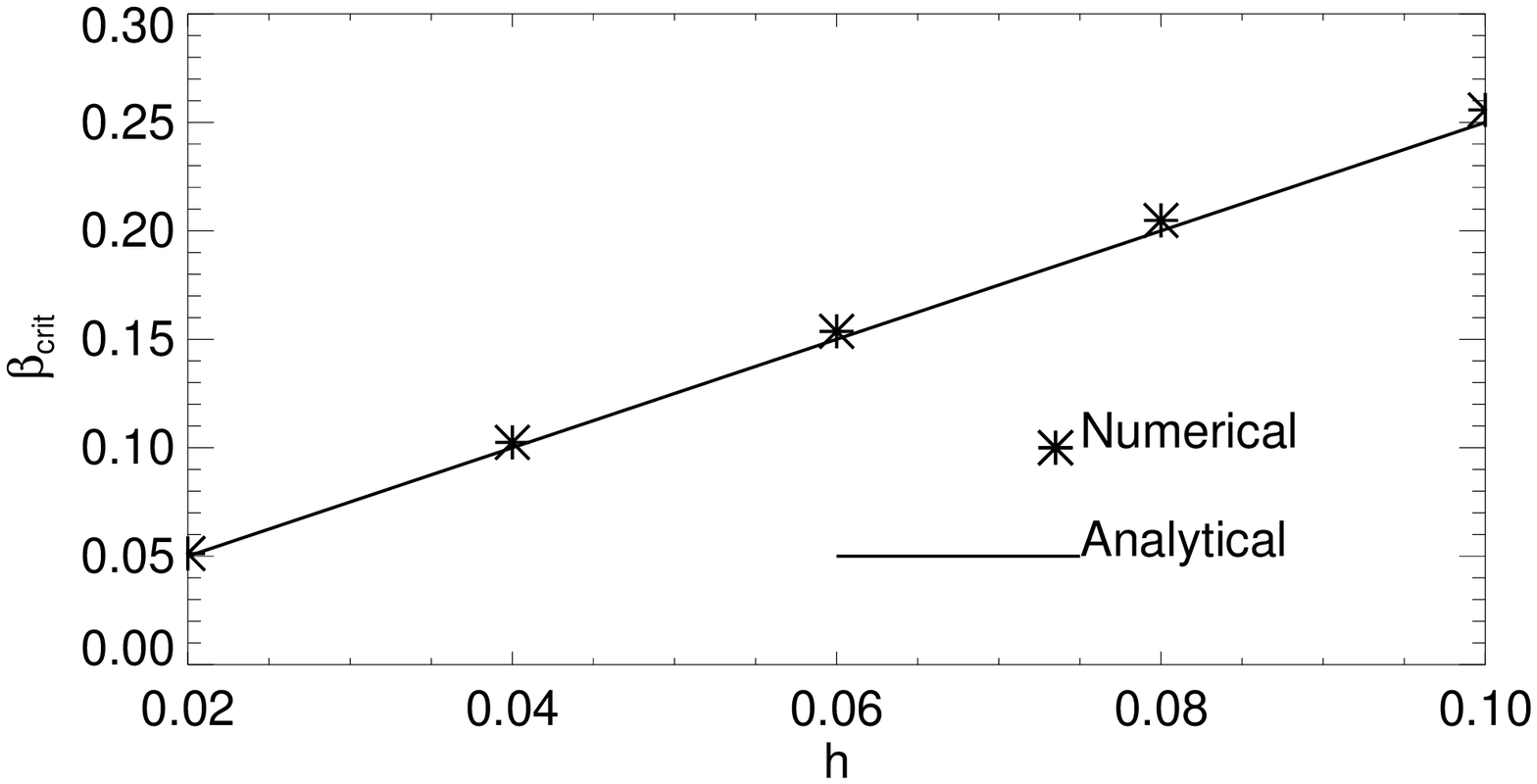} 
  \caption{Dependence of the upper limit to the thermal relaxation timescale
    $\beta_\mathrm{crit}$ for the fundamental VSI mode on disk
    parameters. The fiducial setup is $(\gamma, \Gamma= (1.4, 1.011)$
    and $(p,q, h)=(-1.5,-1,0.05)$. Top: varying 
    $q\in[-1.2,-0.2]$; middle: varying $\gamma\in[1.2,2.0]$; bottom:
    varying $ h\in[0.02,0.1]$. The perturbation wavenumber is
    $\khat=10$.  
    \label{bcrit_compare}}  
\end{figure}

Fig.\ \ref{bcrit_compare} confirms that the numerically determined
$\beta_\mathrm{crit}$ shows the expected scaling with disk parameters $h$, $q$, and $\gamma$.
For this test we fix $\khat = 10$ (an appropriate value for all the reasons discussed above) and measure
the smallest $\beta$ value where growth vanishes.  The agreement with the analytic scaling of 
Eq.\ \ref{iso_vsi_cond} is quite good, confirming the applicability of our critical cooling time in
vertically isothermal disks.

\subsection{Vertically non-isothermal disk model}\label{nonvertiso}
We now generalize our critical cooling time by considering disks which
are not vertically isothermal.  We rely on physical arguments for this generalization.
In disks with weaker vertical stratification, i.e.\ $\Gamma > 1$ for
fixed $\gamma$, we expect $\beta_\mathrm{crit}$ to be larger.
We thus transform $1/(\gamma -1) \to 1/(\gamma - \Gamma)$.  

We also expect $\beta_\mathrm{crit}$ to scale with the vertical shear.   
In general the vertical shear rate $\p_z\Omega\propto s$, the radial
entropy gradient, see Eq.\ \ref{vertical_shear_ex}.  We thus 
 transform $q \to s = q +p(1-\Gamma)$.
 
Our estimate for the generalized cooling 
 time criterion --- valid for both vertically isothermal and non-isothermal disks (cf.\ Eq.\ \ref{iso_vsi_cond}) --- is thus
\begin{align}\label{bcrit_noniso}
 \beta_\mathrm{crit}\to\beta_\mathrm{crit, gen} =
 \frac{h|s|}{\gamma - \Gamma}. 
\end{align}

We test $\beta_\mathrm{crit, gen}$ using a disk model with 
$(p,q, h)=(-0.5,-1,0.05)$ and $(\gamma,
\Gamma)=(1.4,1.3)$. For this setup, $s=-0.85$ and
$\beta_\mathrm{crit,gen}=0.425$, which is $\sim 3$ times larger than
our fiducial, vertically isothermal disk with $\beta_\mathrm{crit}=0.125$. 

We set the vertical domain size just inside the
physical, zero-density disk surface, $\zmax = 0.99H_s\simeq 2.6H$, see Eq.\ \ref{eqm_dens}. 
At this surface, the vertical buoyancy frequency, $N_z$,  diverges as $|z|\to H_s$.  While we might expect
pathological behavior for this model, we fortunately do not find it.

\begin{figure}
  \includegraphics[width=\linewidth,clip=true,trim=0cm 0cm 0cm
  0cm]{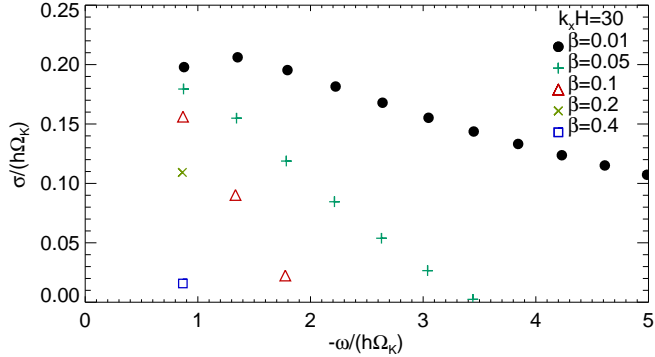}
  \caption{Unstable modes in a vertically non-isothermal disk with
    $\khat=30$ and a range of thermal relaxation timescales.  
    \label{compare_modes_vnoniso_kx10}}
\end{figure}

In Fig. \ref{compare_modes_vnoniso_kx10} we plot the mode diagram for
$\khat=30$ and several values of $\beta$. This plot is 
qualitatively similar to the vertically isothermal case, Fig. \ref{compare_modes_cool_kx10}. As before, 
larger $\beta$ values rapidly
stabilize higher order modes.  For sufficiently large $\beta$ only the fundamental mode is unstable, with a growth rate
that decreases as $\beta$ approaches 
$\beta_\mathrm{crit,gen}$.

\begin{figure}
  \includegraphics[scale=0.4415]{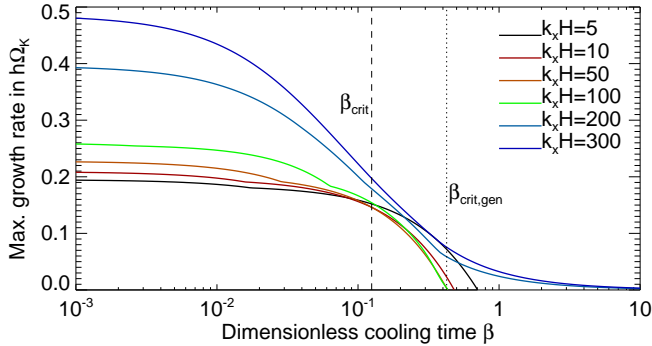} 
  \caption{Maximum VSI growth rates in the vertically non-isothermal disk
    model as a function of the thermal relaxation timescale
    $t_c=\beta\OmK^{-1}$. The vertical dashed line is the critical thermal
    timescale given by Eq. \ref{iso_vsi_cond}. A conjectured critical thermal
    timescale accounting for a non-isothermal background, given by 
     $\beta_\mathrm{crit,gen}= 0.425$ (see Eq. \ref{bcrit_noniso}),   
    is also shown as the vertical dotted line. 
    \label{bcrit_compare2}}   
\end{figure}

Fig.\ \ref{bcrit_compare2} shows how maximum VSI growth
rates vary with $\beta$ in the vertically non-isothermal model. The qualitative
behavior is similar to Fig.\ \ref{bcrit_compare1} for the vertically
isothermal disk. For $\khat=O(10)$, growth rates drop to zero at the
predicted $\beta_\mathrm{crit,gen}$.  As in the vertically isothermal case,
modes with both lower and higher wavenumbers  exhibit slow growth rates 
beyond the critical cooling time. As argued in \S\ref{bcrit_num_test},
the critical cooling time is useful despite not being an absolute stability limit.

The generalized cooling time criterion in Eq.\ \ref{bcrit_noniso}
thus appears to be valid.  If a disk's vertical structure is
not characterized by a single polytropic index, $\Gamma$, we 
expect that a density weighted average of  $\beta_\mathrm{crit,gen}$
should be a good approximation.  This expectation remains to be tested.

%% file: application.tex
\section{The VSI with realistic  protoplanetary disk cooling}\label{application} 
To this point we have treated the cooling time, $\beta$, as a free
parameter that, moreover, is independent of height.  
To understand the applicability of the VSI to PPDs, we now consider
$\beta$ values that are consistent with PPD models. 
Section \ref{cooling_model}  develops our PPD cooling law, which
depends on disk radius, disk height and perturbation wavelength. 
We compare these PPD $\beta$ values to $\beta_{\rm crit}$ in
\S\ref{bcritPPD}.  While this comparison is instructive, it is not
complete since $\beta_{\rm crit}$ was derived assuming a vertically
constant $\beta$.  Thus in \S\ref{vsi_mmsn_grow} we analyze the linear
growth of VSI with our PPD $\beta$ values. 



\subsection{Cooling model} \label{cooling_model}

\subsubsection{Radiative diffusion and Newtonian cooling}
The thermal relaxation of a perturbation depends on the relative sizes of the perturbation's 
lengthscale, $l$, and the photon mean-free-path,
\begin{align}\label{lrad}
  l_\mathrm{rad} \equiv \frac{1}{\kappa_d\rho} 
\end{align}
where $\kappa_d$ is the opacity, specifically the dust opacity appropriate for cold PPDs. 

In the optically thick limit, $l\gg l_\mathrm{rad}$, radiative diffusion smooths out thermal perturbations.
In this regime, the linearized cooling function is

\begin{align}\label{diff_cool_proper}
  \delta \Lambda_\mathrm{diff} = \frac{P}{\rho T C_v} \nabla\cdot\left(k_\mathrm{rad}\nabla\delta
    T\right),  
\end{align}
where $k_\mathrm{rad}$ is the radiative conduction coefficient defined
below, and $C_v$ is the specific heat capacity at constant volume. 

Since the VSI is characterized by vertically-elongated,
radially-narrow disturbances, we retain only the radial derivatives of
the perturbations in Eq. \ref{diff_cool_proper}. 
Thus, in the radially local approximation, we have
\begin{align}\label{diff_cool_approx}
  \delta\Lambda_\mathrm{diff} \simeq 
  -\eta k_x^2 P \frac{\delta T}{T}, 
\end{align}
where the radiative diffusion coefficient is
\begin{align}\label{eta_def}
  \eta \equiv {k_\mathrm{rad} \over \rho C_v} = \frac{16\sigma_s T^3}{3\kappa_d\rho^2 C_v}, 
\end{align}
and $\sigma_s$ is the Stefan-Boltzmann constant. 
The thermal relaxation  timescale (defined in Eq. \ref{thermal_relax}) 
for radiative diffusion is thus 
\begin{align}\label{tc_diff_cool} 
  t_\mathrm{diff} = \frac{1}{\eta k_x^2}.
\end{align}

In the optically thin regime, $l\ll 
l_\mathrm{rad}$, thermal relaxation operates by `Newtonian cooling'.
The cooling time is independent of $l$ and 
inversely proportional to the opacity, $t_\mathrm{thin} \propto 1/\kappa_d$. 
Specifically 
\begin{align}
  t_\mathrm{thin} = \frac{l_\mathrm{rad}^2}{3\eta},
\end{align}
and $t_\mathrm{thin}$ does not depend on $\rho$ (because our adopted
$\kappa_d$ depends on $T$ only, see below). 

Our general cooling time for all perturbations,
\begin{align}\label{tc_def}
  t_c &\equiv t_\mathrm{thin} + t _\mathrm{diff},
\end{align}
is a simple prescription to smoothly match the optically thick and thin limits.

\subsubsection{PPD cooling times}\label{toy_relax}
In a vertically isothermal disk with surface density $\Sigma = \rho_0\sqrt{2\pi}H$, 
the dimensionless thermal
relaxation time $\beta$ becomes 
\begin{align}\label{real_beta}
  \beta(r,z;\khat) \equiv t_c\OmK =
  \frac{\Sigma^2\OmK}{\eta\rho^2}\left[\frac{1}{3\kappa_d^2\Sigma^2} 
    + \frac{\hat{\rho}^2(z)}{2\pi \khat^2}\right].
\end{align}
where $\hat{\rho} = \rho/\rho_0$.  The first and second terms in
square brackets represent the optically thin and thick cooling
regimes, respectively. 

We adopt the Minimum Mass Solar Nebula
(MMSN) disk model of \cite{chiang10} which specifies
\begin{subequations}
\begin{align}
\label{mmsn_sigma}
  \Sigma &= 2200 
 r_\mathrm{AU}^{-3/2} \mathrm{g}\,\mathrm{cm}^{-2},  \\
 T &= 120
  r_\mathrm{AU}^{-3/7} \mathrm{K}, \label{mmsn_temp}  
\end{align}\end{subequations}
with $r_\mathrm{AU}\equiv r/\mathrm{AU}$. This model has $q = -3/7$,
$p = -39/14$ and $h  = 0.022 r_\mathrm{AU}^{2/7}$, using $\mu = 2.33$
and the above relation between $\Sigma$ and $\rho_0 \propto r^p$.   

For the dust opacity we adopt the \citet{bell94} law with $\kappa_d
\propto T^2$, giving 
\begin{align}
 \kappa_d 
    =
   2.88\hat{\kappa}_d r_\mathrm{AU}^{-6/7}\mathrm{cm}^2\,\mathrm{g}^{-1},   
\end{align}
where the opacity normalization factor, $\hat{\kappa}_d$, scales with
the ratio of small dust to gas;  $\hat{\kappa}_d = 1$ is our fiducial
case. 


For this disk and opacity model, the cooling time becomes
\begin{align}\label{beta_mmsn_simp}
  \beta(r,z;\khat)\notag = &3.9\times10^{-3}\hat{\kappa}_d^{-1}
  r_\mathrm{AU}^{9/14}
  \notag\\ &\times \left[1 +
    1.9\times10^7\hat{\kappa}_d^{2}\hat{\rho}^2(z)r_\mathrm{AU}^{-33/7}\khat^{-2}\right]. 
\end{align}
Perturbations are in the optically thin regime for
\begin{align}
  \khat\gtrsim 2.5\times10^3\hat{\kappa}_d\hat{\rho}(z)r_\mathrm{AU}^{-33/14}.  
\end{align}
In the inner disk, e.g. $r_\mathrm{AU}\sim 1$, only
extremely small-scale perturbations are in the optically-thin
regime near the midplane. However, in the outer disk, e.g. $r_\mathrm{AU}\sim 10$, 
 more moderate wavenumbers $\khat\gtrsim 10$ experience optically-thin cooling.

\begin{figure}
  \includegraphics[width=\linewidth,clip=true,trim=0cm 0cm 0cm
  0cm]{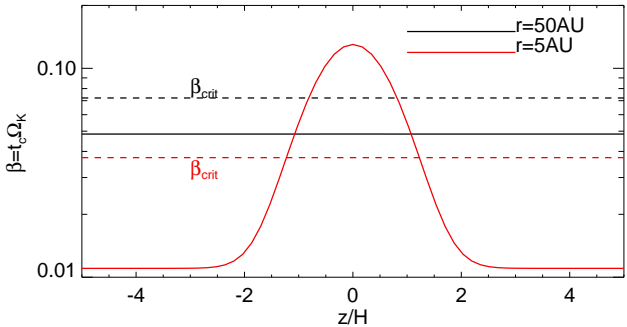}
  \caption{Thermal relaxation timescales in the fiducial MMSN at $r=50$AU
    and $r=5$AU for $\khat=30$ (solid lines). The
    corresponding horizontal dashed lines are the critical thermal
    relaxation timescales derived in linear theory. 
    \label{beta_compare}}
\end{figure}

\begin{figure*}
  \includegraphics[scale=.47,clip=true,trim=0cm 1.8cm 0cm
  0cm]{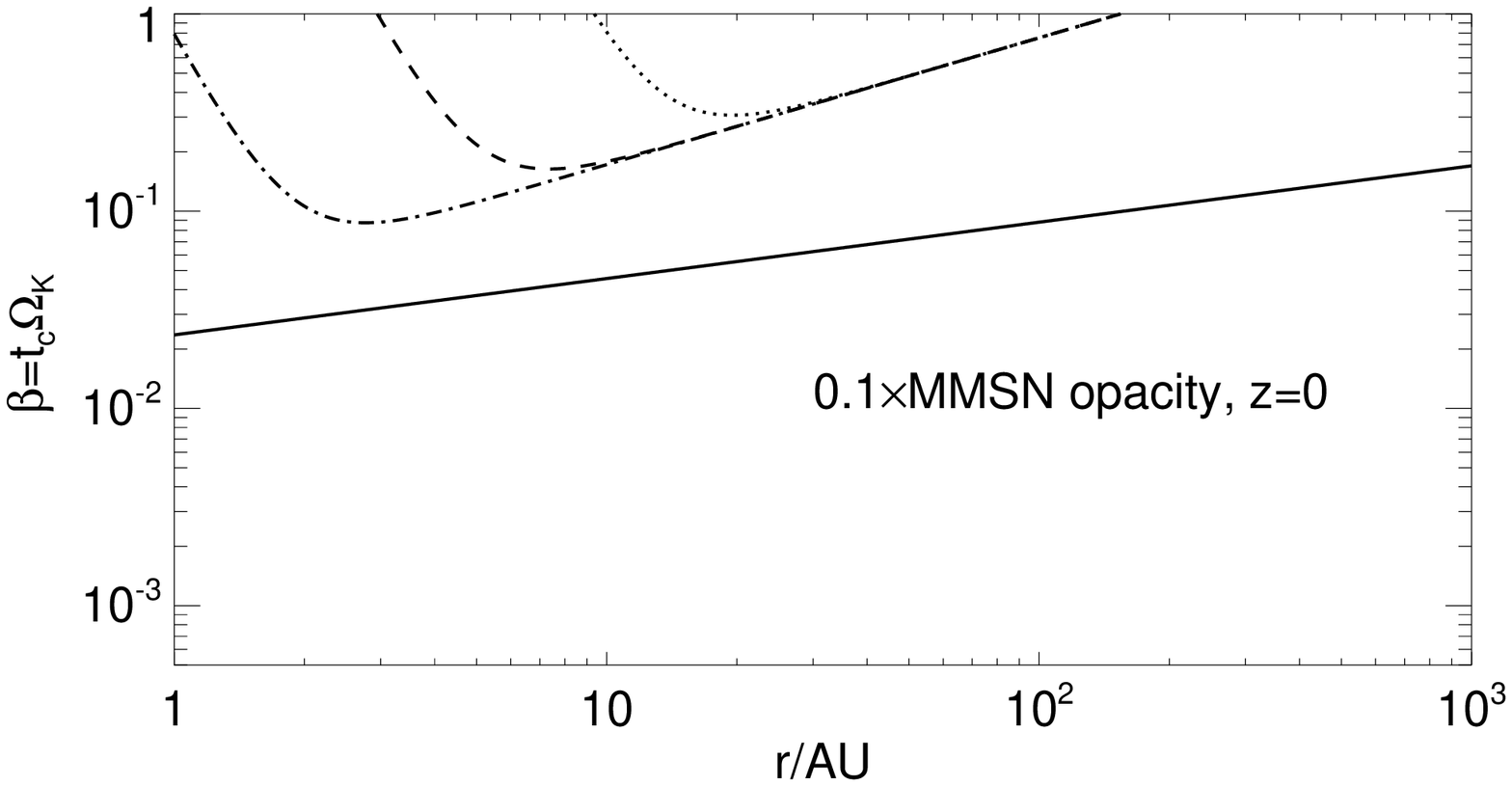}\includegraphics[scale=.47,clip=true,trim=2.5cm 1.8cm 0cm
  0cm]{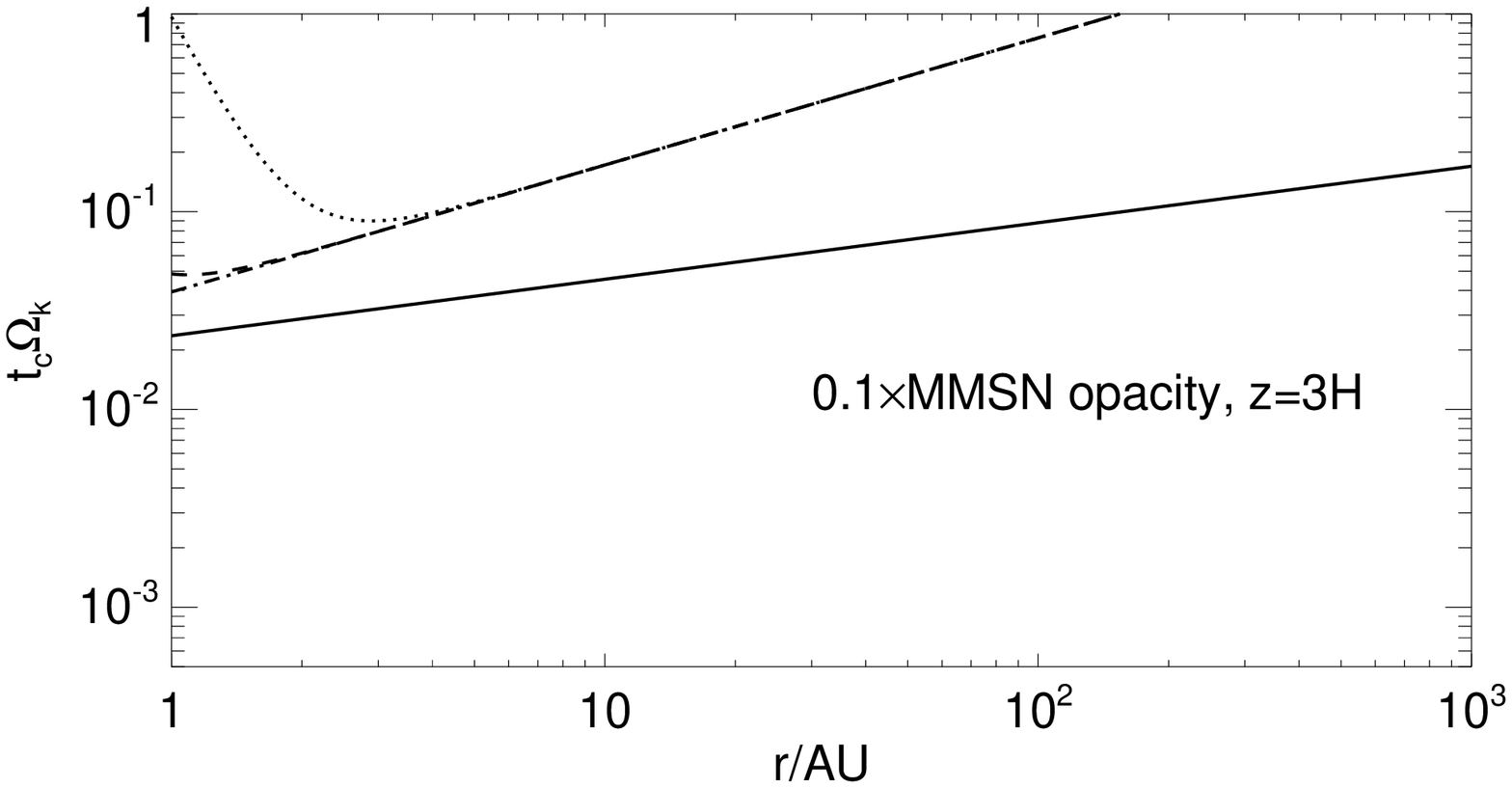}\\ 
  \includegraphics[scale=.47,clip=true,trim=0cm 1.8cm 0cm
  1cm]{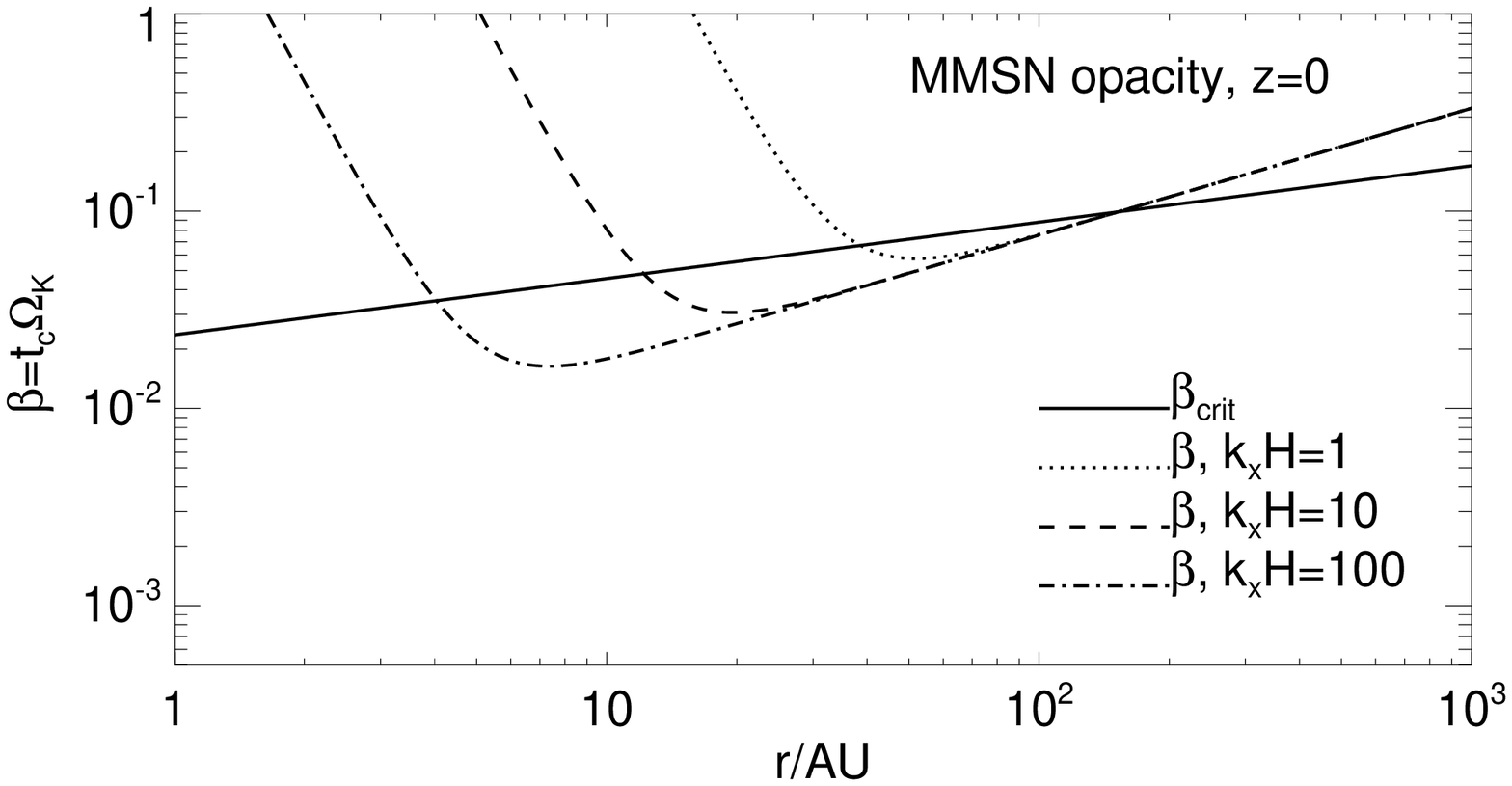}\includegraphics[scale=.47,clip=true,trim=2.5cm
  1.8cm 0cm 1cm]{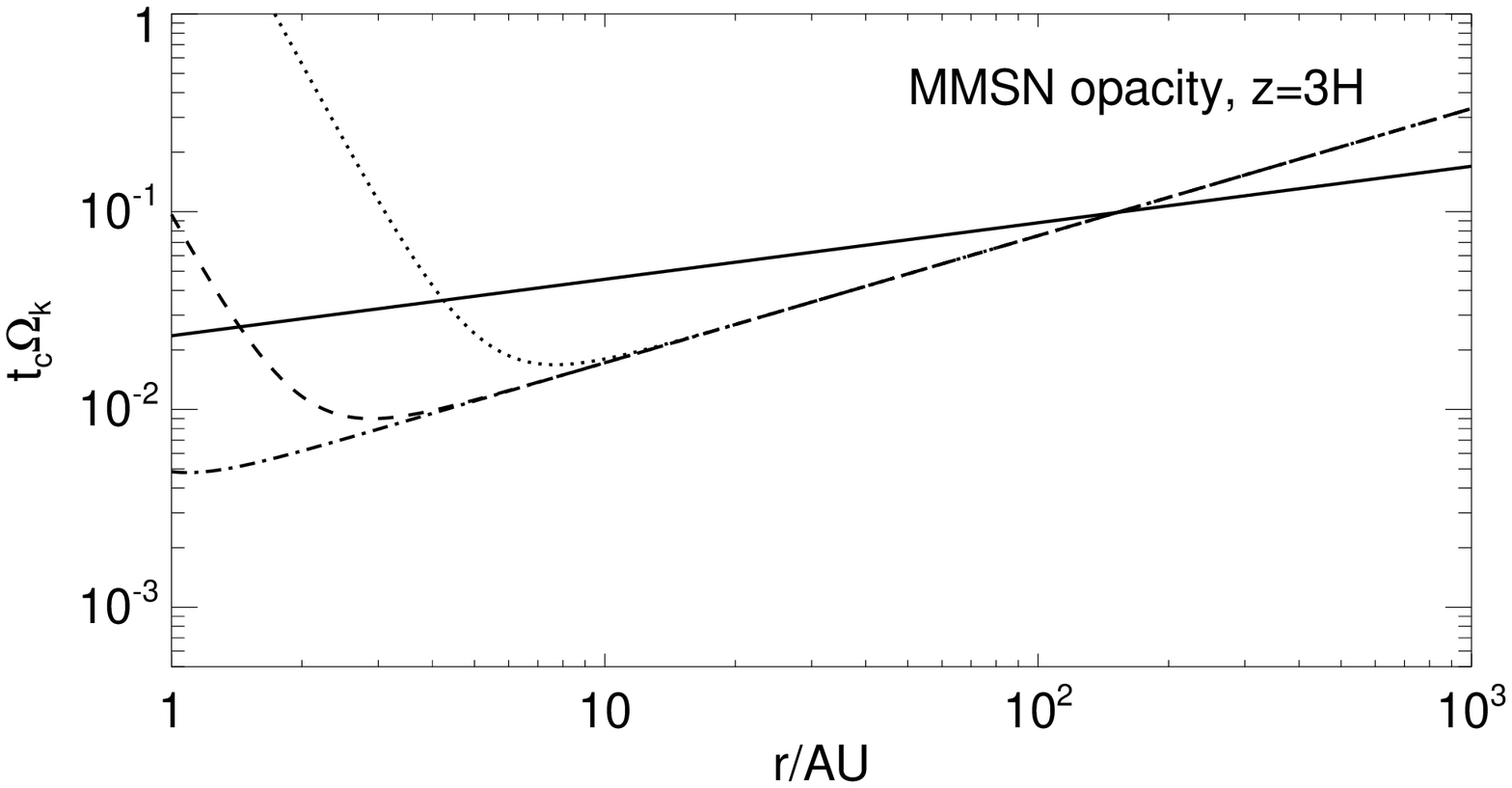}\\
  \includegraphics[scale=.47,clip=true,trim=0cm 0cm 0cm
  1cm]{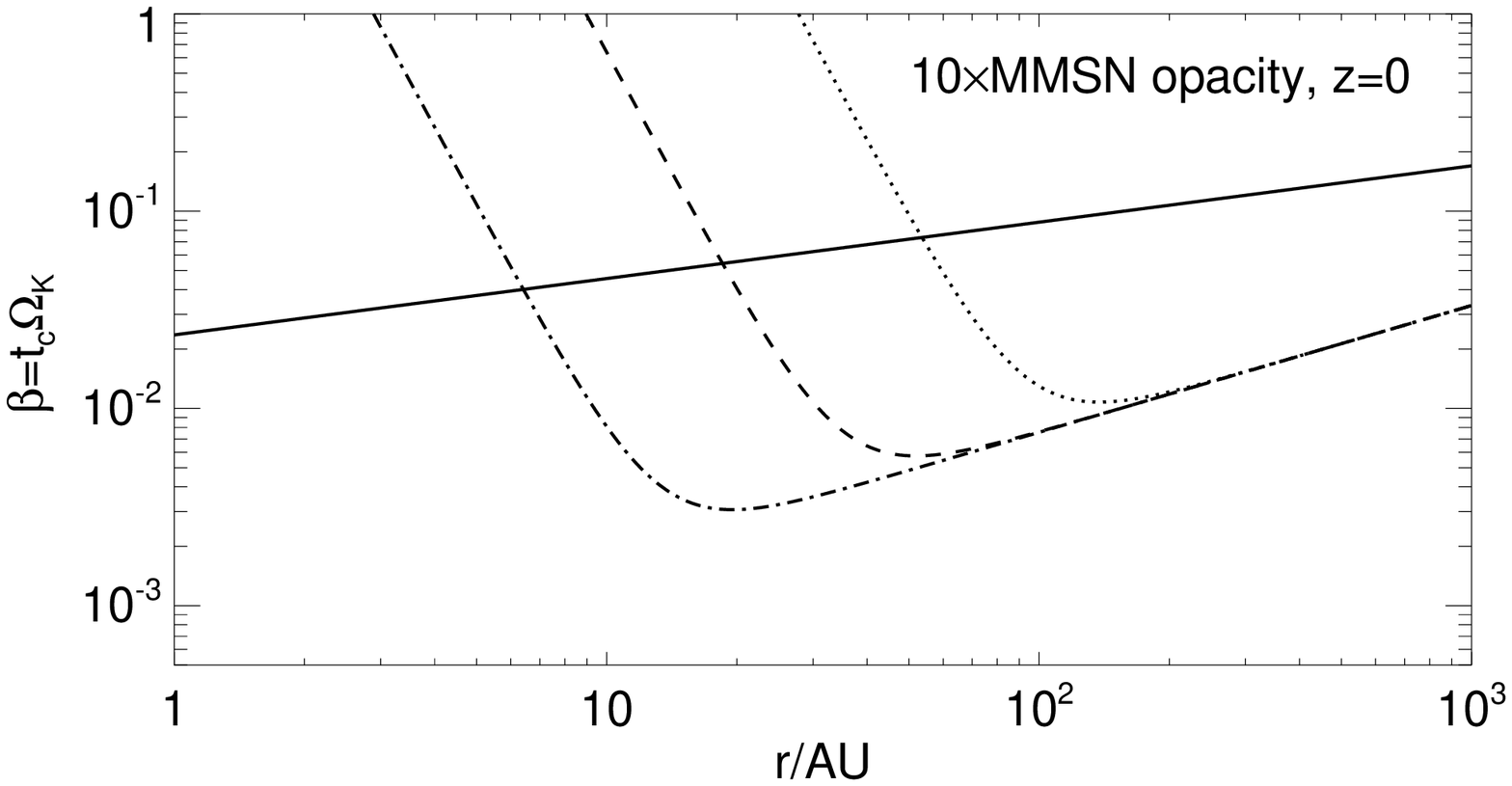}\includegraphics[scale=.47,clip=true,trim=2.5cm 0cm 0cm
  1cm]{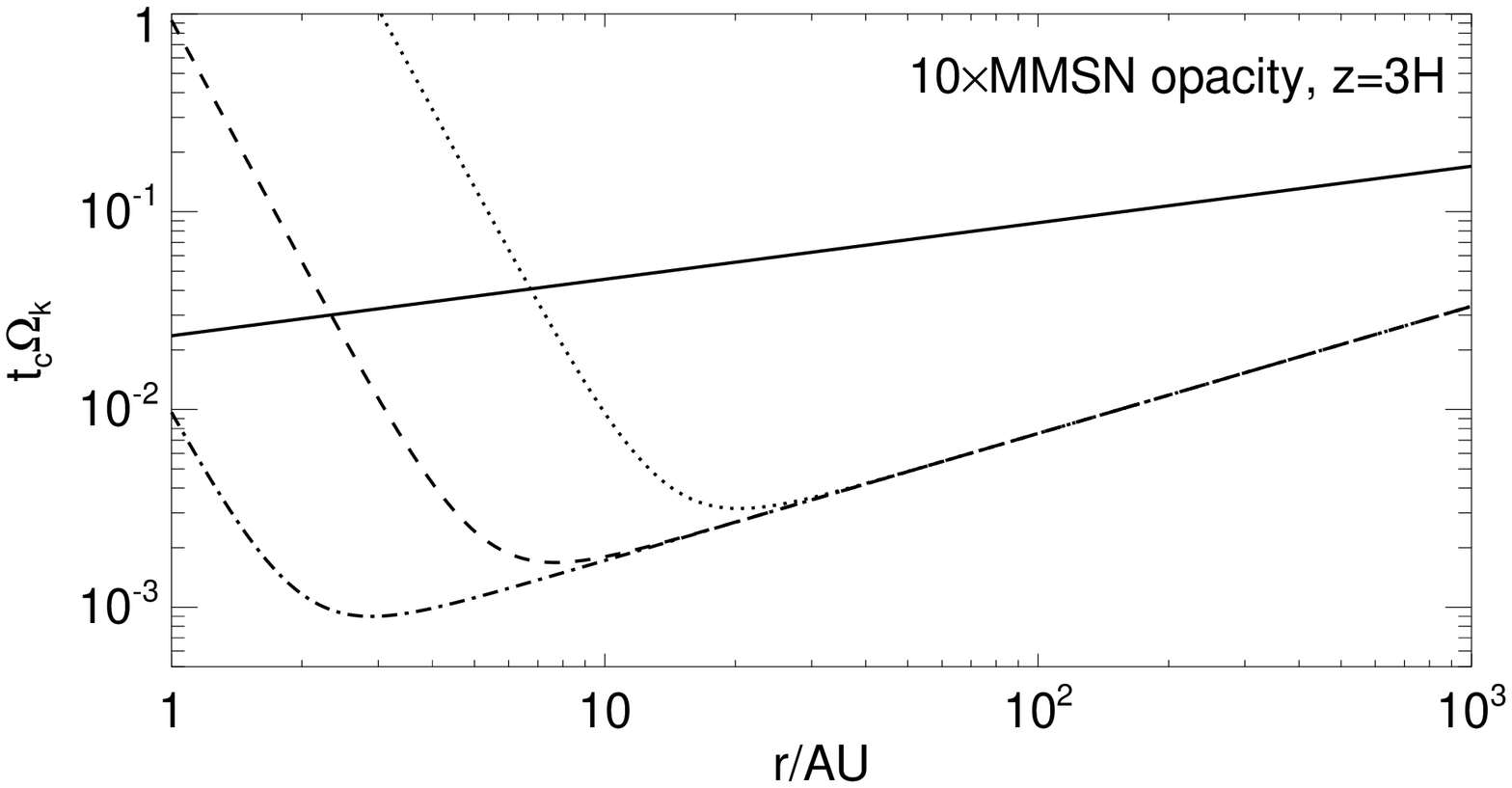} 
  \caption{Dimensionless thermal relaxation timescales $\beta$,
    evaluated at the midplane (left) and at $z=3H$ (right) in the
    fiducial PPD. Eq. \ref{beta_mmsn_simp} is plotted  
    for three values of the 
    perturbation radial wavenumber: $\khat=1$ (dotted), $\khat=10$
    (dashed) and $\khat=100$ (dashed-dot), for three values of the
    opacity relative to the MMSN: $\hat{\kappa}_d=0.1$ (top),
    $\hat{\kappa}_d=1$ (middle) and $\hat{\kappa}_d=10$ (bottom).  
    The solid line is the 
    critical thermal relaxation timescale $\beta_\mathrm{crit}$.  
    \label{mmsn_bcrit_bcool}}   
\end{figure*}  

\subsection{$\beta$ vs.\ $\beta_\mathrm{crit}$ in PPDs}\label{bcritPPD} 
The simplest way to estimate whether the VSI can operate at a given radius in a PPD is to 
compare the cooling time (Eq. \ref{beta_mmsn_simp}) to 
its critical value (Eq. \ref{iso_vsi_cond}), evaluated for the MMSN as    
\begin{align}\label{bcrit_mmsn}
  \beta_\mathrm{crit} = 0.024r_\mathrm{AU}^{2/7}. 
\end{align}

This comparison is strictly valid only for optically thin cooling
which is independent of height, as assumed in the derivation of
$\beta_\mathrm{crit}$.  This condition typically holds in the outer
disk.   Fig. \ref{beta_compare} shows that a  $\khat=30$ perturbation
at 50AU has vertically constant $\beta$. Furthermore, since $\beta <
\beta_\mathrm{crit}$, the disturbance would be unstable. 
 
For optically thick cooling, $\beta$ varies with height, complicating
the comparison with $\beta_\mathrm{crit}$.    
At 5AU, Fig. \ref{beta_compare} shows that cooling times are long in
the midplane with $\beta > \beta_\mathrm{crit}$. 
However, since $\beta < \beta_\mathrm{crit}$ away from the midplane,
we require the analysis of \S\ref{vsi_mmsn_grow} to determine if the
VSI can grow.  (That calculation will show that growth is strongly
suppressed for this case.)  We proceed with the awareness that
optically thick regions in the inner disk are a complication, but that
we can reasonably expect VSI growth if $\beta < \beta_\mathrm{crit}$
at all heights.

%
%

In Fig. \ref{mmsn_bcrit_bcool}, we compare $\beta$ to
$\beta_\mathrm{crit}$ across a range of disk radii for different
heights, opacity values and wavenumbers. 
For optically thin cooling in the outer disk, curves for different
wavenumbers overlap, as expected from Eq.\ \ref{beta_mmsn_simp}. 
Since the (optically thin) slope of $\beta$ is steeper than
for $\beta_\mathrm{crit}$, VSI growth can be suppressed at large radii
(for the chosen opacity law).  This effect is seen for $\hat{\kappa}_d
= 1$ in the central panels of Fig.\ \ref{mmsn_bcrit_bcool}, where VSI
is damped outside $\sim 150$ AU. 

We find that the most important factor for VSI growth is the opacity.
With a smaller opacity, $\hat{\kappa}_d = 0.1$, growth is suppressed
at all radii, as shown in the top panels of Fig.\
\ref{mmsn_bcrit_bcool}.  Since optically thin cooling is too slow in
this case, optically thick cooling --- above the floor set by optically
thin cooling --- is also too slow.   

Larger opacities make optically thin cooling much faster than $\beta_\mathrm{crit}$, as shown in the top panels of Fig.\ \ref{mmsn_bcrit_bcool}.  However with larger opacities, optically thick cooling affects larger disk radii, slowing the cooling.  Remarkably, the adopted MMSN value of opacity hits the a sweet spot where optically thin cooling is fast enough, yet slower optically thick can be restricted to inner disk radii.

At smaller disk radii,  Fig. \ref{mmsn_bcrit_bcool} shows the hallmark of diffusive cooling, that larger wavenumbers can cool faster, but not below the floor set by optically thin cooling.  Optically thick cooling times also rise sharply toward smaller radii (as $\beta \propto r^{-57/14}$) due to high densities and short orbital times.  This effect suppresses VSI growth at small radii, but with a strong wavenumber dependence.

\begin{figure}
  \includegraphics[width=\linewidth]{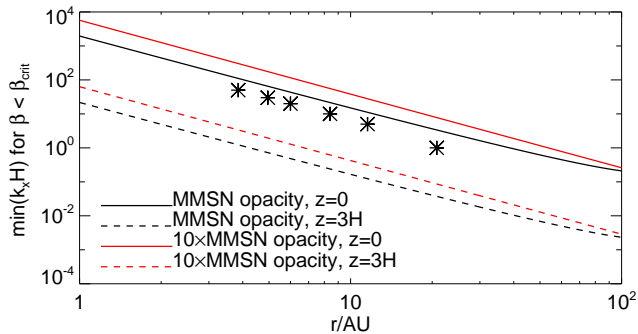} 
  \caption{The minimum perturbation wavenumber $\khat$ in
    the MMSN such that the associated dimensionless thermal
    relaxation time $\beta$ at $z=0$ (solid) and $z=3H$ (dashed) is
    less than the critical timescale $\beta_\mathrm{crit}$. Asterisks
    correspond to the inner cut-off radii shown in
    Fig. \ref{mmsn_overall}.   
    \label{mmsn_bcrit_bcool_mink}}   
\end{figure}  

Fig. \ref{mmsn_bcrit_bcool_mink} highlights this wavenumber dependence by plotting
the wavenumbers for which $\beta = \beta_\mathrm{crit}$.  Above the
solid curves (i.e.\ for larger wavenumbers), $\beta <
\beta_\mathrm{crit}$ at all disk heights, implying linear VSI growth.
Below the dashed curves, VSI growth is strongly suppressed since
$\beta >  \beta_\mathrm{crit}$ for all $|z| < 3H$.  Between the solid
and dashed curves some growth is possible, but only fairly close to
the solid curve (according to \S\ref{vsi_mmsn_grow}).  Thus linear
growth of the VSI near 1 AU is only possible if $k_x \gtrsim 10^3/H$.
As argued elsewhere, such small-scale disturbances may not 
drive significant turbulence or transport. 

We thus doubt that the VSI is significant at 1 AU in MMSN-like PPDs,
for any opacity.  At higher opacities, the required wavenumbers become
shorter and even more problematic, as shown by the red curves in
Fig. \ref{mmsn_bcrit_bcool_mink}.  At lower opacities, even optically
thin cooling (the fastest possible) is too slow, as discussed above.  



\subsection{VSI growth in the MMSN}\label{vsi_mmsn_grow}
We now consider the growth of the VSI in a MMSN disk model.  This numerical calculation is similar to   
\S\ref{numerical} but with different disk parameters and with the
self-consistent cooling times of  Eq. \ref{beta_mmsn_simp}. 

We consider the growth timescales of the fundamental mode for a range of
wavenumbers and disk radii. We focus on the fundamental, i.e. lowest order vertical, mode 
because it is the fastest growing mode except for some surface modes at high 
wavenumbers. We neglect these surface modes for reasons discussed in
\S\ref{numerical} and \S\ref{caveats_visc}. 

\begin{figure}
  \includegraphics[width=\linewidth]{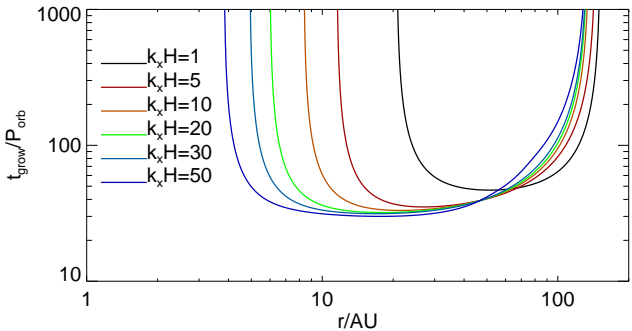}
  \caption{VSI growth times ($t_{\rm grow} = 1/\sigma$) in orbital units ($P_\mathrm{orb} = 2\pi/\OmK$) for
    the reference MMSN disk model with self-consistent dust cooling. 
    \label{mmsn_overall}}    
\end{figure}

Fig. \ref{mmsn_overall} shows that in the MMSN, the VSI  is active
from $r_\mathrm{AU}\sim 5$ to $r_\mathrm{AU}\sim 
50$ with growth timescales $\sim 30$---40 orbits.  A small radius
cut-off exists, inside of which growth is strongly suppressed.  The 
cut-off occurs at smaller radii for larger wavenumbers, as expected
from Fig.\ \ref{mmsn_bcrit_bcool_mink}.  Growth at smaller disk radii
is possible for yet larger wavenumbers, but we remain skeptical about
the non-linear significance of such short lengthscales. 

The growth times rise gradually as $r$ increases towards 100AU in
Fig. \ref{mmsn_overall}.  This trend is expected as the outer radius
cutoff at $\sim 150$AU (from Fig.\ \ref{mmsn_bcrit_bcool}) is
approached. Our numerical results suggest the VSI is efficient at
radii of few tens of AU in the MMSN, consistent with estimates made by
\citetalias{nelson13} using the same disk model.   

%



%% file: summary.tex
\section{Caveats and Neglected Effects}\label{caveats} 

\subsection{Viscosity}\label{caveats_visc}

Our neglect of viscosity is valid in a laminar disk, because molecular viscosity is so small.
However, turbulence could act as an effective viscosity which preferentially damps 
small scale modes.   Since the goal of VSI is to drive turbulent transport,
the most relevant modes for sustaining the VSI should be able to overcome
turbulent damping.

To roughly estimate the wavelengths which are prone to damping, we
consider 
the standard
prescription for the kinematic viscosity $\nu = \alpha c_s H$
\citep{shakura73}, in terms of the dimensionless $\alpha$ 
parameter. The viscous timescale for a perturbation lengthscale  
$l\sim 1/k_x = H/\khat$ is $t_\mathrm{visc} = 1/\alpha\khat^2\Omega $. 
For significant growth, $t_\mathrm{visc}$ should be longer than the
characteristic VSI growth timescale $1/(h|q|\Omega)$, i.e.\ 
$\khat^2\lesssim |q|h/\alpha$. With $h \sim 0.05$, $q = -1$  and
 $\alpha$ from simulations \citepalias{nelson13},
we estimate that growth requires $\khat\lesssim 20$ for
$\alpha\sim 10^{-4}$ or $\khat\lesssim 70$ for $\alpha\sim
10^{-5}$. 

This argument justifies our focus on moderate
wavenumbers, $\khat=O(10)$. Future work should consider
the viscous effects in more detail, to better understand
how the VSI operates in nature and in simulations.


\subsection{Convective overstability}
The VSI bears some similarity to the `convective overstability'
 (CONO, \citealp{klahr14, lyra14}).   Both instabilities rely on thermal relaxation to 
overcome the stability of disks to adiabatic perturbations, i.e.\ the Solberg-Hoiland
criteria (see \S\ref{solberg}).

The key differences are that the CONO neglects vertical buoyancy
and requires an imaginary horizontal buoyancy,
\begin{align}\label{Nr}
N_r^2 \equiv -{1 \over \rho C_P}{\p P \over \p r} {\p S \over \p r} < 0\, .
\end{align}
Since vertical buoyancy is an important stabilizing influence for the VSI, it should
be considered in future studies of CONO.

The sign of $N_r^2$ depends on disk parameters.  To demonstrate that the parameters needed for $N_r^2 <0$
are somewhat extreme, or at least non-standard, we consider the midplane of vertically isothermal disks with 
$\Sigma \propto r^n$ so that $n = p+q/2 + 3/2$. Using Eqs. \ref{dSdr} and \ref{Nr} with $\gamma = 1.4$,
$N_r^2 < 0$ requires $n> 3(1/2 + q)$.  For standard disk temperature laws, this requirement implies
a flat or rising surface density profile, e.g.\ $n = 3/14$ for $q = -3/7$ or $n = 0$ for $q = -1/2$.
Since standard $\Sigma$ profiles decline with radius, CONO is most like to operate at 
special locations, like the outside edges of disk gaps or holes and/or shadowed regions where $q < -1/2$.

Perhaps due to these physical differences, CONO operates in different regimes of parameter
space than the VSI.  The CONO grows best for longer wavelengths $\khat \lesssim 1$ and longer 
cooling times $\beta \sim 1$. 

Future work should aim to understand these related instabilities in the same framework, thereby 
explaining the key differences.

\subsection{Radiative transfer} 
Our relatively simple treatment of cooling with an idealized dust opacity
could certainly be generalized in future works.  In hotter disks, gas phase opacities
must be considered.   In cold disks, there are many choices for the dust opacity, which 
varies with grain sizes and compositions. Changing dust properties would alter 
the viability of the VSI for better or worse.  The radiative transfer itself
could be calculated with higher levels of sophistication, as is already being done in
numerical simulations of the VSI \citep{stoll14}.

\section{Summary and discussion}\label{summary}
In this paper we study the vertical shear instability (VSI) with a focus the 
role of radiative cooling.  In turn we assess the viability of the VSI 
as an angular momentum transport mechanism in protoplanetary disks (PPDs).

Our linear, axisymmetric analysis of the VSI 
considers (uniquely to our knowledge) finite cooling times in a vertically global model.  
In order for vertical shear to drive the VSI, short cooling times are needed to weaken 
the stabilizing effects of vertical buoyancy. Our main analytical finding, which we confirm numerically, 
is the critical cooling  timescale above which VSI growth is suppressed, Eqs.\ \ref{prelim_bcrit} and \ref{iso_vsi_cond}.

Our main focus is irradiated, vertically isothermal disks which have strong vertical buoyancy.
The critical cooling time is thus short, shorter than the orbital time by a factor of the disk
aspect ratio.  This finding is consistent with, and helps explain, the results of recent numerical simulations \citepalias{nelson13}.
We briefly consider  vertically non-isothermal disks in \S\ref{nonvertiso}.

In applying our results to PPDs, we pay particular attention to the transition from 
optically thick radiative diffusion to optically thin Newtonian cooling.  The largest obstacle to VSI occurs in 
high density inner disk, $\lesssim 1$ -- 5AU, where radiative diffusion times are slow.  Shorter wavelength disturbances
 cool faster, so long as they remain optically thick.  In the inner disk, however, our VSI cooling criterion requires 
 wavelengths that are too short to drive significant transport.  The best hope for the VSI in inner disk is a low surface density, which speeds radiative diffusion by lengthening the photon mean free path.  This option naturally begs 
 the question of what accretion mechanism lowered the surface density in the first place. 
 
 In the outer disk, the VSI tends to cool in the optically thin limit.  The main issue is whether the opacity 
 is high enough for optically thin cooling to be sufficiently fast.  Our standard opacity assumes a Solar abundances of small dust.
 In this case, the VSI can operate from $\sim 5$ -- 100AU.  
A factor of 10 reduction in the opacity, for instance by locking
 small dust into planetesimals and planetary cores \citep{youdin13}, makes cooling times too slow for VSI growth.  
 In this case cooling is too slow not just in the outer disk, but into  $< 1$ AU.
 
 An enhanced opacity makes optically thin cooling faster and radiative diffusion slower.  This shift favors VSI growth 
 in the outer disk at the expense of the inner disk, pushing the inner limit of VSI growth further out.
 The standard choice --- corresponding to Solar abundances in a standard MMSN disk --- allows the VSI 
 to grow over the widest range of relevant disk radii.
 
 Our detailed study of the spectrum of VSI modes confirms that some artificial `surface modes' are 
 triggered by imposed vertical boundaries \citepalias{nelson13, barker15}.  Domain size
 convergence studies are thus essential.  Fortunately, our results show that longer cooling times 
  stifle the growth of surface modes.  Thus, at least in some cases, more realistic radiative transfer also produces more reliable dynamics.
 
 The VSI deserves further study as a viable mechanism to drive at least low levels of accretion in cold disks.

%% file: appendix.tex
\section{Derivation of the approximate equations}\label{adia_improve}
Here we detail the derivation of Eq. \ref{nearly_iso_explicit} used in the 
analytical discussion of \S\ref{analytical}.  Starting with Eqs.\ \ref{lin_mass} --- \ref{lin_energy},
we set $\Gamma = 1$ and $\zeta = 0$ for the vertically isothermal limit and the fully-radially-local 
approximation, respectively.  We  eliminate
the horizontal velocity perturbations ($\delta v_x,\, \delta v_y$) to
obtain  
\begin{subequations}\label{A1}\begin{align}
  & \ii\upsilon\delta v_z = \frac{dW}{dz} + \frac{\p\ln{\rho}}{\p z}\left(W-Q\right),\label{ode_vz} \\
  &\frac{\upsilon^2}{c_s^2}Q + \frac{\upsilon^2k_x^2}{D}W =\ii\upsilon \left(\frac{\ii
      k_x r}{D}\frac{\p\Omega^2}{\p z} -
    \frac{\p \ln{\rho}}{\p z}\right)\delta v_z - \ii\upsilon \frac{d\delta v_z}{dz},
  \label{ode_w}\\
  &\upsilon^2W - \gamma\upsilon^2 Q +
  \frac{\ii\upsilon}{t_c}\left(W-Q\right) 
  =\ii \upsilon c_s^2\frac{\p\ln{\rho}}{\p z}\left(\gamma -
    1\right)\delta v_z,
  %
  \label{ode_Q} 
\end{align}\end{subequations}
where
\begin{align}
  D \equiv \kappa^2 - \upsilon^2.
\end{align} 

Reduction to a single ODE requires 
$\p_zD = \p_z\kappa^2$.  At this point we could apply the 
low frequency and Keplerian approximations to set $D \rightarrow
\kappa^2 \rightarrow \OmK^2$, then $D$ is vertically constant, 
and we can obtain Eq. \ref{nearly_iso_explicit} more directly.
However, to demonstrate that the order of approximation is irrelevant,
we will retain $\p_z\kappa^2$ initially. Using 
Eq. \ref{kap2_def}---\ref{vertical_shear},  
\begin{align}\label{dkappa2}
  \frac{\p \kappa^2}{\p z} = 4 \frac{\p\Omega^2}{\p z} + r\frac{\p}{\p
    r}\frac{\p\Omega^2}{\p z} = -
  \frac{d\ln\rho}{d z}\frac{qc_s^2}{r^2} \left(
    \frac{3z^2}{r^2+z^2}-1\right) \equiv - \frac{d\ln\rho}{d z}\frac{qc_s^2}{r^2}F(r,z).
\end{align}
The function $F$ increases monotonically from $F=-1$ at $z=0$ to $F\to2$
as $|z|\to\infty $, so $|F|=O(1)$. 



We  eliminate $W,\, Q$ from Eqs.\ \ref{A1}, using  Eqs. \ref{chi} and \ref{dkappa2} to obtain  
\begin{align}\label{full_ode}
  0 =& \frac{d^2\delta v_z}{dz^2} + \left[1 + \frac{\ii k_x c_s^2
      q}{Dr} - \frac{k_x^2c_s^2}{\left(k_x^2c_s^2 + \chi
        D\right)}\frac{qc_s^2F}{Dr^2}\right]\frac{d\ln\rho}{dz}\frac{d\delta
    v_z}{dz} \notag\\
  &+ \left\{\upsilon^2\left(\frac{k_x^2}{D} +
      \frac{\chi}{c_s^2}\right) + \left(\chi + \frac{\ii k_x c_s^2
        q}{Dr}\right)\frac{d^2\ln\rho}{dz^2} -
    \frac{c_s^2}{D}\left(\frac{d\ln\rho}{dz}\right)^2\left(k_x^2 -
      \frac{\ii k_x q}{r}\right)
   \left[\left(1-\chi\right) +
     \frac{\chi}{\left(k_x^2c_s^2 + \chi D\right)}\frac{qc_s^2 F}{r^2}\right] 
   \right\}\delta v_z, 
\end{align}
where we have replaced $\p/\p z$ by $d/dz$ for the
fully-radially-local treatment. Now we make the low frequency and Keplerian
approximations, setting $D\to 
\OmK^2$, to give  
\begin{align}
  &\delta v_z ^{\prime\prime} + \left[1 + \ii h q\hat{k} -
    \frac{ \hat{k}^2}{
      \left(\hat{k}^2+\chi\right)}q h^2F\right]\ln\rho^{\prime}\delta v_z^\prime +
  \left\{\left(\chi + \ii h q
      \hat{k}\right)\ln\rho^{\prime\prime} - \ln\rho^{\prime
      2}\left(\khat^2 -
      \ii h
      q\hat{k}\right)\left[1 - \chi +
      \frac{\chi}{\left(\hat{k}^2+\chi\right)}qh^2F\right]\right\}\delta v_z \notag\\&=
  -\hat{\upsilon}^2\left(\hat{k}^2+\chi\right)\delta v_z\label{adia_diso3},
\end{align} 
in terms of dimensionless variables of Eq.\ \ref{nondim}. 
Retaining terms to first order in the disk aspect ratio, $h \ll 1$, gives
\begin{align}
  0= \delta v_z ^{\prime\prime} + \left(1 + \ii h
     q\hat{k}\right)\ln\rho^{\prime}\delta v_z^\prime 
   +
   \left[\hat{\upsilon}^2\left(\hat{k}^2+\chi\right) +
     \left(  \chi + \ii h q\hat{k}\right)\ln\rho^{\prime\prime}
     - \ln\rho^{\prime
       2}\left(\khat^2 -
       \ii h
       q\hat{k}\right)\left(1 - \chi\right)\right]\delta v_z.\label{vertiso_gov_nondim}
\end{align}
Approximating the density field by Eq.\ \ref{rhoisothin} then gives 
Eq. \ref{nearly_iso_explicit}. 

We can establish a correspondence between our  Eq. \ref{nearly_iso_explicit} and
Eq. 41 in \cite{lubow93}, which describes adiabatic axisymmetric waves in
a vertically isothermal disk without vertical shear.   Accounting for the required change of variables, 
the correspondence is exact after setting $q=0$ (no vertical shear) and
$\chi=1/\gamma$ (adiabatic flow) in our Eq. \ref{nearly_iso_explicit},
and applying the approximations in our \S \ref{sec:simplified} to \citeauthor{lubow93}.    




\section{Applicability of the fully-radially-local approximation}\label{global_corr}
In the fully-radially-local approximation, background radial
gradients are ignored except where it appears
implicitly in the expression for the vertical shear rate $\p_z\Omega$
(via Eq. \ref{vertical_shear}).  This is done by neglecting the terms proportional to $\zeta$
in Eq. \ref{lin_mass}, \ref{lin_vx} and \ref{lin_energy}, i.e. setting 
$\zeta=0$. (Nominally $\zeta=1$, which is used in our numerical study.)

For a power-law disk, the neglected radial gradients   
are $O(r^{-1})$, and they appear in comparison with terms of
$O(k_x)$. The neglected terms therefore have a relative magnitude of
$O(h/\khat)$, which is small for thin disks ($h\ll1$) 
and/or small radial wavelengths ($\khat\gg 1$).  We show in the
following sections that the fully-radially-local approximation only 
becomes invalid in the adiabatic limit, which is not the relevant
regime for the VSI. 

We comment that this approximation is equivalent to 
that adopted in the vertically global shearing box formalism
\citepalias[VGSB,][]{mcnally14}, which is an 
extension of the standard shearing box \citep{goldreich65} to
background shear flows that are height dependent.

   

\subsection{Spurious growth of adiabatic perturbations when
  $\zeta=0$}\label{analytic_adia}  
A limitation of the reduced model described in Appendix
\ref{adia_improve}, \S\ref{sec:simplified} and used in \S\ref{analytical}, is that it
cannot be employed for adiabatic flow when there is vertical shear, even
if $h/\khat\ll1$.  We explain this by setting  $\beta\to\infty$ and hence $\chi = 1/\gamma$ in 
Eq. \ref{vertiso_gov_nondim} 
to give 
\begin{align}
  0 =\dd v_z^{\prime\prime} + \left(1 + \ii h q
    \hat{k}\right)\left(\ln\rho^{\prime}\delta v_z\right)^\prime
  +\left\{\hat{\upsilon}^2\left(\frac{1}{\gamma}+\hat{k}^2\right) 
    -\left(\frac{\gamma-1}{\gamma}\right)\left[\ln\rho^{\prime\prime}+\hat{k}^2\left(1-\frac{\ii h  
          q}{\hat{k}}\right)\ln\rho^{\prime 2}\right]\right\}\delta v_z.\label{adia_iso3}
\end{align}
We multiply Eq. \ref{adia_iso3} by $\rho\delta v_z^*$ and
integrate vertically, assume boundary terms vanish when integrating by
parts, to obtain
\begin{align}
  \hat{\upsilon}^2\left(\frac{1}{\gamma} +
    \hat{k}^2\right)\int_{\zhat_1}^{\zhat_2}\rho|\delta
  v_z|^2 d\zhat 
  =&  \left(\frac{\gamma-1}{\gamma}\right)
  \int_{\zhat_1}^{\zhat_2}\rho|\delta v_z^\prime|^2 d\zhat
  +\frac{1}{\gamma}\int_{\zhat_1}^{\zhat_2}\frac{1}{\rho}|(\rho\delta
  v_z)^\prime|^2 d\zhat\notag\\
&+
  \left(\frac{\gamma-1}{\gamma}\right)\hat{k}^2\left(1-\frac{\ii h
      q}{\hat{k}}\right) \int_{\zhat_1}^{\zhat_2}\rho\ln\rho^{\prime
    2}|\delta v_z|^2 d\zhat
+ \ii h q \hat{k}
  \int_{\zhat_1}^{\zhat_2}\ln\rho^\prime(\rho\delta v_z^*)^\prime
  \delta v_z d\zhat.\label{adia_integral}
\end{align}
In the presence of vertical shear $q\neq0$, Eq. \ref{adia_integral}
 shows that $\hat{\upsilon}^2$ is complex for real $\khat$, which
indicates instability  for any value of  $\gamma>1$. This stability condition is inconsistent
  with the second Solberg-Hoiland criterion  
(Eq. \ref{solberg2}), which states that instability requires the disk
to be close to neutral stratification (i.e. $|\gamma-1|\ll 1$ for a
vertically isothermal disk). 
This spurious growth in the $\zeta = 0$ model arises 
 because we have retained the global radial temperature 
gradient to obtain vertical shear, 
but have ignored it elsewhere in the linear problem (as well as the
background radial density gradient). Nevertheless, we demonstrate below
that this inconsistency is unimportant for the VSI, which occurs 
for $\beta\ll1$,  not for adiabatic perturbations.   

\subsection{Effect of global radial gradients}
\begin{figure}
  \includegraphics[width=\linewidth,clip=true,trim=0cm 0.0cm 0cm
  0cm]{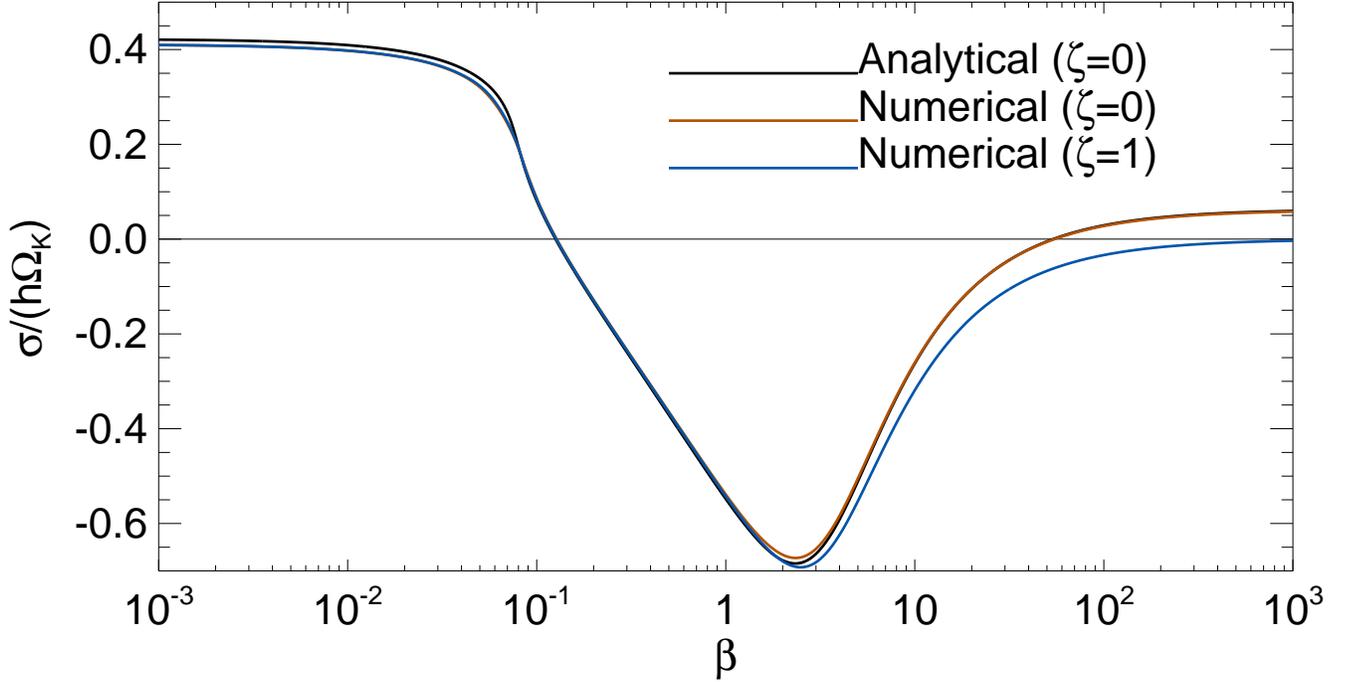} 
  \caption{Growth rate of the fundamental VSI mode as a function of
    the thermal relaxation time $\beta$. The `Analytical ($\zeta=0$)' curve
    is calculated from the dispersion relation Eq. \ref{relax_disp}. 
    The `Numerical ($\zeta=0$)' curve is obtained from the numerical
    solution to Eq. \ref{lin_mass}---\ref{lin_energy} with
    $\zeta=0$,  which 
    neglects explicit dependencies on the radial disk structure.  
    The `Numerical
    ($\zeta=1$)' curve is obtained from Eq. \ref{lin_mass}---\ref{lin_energy} with
    $\zeta=1$, which accounts for the radial disk structure self-consistently.   
    The disk parameters are $(\gamma, \Gamma)=(1.4, 1.011)$ and
    $(p,q,h)=(-1.5,-1,0.05)$. The perturbation 
    wavenumber is $\khat=30$. 
    \label{gcorr_compare}}  
\end{figure}

In  Fig. \ref{gcorr_compare}, we plot the effect of the global radial gradient terms
proportional to $\zeta$ in Eq. \ref{lin_mass}---\ref{lin_energy}
by calculating the
fundamental VSI growth rates using three approaches. We compute growth rates from the dispersion
relation Eq. \ref{relax_disp}, which assumes $\zeta=0$;  from
Eq. \ref{lin_mass}---\ref{lin_energy} with $\zeta=0$; and from
Eq. \ref{lin_mass}---\ref{lin_energy}  with $\zeta=1$.   

All three methods give similar behavior, and growth rates are in close
agreement for $\beta\lesssim 1$. Differences arise for 
$\beta\gtrsim1$, and as $\beta\to\infty$ the fully-radially-local
approximation, where $\zeta=0$, gives a
(spurious) growth rate as expected from the discussion
above. Inclusion of the global radial gradient terms results in the
expected behavior  ($\sigma\to0$ as $\beta\to\infty$). Despite this
caveat, Fig. \ref{gcorr_compare} shows that provided we consider 
$\beta\lesssim O(1)$, then setting $\zeta=0$ does not affect the VSI
significantly.  
  
\section{Linear problem in the isothermal limit}\label{iso_discuss}  
Here we summarize selected results for isothermal
perturbations ($\beta\equiv 0$) in vertically isothermal disks 
($\Gamma=1$) in the fully-radially local approximation
($\zeta=0$). In this case Eq. \ref{ode_Q}
becomes    
\begin{align} 
  Q = W 
\end{align}
(i.e. $\delta P = c_s^2\delta \rho$)
. For isothermal perturbations it is 
simpler to work with an equation for $W$ by substituting $Q=W$ into
Eq. \ref{ode_w} and using Eq. \ref{ode_vz} to eliminate $\delta v_z$. 
In this case, we shall not yet make the low frequency
approximation, but first make the  Keplerian approximation. We
obtain, in terms of dimensionless variables, 
\begin{align}
  0 = W^{\prime\prime} + \left[\ln\rho^\prime - \frac{\ii h q \hat{k} f(\zhat)
      }{1-\hat{\upsilon}^2} \right]W^\prime +
  \hat{\upsilon}^2\left(1+\frac{\hat{k}^2}{1-\hat{\upsilon}^2}\right)W,\label{iso_ode0} 
\end{align}
where for discussion purposes we have defined $f(\zhat)$ such that
\begin{align}\label{fz_shear}
  \frac{d\Omega^2}{d\hat{z}} = h^2q f(\hat{z})\OmK^2.
\end{align}
By comparing Eq. \ref{fz_shear} with Eq. \ref{vertical_shear_ex}, we
see that $f(\zhat)= 
\zhat\left(1+h^2\zhat^2\right)^{-3/2}$. More generally, though,
$f$ may be regarded as a representation of the vertical
shear profile. Physically, we expect there is a maximum value of 
$|d\Omega/dz|$, the existence of which should limit the growth rate of
the VSI,  as remarked in \S\ref{vshear_def}. We explicitly demonstrate
this below.     

\subsection{Maximum growth rate  in the low-frequency limit}
Here we consider the low-frequency limit $|\hat{\upsilon}|\ll 1$ and 
show that the growth rate is limited by the maximum vertical
shear rate in the domain. 
We approximate Eq. \ref{iso_ode0} as 
\begin{align}
  0 = W^{\prime\prime} + \left[\ln\rho^\prime - \ii h q \hat{k}
    f(\zhat)\right]W^\prime +
  \hat{\upsilon}^2\left(1+\hat{k}^2\right)W. \label{iso_ode1}
\end{align}
We multiply Eq. \ref{iso_ode1} by $\rho W^*$ and integrate vertically from
$\zhat=\zhat_1$ to $z=\zhat_2$. We neglect boundary 
terms when integrating by parts, by assuming $W$ or
$W^\prime$ vanishes at the boundaries, or that the boundaries are 
sufficiently far away so that the boundary terms are negligible because of the
decaying background density with increasing height. Then,
\begin{align}
  \hat{\upsilon}^2\left(1+\hat{k}^2\right)\int_{\zhat_1}^{\zhat_2}\rho|W|^2d\zhat
  =\int_{\zhat_1}^{\zhat_2}\rho|W^\prime|^2d\zhat 
  +\ii h q \hat{k}\int_{\zhat_1}^{\zhat_2}\rho f(\zhat) W^*W^\prime d\zhat.\label{integral_relation1}
\end{align}
It follows that for instability ($\imag\hat{\upsilon}>0$), it is necessary to
have $q\neq0$ or more generally $d\Omega/dz\neq 0$, i.e. there must
be vertical shear. 

The real and imaginary parts of 
Eq. \ref{integral_relation1} are
\begin{align}
  \left(\hat{\omega}^2-\hat{\sigma}^2\right)\left(1+\hat{k}^2\right) 
  \int_{\zhat_1}^{\zhat_2}\rho|W|^2d\hat{z} -
  \int_{\zhat_1}^{\zhat_2}\rho|W^\prime|^2d\hat{z}
  \notag
  =\real\left[\ii
    h\hat{k}q \int_{\zhat_1}^{\zhat_2}\rho
    f(\hat{z}) W^*W^\prime d\zhat\right], \\
   2\hat{\omega}\hat{\sigma}\left(1+\hat{k}^2\right)
  \int_{\zhat_1}^{\zhat_2}\rho|W|^2d\hat{z}
  =\imag\left[\ii
    h\hat{k}q \int_{\zhat_1}^{\zhat_2}\rho
    f(\hat{z}) W^*W^\prime d\hat{z}\right],
\end{align}
where we recall $\hat{\omega} = \real \hat{\upsilon}$ and
$\hat{\sigma}=\imag\hat{\upsilon}$. 
Adding the square of these equations give
\begin{align}
  \left[|\hat{\upsilon}|^2\left(1+\hat{k}^2\right)
    \int_{\zhat_1}^{\zhat_2}\rho|W|^2d\hat{z} -
    \int_{\zhat_1}^{\zhat_2}\rho\left|W^\prime \right|^2d\hat{z}\right]^2
  +4\hat{\sigma}^2\left(1+\hat{k}^2\right) 
  \int_{\zhat_1}^{\zhat_2}\rho
  |W|^2d\hat{z}\int_{\zhat_1}^{\zhat_2}\rho\left|W^\prime \right|^2d\hat{z}
  =\left|\ii
    h\hat{k}q\int_{\zhat_1}^{\zhat_2}\rho
    f(\hat{z}) W^*W^\prime d\hat{z}\right|^2.
\end{align}
It is clear that
\begin{align}\label{sigma_finite_domain} 
  4\hat{\sigma}^2\left(1+\hat{k}^2\right) 
  \int_{\zhat_1}^{\zhat_2}\rho
  |W|^2d\hat{z}\int_{\zhat_1}^{\zhat_2}\rho\left|W^\prime
  \right|^2d\hat{z} 
  \leq\left|
    h\hat{k}q\int_{\zhat_1}^{\zhat_2}\rho
    f(\hat{z}) W^*W^\prime d\hat{z}\right|^2.
\end{align}
On the left hand side of this inequality, we apply the Cauchy-Schwarz
inequality to obtain
\begin{align}
  \left( \int_{\zhat_1}^{\zhat_2}\rho
    |W|\left|W^\prime \right|d\hat{z}\right)^2\leq
  \int_{\zhat_1}^{\zhat_2}\rho 
  |W|^2d\hat{z}\int_{\zhat_1}^{\zhat_2}\rho\left|W^\prime \right|^2d\hat{z}.
\end{align}
On the right hand side of Eq. \ref{sigma_finite_domain} we have
\begin{align}
  \left|\int_{\zhat_1}^{\zhat_2}\rho
    f(\hat{z}) W^*W^\prime d\hat{z}\right|\leq \int_{\zhat_1}^{\zhat_2}\rho
  \left|f(\hat{z})W^*W^\prime \right|d\hat{z}
  \leq
  \mathrm{max}\left(|f|\right)\int_{\zhat_1}^{\zhat_2}\rho
  |W|\left|W^\prime \right|d\hat{z},
\end{align}
where $\mathrm{max}(|f|)$ is the maximum value of $|f|$ in
$\zhat\in[\zhat_1,\zhat_2]$. Inserting these inequalities into
Eq. \ref{sigma_finite_domain} gives
\begin{align}\label{max_growth}
  |\hat{\sigma}|\leq
  \frac{h |\hat{k} q|}{2\sqrt{1+\hat{k}^2}}\mathrm{max}(|f|) < \frac{h |q|}{2}\mathrm{max}(|f|) . 
\end{align}
It
follows that the maximum possible growth rate of unstable modes,
satisfying the above boundary conditions, is limited by the maximum
vertical shear rate in the domain considered,
\begin{align}
  |\sigma| < \mathrm{max} \left|r\frac{d\Omega}{dz}\right|, 
\end{align}
as expected on physical grounds. Note that if the thin-disk
approximation is used in an infinite domain, then the maximum growth
rate is unbounded (since in that case $\mathrm{max}|f|\to\infty$).
 However, large growth rates invalidate the low-frequency
  approximation and the above analysis breaks down.


In practice, one might consider a vertical domain of a few scale 
heights in a thin disk with $|q|=O(1)$. Then 
$f\simeq \zhat$, so that $\mathrm{max}|f| = O(1)$, implying a
maximum growth rate $O(h \OmK)$, consistent with numerical
results. 

\subsection{Explicit solutions in the thin-disk limit}\label{iso_explicit}
Here we assume a thin disk ($h\ll1$) so that $\ln\rho\simeq
-\zhat^2/2$ and $f(\zhat)\simeq \zhat$. However, we do not assume 
low frequency from the onset. Then Eq. \ref{iso_ode0} becomes 
\begin{align}\label{iso_ode3}
  W^{\prime\prime} - \left(1 + \frac{\ii qh \hat{k}}{1-\hat{\upsilon}^2}\right)\hat{z}W^\prime  +
  \hat{\upsilon}^2\left(1+\frac{\hat{k}^2}{1-\hat{\upsilon}^2}\right)W = 0.
\end{align}
We remark that taking the low frequency limit and considering
$\hat{k}^2\gg 1$, Eq. \ref{iso_ode3} becomes equivalent to Eq. 39 in
\citetalias{nelson13} or Eq. 28 in \citetalias{barker15}, although we have taken a
different route.    
 

We seek power series solutions to Eq. \ref{iso_ode3}, 
\begin{align}
  W(\zhat) = \sum_{l=0}^\infty a_l\zhat^l. 
\end{align}
Then the coefficients $a_l$ are given by the recurrence relation
\begin{align}
  (l+2)(l+1)a_{l+2} +
  \left[\hat{\upsilon}^2\left(1+\frac{\khat^2}{1-\hat{\upsilon}^2}\right)
    - l\left(1+\frac{\ii h q
        \hat{k}}{1-\hat{\upsilon}^2}\right)\right] a_l = 0. 
\end{align}
We demand the series to terminate  at $l=L$, i.e. a polynomial of
order $L$, so that the vertical kinetic energy density remain bounded as 
$|\zhat|\to\infty$.  Then the eigenfrequency $\hat{\upsilon}$ is given via 
\begin{align}
\hat{\upsilon}^4 - \left(L+1+\khat^2\right)\hat{\upsilon}^2 + L\left(1 +
  \ii h q \khat\right) = 0. \label{simple_growth}
\end{align}

Note that we have applied a regularity condition at infinity, since
the vertically isothermal disk has no surface. If vertical boundaries
are imposed at finite height, as done in numerical calculations, then
the above solution needs to be modified to match the desired boundary
conditions. This enables the `surface modes' seen in numerical
calculations \citepalias{barker15}. 

\subsubsection{Stability without vertical shear}\label{stable_novshear}
In the absence of vertical shear $q=0$, Eq. \ref{simple_growth} can be
written as \begin{align}
  \hat{\upsilon}^2\khat^2 =
  \left(L-\hat{\upsilon}^2\right)\left(1-\hat{\upsilon}^2\right), \label{iso_disp_full}
\end{align}
which is just the dispersion relation for axisymmetric isothermal waves in a
vertically isothermal disk (e.g. \citealt{okazaki87}; \citealt{takeuchi98}; \citealt{tanaka02}; 
\citealt{zhang06}; \citealt{ogilvie13}; \citealt{barker14}; \citetalias{barker15}). In 
this case the solutions are Hermite polynomials, $W\propto
\He_L(\zhat)$.  The eigenfrequency $\hat{\upsilon} = \hat{\omega}$ is real and the disk
is stable. The low frequency branch of Eq. \ref{iso_disp_full} are 
inertial waves \citep{balbus03}. For 
$|\hat{\omega}|\ll 1$ and $L\geq 1$ the dispersion relation is
$\hat{\omega}^2\khat^2 = L$, or $\hat{\omega}\propto \khat^{-1}$ for
fixed $L$. This inverse relation has been qualitatively
observed in numerical simulations of \cite{stoll14}. 
Note that $L=M+1$, where $M\geq 0$ is the mode number used for analytical
discussion (\S\ref{analytical}) based on the reduced equation for
$\delta v_z$, rather than for $W$ as considered here. 

\subsubsection{Instability with vertical shear}
The VSI corresponds to unstable inertial waves. This
is readily obtained for large $\khat^2$ by balancing
the last two terms of Eq. \ref{simple_growth} to give the low
frequency branch. Then 
\begin{align}
  \hat{\upsilon}^2 \simeq L\left(\frac{1+\ii q h
       \hat{k}}{L+1+\hat{k}^2}\right), \label{simple_growth2}
\end{align}
which is equivalent to Eq. 34 of \citetalias{barker15} in the limit
$\khat^2\gg L,\,1$. This signifies instability for
$L\geq1$ since we can choose the sign 
of the square root to make $\imag\hat{\upsilon}>0$.  These
are the low frequency unstable modes seen in
Fig. \ref{compare_modes_iso_kx10} and
Fig. \ref{compare_modes_iso_kx30}, for which $\sigma\propto|\omega|$.  

\section{Analytic dispersion relation
with thermal relaxation}
\subsection{Coefficients}\label{relax_coeff}
The coefficients of the dispersion relation Eq.\ \ref{relax_disp} are:
\begin{subequations}\label{coeffs}\begin{align}
  &c_0 = M(M+1)A^2,\\
  &c_1 = \ii\beta\left\{\left(1-\gamma\right)\left[1 +
      \khat^2\left(1+2M\right)^2 - 4 \ii h q\khat M (M+1)\right] 
    - 2A^2\gamma M (M+1)\right\},\\
  &c_2 = \left(\khat^2 + 1\right)A + \beta^2\left\{(1-\gamma)\left[1
      + \gamma \khat^2(1+2M)^2 - 4\ii h q \khat \gamma M(M+1)
    \right]
    -\gamma^2 A^2 M(M+1)
  \right\},\\
  &c_3 = \beta\left\{h q \khat + \gamma \left[\ii + h q
      \khat \left(1+2\khat^2\right)\right] - 3\ii - 2\ii
    \khat^2\right\},\\
  &c_4 =
  \beta^2\left(1+\gamma\khat^2\right)\left[\gamma\left(1-\ii h q
    \khat\right)-2\right] - \left(1+\khat^2\right)^2,\\
&c_5 = 2\ii\beta\left(1+\khat^2\right)\left(1+\gamma\khat^2\right),\\
&c_6 = \beta^2\left(1+\gamma\khat^2\right)^2.
\end{align}\end{subequations}

\subsection{Finding marginal stability}\label{disp_neut_limit}
To investigate marginal stability we set $\hat{\sigma} = 0$ so the
frequency, $\hat{\upsilon}=\hat{\omega}_c$, is real; and $\beta =
\beta_c$,  the cooling time for marginal stability. 
We take the short radial wavelength limit, $\khat^2\gg 1$, of the
coefficients in Eq.\ \ref{coeffs}. We consider $M \lesssim O(1)$ and
assume $\beta_c \ll \khat$.  The real and 
imaginary parts of Eq. \ref{relax_disp} then give relations for $\hat{\omega}_c$ and $\beta_c$ as 
\begin{align}
  0 =&M(M+1)(1-h^2 q^2 \khat^2) + 4\beta_c h q M(M+1)\left(\hat{\omega}_c\khat\right) 
 +\left[1 +
    \gamma\beta_c^2\left(1-\gamma\right)(1+2M)^2\right]\left(\hat{\omega}_c\khat\right)^2 \label{relax_cond1} \\
 &+ 2h q \gamma\beta_c \left(\hat{\omega}_c\khat\right)^3 -  \left(\hat{\omega}_c\khat\right)^4 
  + \beta_c^2\gamma^2\hat{\omega}_c^2
  \left(\hat{\omega}_c\khat\right)^4, \notag \\
   0=& 2h q M (M+1)\khat +
   \beta_c(1-\gamma)(1+2M)^2\khat\left(\hat{\omega}_c\khat\right) 
   + h q \khat \left(\hat{\omega}_c\khat\right)^2 -
   2\beta_c\hat{\omega}_c\left(\hat{\omega}_c\khat\right)^2
   - h q
   \gamma^2\beta_c^2\hat{\omega}_c\left(\hat{\omega}_c\khat\right)^3 \label{relax_cond2} \\
   &+
   2\beta_c\gamma\hat{\omega}_c\left(\hat{\omega}_c\khat\right)^4 \notag. 
\end{align}
We recall that in the low frequency and thin-disk approximations, $h,
|\hat{\omega}_c| \ll 1$. We note that $|q|=O(1)$ and 
$\gamma>1$ but is $O(1)$. We assume $|\hat{\omega_c}\khat|=O(1)$,
since for inertial waves $|\hat{\omega}\khat|\simeq \sqrt{1 + M}$. Finally, we further assume 
$\beta_c \ll 1$, to be justified \emph{a posteriori}, to give 
Eq. \ref{relax_cond_simp1}---\ref{relax_cond_simp2}.   

\subsection{Maximum critical cooling time}\label{max_cool}   
Here we show that  for sufficiently thin disks the longest critical 
cooling time is that for the $M=0$ or fundamental mode. This allows
us to focus on the fundamental mode to obtain an overall cooling
requirement for the VSI.      

Consider the simplified dispersion relations for marginal stability,
Eq. \ref{relax_cond_simp1}---\ref{relax_cond_simp2}. We write 
$X= \hat{\omega}\khat$, $\theta = (hq\khat)^2$ and  
treat $M$ as a continuous variable. We find from
Eq. \ref{relax_cond_simp1} that  
\begin{align}
  2 X \frac{d X}{dM} = \frac{X^2(1-\theta)(2M+1)}{X^2 +
  2M(M+1)(1-\theta)}, 
\end{align}
and from Eq. \ref{relax_cond_simp2} that
\begin{align}
  2X\frac{dX}{dM} = \left[X^2 +
  2M(M+1)\right]\left(\frac{d\ln{\beta_c}}{dM} + \frac{4}{2M+1} +
  \frac{d\ln{X}}{dM}\right) - 2(2M+1). 
\end{align}
We eliminate $dX/dM$, making use of Eq. \ref{relax_cond_simp1} in the process, to obtain 
\begin{align}
  (2M+1)\frac{d\ln{\beta_c}}{dM} 
  = -\frac{(1-\theta)M(M+1)\left[ 2(2M+1)^2 + 12X^2\right]
  + (3+\theta)X^2}{2 \left[X^2 +
  2M(M+1)(1-\theta)\right]\left[X^2 + 2M(M+1)\right]}.  
\end{align}
Hence,
\begin{align}\label{dbeta_dm}
  \frac{d\beta_c}{dM}<0, 
\end{align}
for all $M\geq 0$ if $\theta\leq 1$, which imply 
$\mathrm{max}(\beta_c)$ occurs at $M=0$. This conclusion may also be reached by explicitly solving 
Eq. \ref{relax_cond_simp1}---\ref{relax_cond_simp2} with $\theta$ as a small parameter. 
For fixed $\khat$ the condition $\theta\leq 1$ can be satisfied for 
sufficiently small $h$. All such modes are stabilized if
$\beta>\beta_c(M=0)\equiv\beta_\mathrm{crit}$. 

This result highlights the importance of the fundamental mode  
--- it is the most difficult mode to stabilize with finite
cooling. Furthermore, for $M=0$ we may obtain the expression  
for $\beta_\mathrm{crit}$ from the dispersion relations 
Eq. \ref{relax_cond1}---\ref{relax_cond2} without  
assuming it is much less than unity at the outset or place
restrictions on $\theta$.